\documentclass[prd,tightenlines,nofootinbib,superscriptaddress]{revtex4}

\usepackage{amsfonts,amsmath,amssymb,amsthm}
\usepackage{hyperref}
\usepackage{color,psfrag}
\usepackage[dvips]{graphicx}
\usepackage{psfrag}

\newcommand{\C}{{\mathbb C}}
\newcommand{\N}{{\mathbb N}}
\newcommand{\R}{{\mathbb R}}
\newcommand{\Z}{{\mathbb Z}}

\newcommand{\cA}{{\mathcal A}}
\newcommand{\cE}{{\mathcal E}}

\newcommand{\cG}{{\mathcal G}}

\newcommand{\cK}{{\mathcal K}}

\newcommand{\cM}{{\mathcal M}}
\newcommand{\cN}{{\mathcal N}}

\newcommand{\cC}{{\mathcal C}}
\newcommand{\cQ}{{\mathcal Q}}
\newcommand{\cS}{{\mathcal S}}
\newcommand{\cO}{{\mathcal O}}
\newcommand{\cR}{{\mathcal R}}

\newcommand{\cW}{{\mathcal W}}
\newcommand{\SU}{\mathrm{SU}}
\newcommand{\Spin}{\mathrm{Spin}}

\newcommand{\U}{\mathrm{U}}

\newcommand{\w}{\wedge}

\newcommand{\be}{\begin{equation}}
\newcommand{\ee}{\end{equation}}
\newcommand{\beq}{\begin{eqnarray}}
\newcommand{\eeq}{\end{eqnarray}}
\newcommand{\bea}{\begin{eqnarray}}
\newcommand{\eea}{\end{eqnarray}}
\newcommand{\nn}{\nonumber}

\newcommand{\mat} [2] {\left ( \begin{array}{#1}#2\end{array} \right ) }

\renewcommand{\u}{{\mathfrak u}}

\newcommand{\la}{\langle}
\newcommand{\ra}{\rangle}

\newcommand{\tr}{{\mathrm Tr}}
\newcommand{\f}{\frac}

\def\bZ{{\overline{Z}}}
\newcommand{\bz}{\overline{z}}
\newcommand{\bw}{\overline{w}}

\def\nn{\nonumber}
\def\pp{\partial}
\def\arr{\rightarrow}

\def\ka{\kappa}
\def\vphi{\varphi}
\def\eps{\epsilon}
\def\om{\omega}

\newcommand{\id}{\mathbb{I}}
\def\vV{\vec{V}}
\def\vsigma{\vec{\sigma}}
\def\vLambda{\vec{\Lambda}}
\def\hu{\hat{u}}
\def\tOmega{\widetilde{\Omega}}
\def\bE{\bar{E}}
\def\bF{\bar{F}}
\def\tU{\widetilde{U}}
\def\tV{\widetilde{V}}
\def\tW{\widetilde{W}}
\def\tz{\widetilde{z}}
\def\tw{\widetilde{w}}
\def\cz{{\cal Z}}
\def\bcz{\bar{{\cal Z}}}
\def\bU{\bar{U}}
\def\hcC{\hat{{\cal C}}}
\def\tcC{\widetilde{{\cal C}}}

\def\Ea{E^{\alpha}}
\def\Fa{F^{\alpha}}
\def\Eb{E^{\beta}}
\def\Fb{F^{\beta}}
\def\hEa{\hat{E}^{\alpha}}
\def\hFa{\hat{F}^{\alpha}}
\def\hEb{\hat{E}^{\beta}}
\def\hFb{\hat{F}^{\beta}}
\def\Ua{U^{\alpha}}
\def\Ub{U^{\beta}}
\def\hcD{\widehat{{\cal D}}}
\def\hcQ{\widehat{{\cal Q}}}
\def\te{\tilde{e}}
\def\tf{\tilde{f}}
\def\htcQ{\widehat{\tilde{{\cal Q}}}}

\begin{document}

\title{Classical Setting and Effective Dynamics for Spinfoam Cosmology}

\author{{\bf Etera R. Livine}}\email{etera.livine@ens-lyon.fr}
\affiliation{Laboratoire de Physique, ENS Lyon, CNRS-UMR 5672, 46 All\'ee d'Italie, Lyon 69007, France}
\affiliation{Perimeter Institute, 31 Caroline St N, Waterloo ON, Canada N2L 2Y5}
\author{{\bf Mercedes Mart\'\i n-Benito}}\email{mercedes@aei.mpg.de}
\affiliation{MPI f. Gravitational Physics, Albert Einstein Institute, Am M\"uhlenberg 1, D-14476 Potsdam, Germany}

\date{\today}

\begin{abstract}
We explore how to extract effective dynamics from loop quantum gravity and spinfoams truncated
to a finite fixed graph, with the hope of modeling  symmetry-reduced gravitational
systems. We particularize our study to the 2-vertex graph with $N$ links. We describe the canonical
data using the recent formulation of the phase space in terms of spinors, and implement a
symmetry-reduction to the
homogeneous and isotropic sector. From the canonical point of view, we construct a
consistent Hamiltonian for the model and discuss its relation with Friedmann-Robertson-Walker
cosmologies. Then, we analyze the dynamics from the spinfoam approach.
We compute exactly the transition amplitude between initial and final coherent spin
networks states with support on the 2-vertex graph, for the choice of the simplest two-complex
(with a single space-time vertex).
The transition amplitude verifies an exact differential equation that agrees with the Hamiltonian
constructed previously. Thus, in our simple setting we clarify the link between the canonical and
the covariant formalisms.

\end{abstract}

\maketitle

\setcounter{tocdepth}{2}
\tableofcontents


\section*{Introduction}

Loop quantum gravity \cite{lqg} and spinfoams \cite{SFreview} form together a proposition for a
well-defined framework for quantum gravity. While loop gravity is the canonical definition of the
theory describing the evolution of quantum states of space geometry, the spinfoam approach provides
the covariant formulation of the theory, such that it describes the quantum structure of space-time.
More precisely, in both loop quantum gravity and spinfoams the quantum states of geometry are given
by spin network states which have support on some graphs. The space of spin networks living on
all possible graphs (up to diffeomorphisms) provide a basis for the kinematical Hilbert space of the
theory. In the canonical framework, the evolution is implemented by a Hamiltonian operator. There
exist both graph-changing and non-graph-changing proposals for this operator depending on the
precise regularization scheme and implementation of the Hamiltonian constraint at the quantum
level, although it is usually assumed that it acts on the states by changing the underlying graph.
The spinfoam approach, on the other hand, defines transition amplitudes between spin network states
living on arbitrary graphs through the construction of a covariant discretized path integral.
This program towards quantum gravity faces three main issues: a clear definitive
definition of the dynamics, the derivation and analysis of the semi-classical regime of the theory
where we should recover fluctuations of the gravitational field around flat space-time, and a
consistent method for extracting quantum gravity corrections and predictions.

\medskip

Here, we propose to discuss these topics in the context of loop quantum gravity and spinfoams
truncated to a finite fixed graph. Of course, there are in principle two possible scenarios: a fixed graph dynamics and a graph changing dynamics. We believe that the quantum gravity dynamics will in the end mix these two scenarios; but we nevertheless think that it would be enlightening to explore to what kind of phenomenology does each approach lead separately, in order to distinguish their effects and later understand the appropriate mix of these two ingredients involved in the various quantum gravity regimes. In the present work, we focus on the fixed graph dynamics postponing the investigation of graph changing dynamics to future work.

From the canonical point of view, this requires defining a Hamiltonian on a fixed graph (without assuming that it comes from the truncation of a graph-changing or a non-graph-changing Hamiltonian). From the covariant point of view, this requires considering transition amplitudes between initial and final spin network states with support on the same graph. The theory restricted to spin network states living on this given graph, is thus truncated to a finite number of degrees of freedom. Our hope is that restricting the theory to a finite fixed graph would allow to formulate physically relevant mini-superspace models for (loop) quantum gravity. Indeed, mini-superspace models in general relativity are restrictions to certain families of 4-metrics parameterized by a finite number of parameters and satisfying certain
symmetries or properties which made them relevant to some particular physical context. We expect
that the development of such mini-superspace models of loop quantum gravity will lead to realistic
models for quantum cosmology and thus allow precise cosmological predictions from loop gravity and
spinfoam models.

Let us emphasize here that we do not yet have a full consistent theory of (loop) quantum gravity\footnotemark.
\footnotetext{There nevertheless exists a mathematically well-defined formulation of the EPRL-FK spinfoam models \cite{EPRL,FK} with a proper definition of the quantum states of geometry and the transition amplitudes between them. However it can not yet be considered as a fully well-defined theory of quantum gravity since we do not fully understand its physical meaning (summing over what kind of geometries?), its renormalization flow, how to localize or implement some symmetry-reduction, how to couple matter fields or how to consistently extract the quantum corrections to general relativity.}
We can not identify mini-superspaces as families of appropriately symmetric metrics satisfying the quantum Einstein equations (i.e the Hamiltonian constraints in our framework) since we do not have such definite equations at our disposal. Instead we try to explore some sectors of loop quantum gravity and spinfoam models that look similar to classical mini-superspaces of general relativity, both at the level of the classical phase space and degrees of freedom and their geometrical interpretation, and see how to define an appropriate dynamics corresponding to their guessed classical counterpart or check if the existing spinfoam models give in this truncation some transition amplitudes comparable to the expected classical dynamics. We hope that such an approach will lead to some insights in how to define the dynamics of the full theory or might even lead to the derivation of physically relevant mini-superspaces of quantum gravity without need to appeal to the full theory but nevertheless phenomenologically interesting (such like loop quantum cosmology \cite{LQCreview}).

Following this logic, investigating the loop quantum gravity dynamics on a simple fixed graph is the simplest possibility in order to define mini-superspace models and it needs to be investigated before moving on to more complex constructions.

\medskip

Our strategy is to choose some simple graph $\Gamma$, to analyze and describe both
the classical phase space  and the space of quantum states living on this graph, to define and study
the dynamics both at the classical and quantum levels using the loop gravity ansatz for the
Hamiltonian or the transition amplitudes of spinfoam models, and finally to understand how it can be
mapped (or not) on some cosmological models (or other interesting situations).
Working on such simplified setting with a fixed underlying graph allows to define rigourously the
dynamics of the classical data and quantum states and more generally to investigate the
possible dynamics that one can define. We also hope that studying such toy models for (loop)
quantum gravity will allow to understand more about the geometrical interpretation of the quantum
states and the construction  of coherent states, about the transition to the semi-classical regime,
and about the structure of spinfoam amplitudes for the evolution of spin networks. From this point
of view, the fact that such models could lead to realistic cosmological models or to the dynamics of
other symmetry-reduced geometries, and that we could possibly extract quantitative quantum gravity
effects in this context would be a bonus.

So what are the various ways to define the dynamics on a fixed graph? Here is a list of the various possibilities:
\begin{enumerate}
\item Discretizing appropriately and regularizing the Hamiltonian constraints of loop quantum gravity: this is the standard method.
\item Combine all the gauge-invariant and appropriately local interactions that can be defined on the graph, see their various actions and select the ones which correspond the best to the expected space diffeomorphisms and evolution in time: this is the natural extension of the standard method, where we also include the possible terms and effective corrections that arise from renormalization or coarse-graining of the originally defined discretized Hamiltonian.
\item Extract an effective classical Hamiltonian from the spinfoam transition amplitudes between coherent spin network states peaked on classical phase space points: typically after computing the transition amplitudes for a given space-time triangulation, one can identify the differential equations that they satisfy and interpret them as the quantum Hamiltonian constraints defining the physical states, then we can finally compute the corresponding classical  Hamiltonian (evaluating the quantum operator on coherent states).
\item Identify (a sector of) the phase space on a given fixed graph  with a classical mini-superspace sector of general relativity on the basis on the geometrical interpretation of the degrees of freedom and use the symmetry-reduced dynamics of general relativity adapted to our variables.
\end{enumerate}
We will discuss the generic procedures and methods behind this fixed finite graph approach, but we focus in practice on the case of the 2-vertex graph, which has been shown to be somewhat related to Friedmann-Robertson-Walker cosmology in earlier works \cite{LQGcosmo1,LQGcosmo2,un3,un5,SFcosmo,SFcosmoLambda}.
In this context, we will show that these four ways of defining the dynamics on the 2-vertex graph all lead to the same answer, which confirms the interpretation of the resulting model as an effective quantum FRW cosmology (in vaccuum or coupled to a massless scalar field).  We hope in the future to be able to investigate more complex graph and generalize our methods to derive more realistic cosmological models with matter fields and inhomogeneities.

\medskip

Let us insist on the fact that our strategy is different from the more usual approach of loop quantum cosmology. Indeed, in loop quantum cosmology, one starts from the full phase space of general relativity, formulated in terms of the triad-connection variables of loop gravity, and defines the reduction to cosmological metrics through the implementation of homogeneity and isotropy (according to the considered model) using appropriate distributions on the phase space \cite{lqc_def}. On the other hand, we are starting here from a finite dimensional phase space of the loop gravity's degrees of freedom on a fixed finite graph and investigating if it is possible to implement an equivalent of the requirements of homogeneity and isotropy and define the equivalent of a cosmological setting. The goal is to address the issue of whether or not it is possible to recover (loop) quantum cosmology from a truncation of loop quantum gravity to a fixed finite graph (without considering complex graphs with many vertices or using graph changing dynamics).

We will see that it is possible to partly recover the FRW loop quantum cosmology from the loop gravity dynamics on the simplest possible graph with two vertices: we will indeed recover the old loop cosmology dynamics and not the improved dynamics (which gives more physically-plausible results especially about the singularity resolution at the Big Bang).
We point out that a similar approach has also been proposed by other authors in \cite{italian}, but they are focusing on the use of cubic lattices as graphs.

\medskip

The first section describes in detail the
classical kinematical structures of loop gravity on an arbitrary fixed graph $\Gamma$. We review the
recently developed approach  of parameterizing the classical phase space with spinor variables
\cite{twisted2,un5,spinor1,spinor2} and discuss its relation to the other parameterizations in terms
of the standard loop gravity holonomy-flux variables, in terms of twisted geometries
\cite{twisted1,twisted2}, and finally in terms of $\U(N)$-covariant variables
\cite{un1,un2,un3,un4,un5,spinor1}. Each set of variables allows to insist on certain aspect of the
kinematics and clarifies the geometrical interpretation of the phase space. This is necessary in order to introduce the relevant definitions and notations for the rest of the paper.

In the second section, we apply the generic method to the particular case of the graph with 2
vertices and $N$ edges linking them, which is the simplest graph on which one can formulate the
theory. We describe the phase space on this 2-vertex graph and define the symmetry reduction to the homogeneous and isotropic sector, following the $\U(N)$-symmetry proposal of \cite{un3,un5}.

Then in section three, we define and implement classical dynamics on this 2-vertex graph consistent with the reduction to the homogeneous and isotropic sector.
We provide a generic $\U(N)$-invariant ansatz for Hamiltonian and prove that the loop quantum gravity Hamiltonian constraint particularized to the 2-vertex graph (as constructed by Rovelli and Vidotto in \cite{LQGcosmo1}) is a special case of that ansatz.
Furthermore, we show the relation between this truncated loop gravity classical dynamics, the (geometrical part of the) Hamiltonian for Friedmann-Robertson-Walker (FRW) cosmology and the effective dynamics derived from loop quantum cosmology. Let us remark that here we focus on the vacuum case (and on the simplest case of the coupling to a massless scalar field). In the future we will need to include matter in order to get true models for cosmology. We further discuss and explain the limitations of the similarities between our model and cosmology, and the failures of our na\"\i ve 2-vertex graph Hamiltonian at large scales when we take into account a non-vanishing curvature or cosmological constant.

In the fourth and final section, we investigate the effective classical dynamics on the 2-vertex
graph induced by the quantum transition amplitudes of spinfoam models applied to coherent spin
network states on the 2-vertex graph. This clarifies and extends the previous results obtained by
Bianchi, Rovelli and Vidotto in \cite{SFcosmo}. We compute exactly the spinfoam transition
amplitudes and identify the exact differential equations that they satisfy. Then we show how these
differential equations lead back to the 2-vertex Hamiltonian defined previously at the classical
level. We discuss how to generalize these results beyond the 2-vertex graph and how our procedure is
related to the study of recursion relations and invariance of spinfoam amplitudes
\cite{SFrecursion_simone,SFrecursion_valentin,SFrecursion_final}. Indeed, from the spinfoam point of
view, such recursion relations are understood to be equivalent to the dynamics of the theory and
have been shown to translate to differential equations when applied to coherent spin network states.
These are exactly the differential equations that we recover in our 2-vertex graph setting and that
we show to be related to the Hamiltonian constraint of flat FRW cosmology. We also discuss how
to modify the spinfoam amplitude to take non-vanishing curvature into account. This final step
finally shows the coherence of the spinfoam cosmology approach initiated in \cite{SFcosmo} with the
canonical point of view, although much work is needed to go beyond our na\" ive truncation to the
2-vertex graph and its homogeneous and isotropic sector.

\section{Classical Phase Space of Loop Gravity}
\label{sec1}

Loop quantum gravity is formulated in terms of spin network states leaving on graphs. A spin network state on a graph $\Gamma$ is defined as a gauge-invariant function of $\SU(2)$ group elements $g_e$ leaving on the edges $e\in\Gamma$. The $\SU(2)$ group elements are physically the holonomies of the Ashtekar-Barbero connection along the edges of the graph. The Hilbert space of these wave-functions $\varphi_\Gamma(g_e)$ provides a quantization of the  phase space of holonomy-flux variables $(g_e,X_e)$, where the holonomies $g_e$ act by multiplication and fluxes $X_e$ act as derivation operators. Following recent developments on the $\U(N)$ formalism for intertwiners \cite{un0,un1,un2,un3,un4,un5} and twisted geometries \cite{twisted1,twisted2}, it has been understood that the holonomy-flux algebra defined in terms of the $(g_e,X_e)$ variables on a given graph $\Gamma$ can be re-written using spinor variables $z_e^v$ living around each vertex $v$ on the edges $e$ \cite{un5,twisted2,spinor1,spinor2}. Then wave-functions will be holomorphic functions of these spinor variables.

In this section, we will quickly review this construction, adding some new material especially on the repackaging of the phase space structure in suitable action principles, and introduce all the relevant notations for the rest of the paper. We will define the phase space of the spinors $z_e^v$, provide an action principle encoding the canonical Poisson structure and constraints generating the $\SU(2)$ gauge-invariance, and explain how to recover the standard holonomy and flux observables from these variables. Finally, we will discuss how to endow this kinematical structure with dynamics, thus defining effective classical dynamical models  of loop quantum gravity on fixed graphs.

Finally, the explicit relation between the spinor variables and the twisted geometry parameterization can be found in appendix \ref{twisted_app}.

\subsection{Spinor Networks and Phase Space on a Fixed Graph}

In loop quantum gravity the spin network states provide an orthonormal basis for the kinematical
Hilbert space of the theory.
%
%
A spin network state has support on a given closed graph and consists in  the coloring of the graph's edges and vertices with appropriate quantum numbers. More precisely, every edge is colored with an irreducible representation of $\SU(2)$ and every vertex with an intertwiner, namely a $\SU(2)$-invariant state living in the tensor product of the $\SU(2)$ representations meeting at that vertex.
This kinematical Hilbert space is usually formulated as the quantization of the phase space of holonomy-flux variables, which are discretized observables for the connection and triad field of loop quantum gravity.
Recently, new descriptions for the kinematical phase space of loop gravity have been devised in order to understand better the geometrical interpretation of the spin network states, also with the aim of constructing suitable coherent states of discrete geometry. Such descriptions, as the twisted geometries and the $\U(N)$ formalism for intertwiners, have converged to a description of the phase space in terms of spinor networks.

Let us start with a closed oriented graph $\Gamma$, with $E$ oriented edges and $V$ vertices.
For simplicity's sake, we choose it connected, else all the definitions will still apply to each
connected component of the graph. Now around each vertex $v$, we associate a spinor variable
$z^v_e\in\C^2$ to each edge $e$ attached to $v$. Equivalently, this amounts to associating to each
edge $e$ two spinors, $z_e^s\equiv z_e^{s(e)}$ and $z_e^t\equiv z_e^{t(e)}$, the former attached to
the source vertex $s(e)$ of the edge and the latter attached to its target vertex.

The phase space is defined by the canonical bracket on the spinor variables, postulating that each spinor $z$ is canonically conjugated to its complex conjugate:
\be
\{z_a, \bz_b\}=-i\delta_{ab}\,,
\ee
where we have dropped the $e,v$ indices and $z_a$ with $a=0,1$ stand for the two components of the spinor $z$.
Then we impose two sets of constraints on these sets of spinors:
\begin{itemize}

\item {\it Closure constraints} at each vertex $v$:

\be
\forall v,\quad
\cC_v\equiv\sum_{e\ni v} |z^v_e\ra\la z^v_e| -\f12 \la z^v_e|z^v_e\ra\id =0,
\ee
where $|z\ra\la z|$ and $\id$ are 2$\times$2 matrices and the linear combination $|z\ra\la z|-\f12\la z|z\ra$ is the traceless part of $|z\ra\la z|$.

\item {\it Matching constraints} along each edge $e$:

\be
\forall e,\quad
\cM_e\equiv\la z^s_e|z^s_e\ra - \la z^t_e|z^t_e\ra=0\,.
\ee

\end{itemize}

It is direct to check that these two sets of constraints are first class. The closure constraints generate $\SU(2)$ transformations at each vertex:
\be
|z^v_e\ra
\,\longrightarrow\,
g_v\,|z^v_e\ra\,
\quad\textrm{with}\quad
g_v\in\SU(2)\,,
\ee
while the matching constraints generate $\U(1)$ transformations on each edge:
\be
|z^s_e\ra
\,\longrightarrow\,
e^{+i\theta_e}\,|z^s_e\ra,
\quad
|z^t_e\ra
\,\longrightarrow\,
e^{-i\theta_e}\,|z^t_e\ra,
\quad\textrm{with}\quad
e^{+i\theta_e}\,\in\,\U(1)\,.
\ee
Moreover, one easily checks that closure and matching constraints commute with each other. A {\it spinor network} is a set of spinors $\{z^v_e\}$ satisfying both sets of constraints and up to $\SU(2)$ and $\U(1)$ transformations, i.e an element in $\C^{4E}//(\SU(2)^V\times\U(1)^E)$, where $//$ stands for the symplectic reduction (i.e both solving the constraints and quotienting by their action).

This constrained phase space structure can be summarized by an action principle:
\be
\label{kin_z}
S^{(0)}[z^v_e]
\,=\,
\int dt\sum_v \sum_{e\ni v} \big(-i\la z^v_e|\partial_t
z^v_e\ra+\la z^v_e|\Lambda_v|z^v_e\ra\big)
+\sum_e\rho_e\big(\la z^s_e|z^s_e\ra-\la z^t_e|z^t_e\ra\big),
\ee
where the 2$\times$2 matrices $\Lambda_v$ with $\tr\,\Lambda_v=0$ and the real variables $\rho_e$ are Lagrange multipliers imposing the closure and matching constraints. This action $S^{(0)}$ defines the kinematical structure of spinor networks and the phase space  on the fixed graph $\Gamma$. We will later add a Hamiltonian term to this action in order to define (classical) dynamics for these spinor networks, the goal being to construct the Hamiltonian in order to produce effective dynamics for loop quantum gravity on fixed graphs $\Gamma$ which can be relevant for symmetry-reduced physical situations such as cosmology.

\medskip

In order to understand the geometrical meaning of spinor networks, the best is to translate spinors into 3-vectors. Indeed, each spinor $z\in\C^2$ determines a 3-vector $\vV\in\R^3$ through its projection onto the Pauli matrices:
\be
\vV=\la z|\vsigma|z\ra,
\,\qquad\,
|z\ra \la z| = \f12 \left( {\la z|z\ra}\id  + \vV\cdot\vsigma\right),
\ee
where the Pauli matrices are normalized such that $\sigma_i^2=\id,\,\forall i$ and the norm of the 3-vector is $|\vV|=\la z|z\ra$. Reversely, the spinor $z$ is entirely determined by its projection $\vV$ up to a global phase (see in appendix for more details). Swapping all the spinors $z^v_e$ for their projections $\vV^v_e$, it is  straightforward to translate the constraints in terms of the 3-vectors:
\begin{itemize}

\item { Closure constraints} at each vertex $v$:

\be
\forall v,\quad
\sum_{e\ni v} \vV^v_e=0
\ee

\item { Matching constraints} along each edge $e$:

\be
\forall e,\quad
|\vV^s_e|=|\vV^t_e|
\ee

\end{itemize}

The geometrical interpretation then appears clearly. The closure constraints mean that each vertex $v$ is dual to a (closed) polyhedron with each edge $e$ dual to a face of the polyhedron. The 3-vector $\vV^v_e$ becomes the normal vector to the face dual to $e$ and the norm $|\vV^v_e|$ gives the area of that face. Every set of vector satisfying the closure constraint automatically defines a unique such polyhedron. A detailed reconstruction of the polyhedron from the normal vectors is achieved through Minkowski theorem and Lasserre's algorithm \cite{polyhedron}.

Thus we have one polyhedron (embedded in the flat 3d Euclidean space $\R^3$) around each vertex $v$. Then the matching constraints impose that the areas of their matching faces along each edge $e$ be equal. Note that this does not mean that the shape of the faces will match. This would require further constraints (see e.g. \cite{gluing_bianca}).
This translation of the closure and matching constraints in terms of 3-vectors provides spinor networks with a clear interpretation as discrete 3d geometries.

\medskip

Now, we would like to define observables and wave-functions over the classical spinor phase space
$\C^{4E}//(\SU(2)^V\times\U(1)^E)$. Since we have two sets of constraints, there are two natural paths to solving them depending if we first implement the $\U(1)$-invariance or the $\SU(2)$-invariance. To start with, on the initial unconstrained phase space $\C^{4E}$, we have wave-functions $\varphi(z^v_e)$, which can be defined simply as holomorphic functions of the spinor variables. Then we have two alternatives:

\begin{itemize}
\item {\bf We first impose the matching constraints $\cM_e$:}

This is the path to the standard formulation of loop (quantum) gravity and to twisted geometries. Natural $\U(1)$-invariant observables, constructed from the spinors, are group elements $g_e\in\SU(2)$ attached to each edge $e$. Together with the 3-vectors $\vV^v_e$, they allow to entirely parameterize  the phase space $\C^{4E}//\U(1)^E$. These $g_e$ define the $\SU(2)$ holonomies along edges of loop gravity and can be taken as the configuration space coordinates. We will review the reconstruction of these group elements from the spinors in the next subsection \ref{U1}.

Then we would work with wave-functions $\varphi(g_e)$, which are $\U(1)$-invariant. Finally imposing the closure constraints on these wave-functions, we would obtain a Hilbert space of $\SU(2)$-invariant wave-functions $\varphi(g_e)$, with the standard spin network functionals as a basis.

This scheme seems to localize the  degrees of freedom (at the kinematical level) on the edges of the graph.

\item {\bf We first impose the closure constraints $\cC_v$:}

This is the path taken by the $\U(N)$ formalism for $\SU(2)$ intertwiners \cite{un1,un2,un5}. Imposing the closure constraints at each vertex $v$, one defines natural $\SU(2)$-invariant observables $F^v_{ef}$, depending on pair of edges $e,f$ attached to $v$, and holomorphic in the spinor variables $z^v_e$. Their definition will be reviewed below in subsection \ref{SU2}.

Then we would work with wave-functions $\varphi(F^v_{ef})$, which are $\SU(2)$-invariant. Finally imposing the matching constraints on these wave-functions, we obtain a Hilbert space of $\U(1)$-invariant wave-functions $\varphi(F^v_{ef})$ on the graph $\Gamma$, which is exactly isomorphic to the standard Hilbert space of spin networks obtained by the other path of first imposing $\U(1)$-invariance and then $\SU(2)$-invariance \cite{un5,spinor2}.

Compared to the previous possibility, this scheme seems to localize the  degrees of freedom  at the vertices of the graph.

\end{itemize}
\vspace{2mm}

These two paths provide interesting parameterizations  of spinor networks, relevant to discuss their (gauge-invariant) dynamics and to impose further symmetries such as homogeneity or isotropy, and we will quickly overview these constructions in the following subsections.

\subsection{Reconstructing Group Elements and Spin Networks}
\label{U1}

Let us start by identifying $\U(1)$-invariant variables. First, one notices that the 3-vectors commute with the matching constraints since they are invariant under the action of phases on the spinors:
\be
\vV^{s,t}_e=\la z^{s,t}_e|\vsigma|z^{s,t}_e\ra,\qquad
\{\cM_e,\vV^{s,t}_f\}=0,\,\forall e,f.
\ee
Then, in order to make the link, we reconstruct the $\SU(2)$ holonomies  along the edges following \cite{twisted2,un5}. We define the unique group element $g_e\in\SU(2)$ which maps the spinor $|z^s_e\ra$ at the source vertex to the dual spinor $|z^t_e]$ living at the target vertex:
\be
g_e |z^s_e\ra= |z^t_e],\quad
g_e |z^s_e]= -|z^t_e\ra,\qquad
g_e\,\equiv\,
\f{|z^t_e]\la z^s_e|-|z^t_e\ra[ z^s_e|}{\sqrt{\la z^s_e|z^s_e\ra\la z^t_e|z^t_e\ra}}\,.
\ee
The dual spinor $|z]$ is defined through the following anti-unitary map (see in appendix for more details):
$$
|z\ra=\mat{c}{z_0\\ z_1}
\,\longrightarrow\,
|z]=|\varsigma z\ra
\,\equiv\,
\mat{c}{-\bz_1\\ \bz_0}=\eps|\bz\ra,
\qquad
\eps\equiv\mat{cc}{0 & -1 \\ +1&0}\,.
$$
It is clear that the group element $g_e$ defined as above is invariant under multiplication of the spinors $z^{s,t}_e$ by opposed phases $\exp(\pm i \theta_e)$:
\be
\{\cM_e,g_f\}=0,\,\forall e,f.
\ee
Following the definition of this group element, it is straightforward to check that it maps the source vector on the opposite of the target vector:
\be
g_e\,|z^s_e\ra\la z^s_e|\, g_e^{-1}
\,=\,
|z^t_e][ z^t_e|\,,
\qquad
g_e\vartriangleright \vV^s_e=-\vV^t_e,
\ee
where $g\vartriangleright$ denotes the action of $\SU(2)$ group elements as three-dimensional rotations acting on $\R^3$.

Counting degrees of freedom, the phase space $\C^{4E}//\U(1)^E$ of spinor variables after imposing the matching condition has dimension $8E-2E= 2\times 3E$, which matches exactly the number of $\U(1)$-invariant variables $\{g_e,\vV^s_e\}$ that we have defined. Furthermore, one can check that it is possible to write the action $S^{(0)}[z^v_e]$ defined by \eqref{kin_z} in terms of these new variables. Indeed, we first compute:
\be
\la z^s_e|\pp_t z^s_e\ra+\la z^t_e|\pp_t z^t_e\ra
\,=\,
\la z^s_e|\pp_t z^s_e\ra+[ z^s_e|\pp_t z^s_e]+[ z^s_e|g_e^{-1}\pp_t g_e|z^s_e]
\,=\,
[ z^s_e|g_e^{-1}\pp_t g_e|z^s_e]
+\pp_t\la z^s_e| z^s_e\ra\,.
\ee
Since $g_e\in\SU(2)$, then the derivative $g_e^{-1}\pp_t g_e$ decomposes on the Pauli matrices $\vsigma$. So its trace vanishes, $\tr\,g_e^{-1}\pp_t g_e\,=0$, and we can decompose the matrix $|z^s_e][z^s_e|$ on the Pauli matrices in terms of the vector $\vV^s_e$:
$$
[ z^s_e|g_e^{-1}\pp_t g_e|z^s_e]\,=\,-\vV^s_e\cdot\f12\tr\vsigma g_e^{-1}\pp_t g_e\,.
$$
Then discarding the total derivative term $\pp_t\la z^s_e| z^s_e\ra$ from the action principle, we finally obtain:
\be
S^{(0)}[z^v_e]\,=\,
\int dt\,\left[
\f i 2\,\vV^s_e\cdot\tr\vsigma g_e^{-1}\pp_t g_e
\,+\,
\sum_e \vec{\rho}_e\cdot(\vV^t_e+g_e\vartriangleright \vV^s_e)
\,+\,
\sum_v \vLambda_v\cdot\sum_{e\ni v} \vV^v_e
\right]\,.
%
\ee
So solving the matching constraints, we can indeed re-express our (kinematical) action principle in terms of the holonomy-vector variables at the $\U(1)$-invariant level.

Parameterizing explicitly the group elements $g=\exp(i\alpha\hu\cdot\vsigma)$ with the class angle $\alpha\in[0,2\pi]$ and the unit vector $\hu\in\cS_2$, we can express the kinematical term $i\vV^s_e\cdot\f12\tr\vsigma g_e^{-1}\pp_t g_e$ in terms of the parameters $(\alpha_e,\hu_e)$ by computing the derivative:
$$
g_e^{-1}\pp_t g_e
\,=\,
i\vsigma\cdot\bigg{(}
(\pp_t\alpha_e)\hu_e+\cos\alpha_e\sin\alpha_e\pp_t\hu_e+\sin^2\alpha_e\hu_e\w\pp_t\hu_e
\bigg{)},
$$
where $(\hu,\pp_t\hu,\hu\w\pp_t\hu)$ form an orthogonal basis of $\R^3$ since $\hu$'s norm is fixed. To derive this formula, we have used the expression $g=\cos\alpha\id+i\sin\alpha\,\hu\cdot\vsigma$.
Let us point out that this is not equal to the na\"ive expression given by the derivative of the Lie algebra element, $i\pp_t(\alpha\hu)\cdot\vsigma$\,.

\medskip

Furthermore, we can compute the Poisson brackets of the $g_e,\vV^{s,t}_e$ variables with each other and we get after some slightly tedious but straightforward calculations:
\be
\{g_e,g_e\}\approx0,\quad
\{(V^{s,t}_e)_i,(V^{s,t}_e)_j\}\,=\,2\eps_{ijk}(V^{s,t}_e)_k,\quad
\{\vV^{s}_e,g_e\}=-i\,g_e\vsigma,\quad
\{\vV^{t}_e,g_e\}=+i\,\vsigma g_e\,.
\ee
where $\approx$ refers to weak equality (i.e imposing the matching condition). All brackets between variables attached to different edges $e\ne f$ vanish trivially.
Finally, note that $\{g_e,g_e\}$ amounts to the commutators of all the components of the group element $g_e$ with each other.

This means that this reproduces the usual holonomy-flux algebra on the graph $\Gamma$. It also means that it is legitimate to consider wave-functions of the group elements $\vphi(g_e)$ at the quantum level with the 3-vectors $\vV^{s,t}_e$ acting as the left and right derivative with respect to the $g_e$'s. This leads back to the standard quantization scheme used in loop quantum gravity.

However we would like to point out that such wave-functions $\vphi(g_e)$ are not holomorphic functions of our spinor variables. Indeed, the group elements $g_e$ contains one holomorphic term in the $z$'s and one anti-holomorphic component. We can nevertheless define holomorphic holonomy variables:
\be
\cG_e\,\equiv\,
|z^t_e\ra[ z^s_e|\,.
\ee
The matrix $\cG_e$ still transports the spinors, but is of rank one: it maps $|z^s_e]$ to $|z^t_e\ra$ but sends $|z^s_e\ra$ to 0. These holomorphic variables $\cG_e$ are still clearly $\U(1)$-invariant. They are not $\SU(2)$ group elements anymore, but still satisfy the right Poisson algebra with the 3-vectors $\vV^{s,t}_e$:
\be
\{\cG_e,\cG_e\}=0,\quad
\{(V^{s,t}_e)_i,(V^{s,t}_e)_j\}\,=\,2\eps_{ijk}(V^{s,t}_e)_k,\quad
\{\vV^{s}_e,\cG_e\}=-i\,\cG_e\vsigma,\quad
\{\vV^{t}_e,\cG_e\}=+i\,\vsigma \cG_e\,,
\ee
where the first commutator vanishes exactly and not only on the constrained surface. As it was implicitly implied in \cite{un5} and explicitly shown in \cite{spinor2}, the Hilbert space of holomorphic wave-functions $\varphi(\cG_e)$ is isomorphic to the standard Hilbert space of spin network functionals $\varphi(g_e)$, and the isomorphism can be realized through a non-trivial kernel $\cK(g_e,\cG_e)$ allowing back and forth between the two polarizations.

\subsection{Solving the $\SU(2)$-invariance: $\U(N)$ Formalism}
\label{SU2}
\label{UN_def}

We have up to now reviewed the path of first implementing the $\U(1)$-invariance on the spinor phase space. This has lead us to the standard holonomy-flux algebra of loop gravity and to the twisted geometry parameterization. In this subsection, we will give a quick overview of the other alternative of first implementing the $\SU(2)$-invariance. This approach has been developed in \cite{un0,un1,un2,un3,un4,un5} and been referred to as the $\U(N)$ formalism for $\SU(2)$ intertwiners. Here, we will not review this whole framework, but will focus on the definition of the classical $\SU(2)$-invariant observables and how to write our action at the kinematical level in terms of them.

Focusing on the closure constraints at vertices, a natural set of $\SU(2)$-invariant observables at each vertex $v$ is given by:
\be
E^v_{ef}=\la z^v_e|z^v_f\ra,\quad
E^v_{fe}=\bE^v_{ef},\qquad
F^v_{ef}=[ z^v_e|z^v_f\ra,\quad
\bF^v_{ef}{}=\la z^v_f|z^v_e],\quad
F^v_{fe}=-F^v_{ef}\,.
\ee
Calling $N_v$ the valency of the vertex $v$ (number of edges attached to $v$), the $N_v\times N_v$ matrix $E^v$ is Hermitian while the matrix $F^v$ is anti-symmetric and holomorphic in the spinor variables.
It is clear that these scalar product between spinors are invariant under $\SU(2)$ transformations at the vertex $v$. Moreover the Poisson algebra of these observables closes:
\beq
{\{}E^v_{ef},E^v_{gh}\}&=&
-i\left(\delta_{fg}E^v_{eh}-\delta_{eh}E^v_{gf} \right)\\
{\{}E^v_{ef},F^v_{gh}\} &=& -i\left(\delta_{eh}F^v_{fg}-\delta_{eg}F^v_{fh}\right),\qquad
{\{}E^v_{ef},\bF^v_{gh}\} = -i\left(\delta_{fg}\bF^v_{eh}-\delta_{fh}\bF^v_{eg}\right),\nn \\
{\{} F^v_{ef},\bF^v_{gh}\}&=& -i\left(\delta_{eg}E^v_{hf}-\delta_{eh}E^v_{gf} -\delta_{fg}E^v_{he}+\delta_{fh}E^v_{ge}\right), \nn\\
{\{} F^v_{ef},F^v_{gh}\} &=& 0,\qquad {\{} \bF^v_{ef},\bF^v_{gh}\} =0.\nn
\eeq
Actually, the algebra of the $E$-observables alone closes and the corresponding Poisson brackets actually define a $\u(N_v)$ algebra, thus the name ``$\U(N)$-formalism". One can go further with the $\U(N)$ idea and write the $E$ and $F$ observables in terms of a single unitary matrix $U^v\in\U(N_v)$:
\beq
&&E^v=\lambda_v\,\overline{U^v}\,\Delta \,{}^tU^v,
\qquad\,\, \Delta=\mat{cc|c}{1 & & \\ &1 & \\ \hline && 0_{N_v-2}}\,, \\
&&F^v=\lambda_v\,U^v\Delta_\eps \,{}^tU^v,
\qquad \Delta_\eps=\mat{cc|c}{ & 1& \\ -1& & \\ \hline & &0_{N_v-2}}\,, \nn
\eeq
where $\lambda_v\in\R^+$. It is direct to write the spinors in terms of the couple of variables $(\lambda_v,U_v)$:
\be
\label{zU}
(z^v_e)_0
\,=\,
\sqrt{\lambda_v}\,U^v_{e1},
\qquad
(z^v_e)_1
\,=\,
\sqrt{\lambda_v}\,U^v_{e2},
\qquad
\lambda_v=\f12\sum_{e\ni v}\la z^v_e|z^v_e\ra
\,=\,
\f{A_v}2
\,.
\ee
The parameter $\lambda_v$ is a global scale factor and  measures in fact the total area around the vertex $v$.
The advantage of this formulation is that the closure constraints on the spinors at the vertex $v$ gets simply encoded in the unitarity of the matrix $U^v$, as shown in \cite{un5}:
\be
\cC_v=\sum_{e\ni v}|z^v_e\ra\la z^v_e|-\f12\la z^v_e|z^v_e\ra=0
\quad\Leftrightarrow\quad
(U^v)^\dag U^v=\id\,.
\ee
One can check that we have the right number of degrees of freedom. Around a fixed vertex $v$, we had started with $N_v$ spinors satisfying the closure constraints and up to $\SU(2)$-transformations, which makes $4N_v-6$ variables. Working with the unitary matrix, we must first notice that the definition of the matrices $E$ and $F$ are invariant under $\U(N_v-2)\times\SU(2)$ (which is the stabilizer group of both $\Delta$ and $\Delta_\eps$ matrices). Thus imposing this symmetry and not forgetting the degree of freedom encoded in $\lambda_v$, we have $1+N_v^2-(N_v-2)^2-3=4N_v-6$.

One can finally write the action defining the kinematics on the graph $\Gamma$ in terms of these $\U(N)$ variables:
\be
S^{(0)}[z^v_e]
\,=\,
\int dt\,
\sum_v +i\lambda_v\tr U^v\Delta\pp_t U_v^{-1}
+\sum_v \tr\Theta_v ((U^v)^\dag U^v-\id)
+\sum_e \rho_e(E^s_{ee}-E^t_{ee}),
\ee
where the new Lagrange multiplier $\Theta_v$ imposes the unitarity of $U^v$ and the matching constraints generate multiplications by phases on the matrix elements of the unitary $U^v$.

Here, we have replaced the closure constraints on the spinors by the unitarity constraints on the matrices $U^v$. We can actually go further and formulate everything entirely in terms of $\SU(2)$-invariants:
\be\label{SkinF}
S^{(0)}[z^v_e]
\,=\,
\int dt\,
\sum_v \f {-i}{2\lambda_v}
\tr (F^v)^\dag\pp_tF^v
+\sum_e \rho_e(E^s_{ee}-E^t_{ee}),
\ee
where both the $E^v$  and $\lambda_v$ should be considered as functions of the $F$-variables:
\beq
\lambda_v
&=&
\f12\sum_{e\ni v}\la z^v_e|z^v_e\ra=\,\sqrt{\f12\,\tr (F^v)^\dag F^v},\nn\\
E^v=
&=&
\f1{\lambda_v}(F^v)^\dag F^v\,,\nn
\eeq
where the equation giving the $E$-matrix in terms of the $F$-matrix is part of a system of quadratic equations relating the $E^v_{ef}$ and $F^v_{ef}$ observables \cite{un5}.
This means that we write our action principle entirely in the $F^v$-variables, which are $\SU(2)$-invariants.
However, if we consider the $F^v_{ef}$ as our basic variables, the drawback is that they satisfy specific constraints. Of course, one must not forget that the matrices $F^v$ are anti-symmetric, but furthermore they satisfy the Pl\"ucker relations (e.g. \cite{un2,un4,un5}):
$$
F^v_{ij}F^v_{kl}=F^v_{il}F^v_{kj}+F^v_{ik}F^v_{jl}\,.
$$
Nevertheless, keeping this in mind, the $F^v$-variables can be very useful, since they were shown to be the natural variables when quantizing the $\SU(2)$-invariant phase space in order to recover the Hilbert space of $\SU(2)$ intertwiners  and to construct coherent intertwiner states \cite{un2,un4,un5}. From this perspective, the Pl\"ucker relations define the basic recoupling relations for the spin-$\f12$ representation.

The insights that one has to keep in mind from this formalism are:
\begin{itemize}

\item At each vertex $v$, the unitary matrix $U^v$ up to $\U(N_v-2)\times\SU(2)$ transformations encodes the $\SU(2)$-invariant information about the spinors, i.e the shape of the dual polyhedron at the classical level and the intertwiner state at the quantum level.

\item When quantizing, we can replace the wave-functions $\varphi(z^v_e)$ by wave-functions $\vphi(U^v)$ or actually by wave-functions $\vphi(F^v)$ which already implement the $\SU(2)$-gauge invariance at every vertex. The equivalence between these formulations have already been studied in \cite{un1,un2,un5}.

\item There is a natural $\U(N_v)$ action on the spinors at each vertex. Dropping the index $v$ for simplicity's sake, it reads for a transformation $V\in\U(N)$:
$$
U\arr \tU\equiv VU,\qquad
z_e\arr \tz_e\equiv V_{ef}z_f.
$$
The closure constraints commute with this $\U(N)$ action. These $\U(N)$ transformations deform the shape of the dual polyhedron (or equivalently the intertwiner at the quantum level) while keeping the total boundary area unchanged. And one can define coherent intertwiners which transforms covariantly under such transformations \cite{un2,un4}.

\end{itemize}

\subsection{Defining LQG Dynamics on a Fixed Graph}

%

Up to now, we have described the loop gravity kinematics on a fixed graph $\Gamma$, either in terms of spinors or $\U(1)$-invariants or $\SU(2)$-invariants. We have discussed the physical interpretation of these variables as defining a discrete geometry and we have introduced an action principle $S^{(0)}[z^v_e]$ encoding the kinematical phase space  structure together with the closure and matching constraints generating the $\U(1)$ and $\SU(2)$ gauge symmetries.
The next step is to endow these kinematical structures with dynamics. The natural way to do so is to add a Hamiltonian term of our action and define:
\be
S[z^v_e]=S^{(0)}[z^v_e]-\int dt\, H\,,
\ee
where the Hamiltonian $H$ will be a gauge invariant functional of the spinors (and possibly depending on the time $t$). This is the nice advantage of the spinorial formalism: the kinematics and phase space structure, and thus the dynamics, can be simply written in an action principle.

The dynamics defined by $H$ will either be a Hamiltonian constraint or the evolution with respect to an external or gauge-fixing time parameter. We can construct $H$ by different ways. We could discretize general relativity's Hamiltonian constraint on a fixed graph (e.g. \cite{AQG}), we could truncate the loop quantum gravity dynamics to $\Gamma$ (e.g. \cite{LQGcosmo1,LQGcosmo2}), we could also extract an effective Hamiltonian evolution from spinfoam transition amplitudes (e.g. \cite{SFcosmo}), but we can also consider all the possibilities of gauge-invariant dynamics (compatible with some notion of homogeneity and isotropy of the chosen graph $\Gamma$) and investigate which of these dynamics can be interpreted as the classical evolution (plus potential quantum corrections) in general relativity of certain 3-metrics.

We see our construction of a Hamiltonian as defining an effective dynamics on the fixed graph from various viewpoints. First, since we believe that we are only studying a truncation of the full theory, we are then studying the dynamics induced on the restricted state space that we are considering by projecting the full Hamiltonian on that smaller space. Second, from the spinfoam point of view, we will evaluate the transition amplitudes between coherent states peaked on classical phase space points and extract from it an effective classical Hamiltonian living on the phase space but taking into account the quantum corrections involved in the evolution of the coherent states. Finally, we follow the approach of considering all possible terms in the dynamics compatible with the $\SU(2)$ gauge invariance and the expected symmetries of our restricted model (based on the choice of graph). This can be understood from an effective field theory point of view as considering all possible terms that can appear as corrections coming from the renormalization or coarse-graining of the theory, then we can investigate which ones are physically-relevant or not and how they affect the evolution of our system.

\medskip

From this perspective, it is very important to understand what kind of 3-metrics and 4-metrics can be generated by restricting oneself to a fixed graph $\Gamma$ in loop gravity. Our point of view is that working on a fixed graph does not necessarily need to be interpreted as describing the evolution of discrete geometries. Indeed, one can view the fixed graph and the data living on it as a triangulation of a continuous geometry, a sampling from which we can reconstruct the whole geometry if the triangulation is refined enough and the geometry is smooth enough to interpolate between the discrete sampling that the triangulation has defined.
The same perspective has been developed recently independently in \cite{new_laurent}, where they describe the class of continuum space geometries compatible with the finite-dimensional phase space structure of loop quantum gravity on a fixed graph.
In other words, we consider working on a fixed graph as defining a mini-superspace model of (loop) quantum gravity\footnotemark , or possibly a midi-superspace model in the case where we consider a certain class of graphs (e.g. 2-vertex graphs as in \cite{un3}) and look at the limit with infinite number of edges. In this context, families of geometries defined by a finite number of parameters, as in mini- and midi- superspace models,  can be effectively represented by the data living on a fixed finite graph, even if these geometries are not intrinsically discrete and even if they are not compact. For example, as we will see in the next sections, following the original ideas developed in \cite{LQGcosmo1,LQGcosmo2,un3}, 2-vertex graphs are well-suited to describe homogeneous and isotropic cosmology of the FRW type (and possibly also some of Bianchi's models). Then, having fixed the graph $\Gamma$ defining the geometry on spatial slices and having identified the geometries that we wish to describe, we would like to define a proper dynamics of the spinor variables living on $\Gamma$ in order to reproduce the correct dynamics of general relativity for these geometries, with possibly effective quantum gravity corrections.

Thus, the interpretation that we propose of truncating the loop quantum gravity dynamics to a fixed graph definitely lies within mini-superspace models. We do not insist that the classical spinor data or the spin network state living on this graph defines the entire space itself and its geometry as in standard loop quantum gravity. But we mean that the kinematics and dynamics of some restricted class of geometries can be effectively modeled by the dynamics of classical spinor networks on some fixed graph.
\footnotetext{
By mini-superspace models, we mean here a truncation of the gravity phase space to a finite number of (kinematical) degrees of freedom. In the context of cosmology, this is usually achieved by an appropriate symmetry reduction to a homogeneous sector of general relativity.
But it can be more generally understood  as the truncation of general relativity to a certain  metric ansatz defined by a finite number of parameters, which then become the degrees of freedom of the mini-superspace model. That's what is achieved by working on a fixed graph in our context. Nevertheless, once having fixed the background graph, one can further work out some symmetry reduction, like the restriction to the homogeneous and isotropic sector for the 2-vertex model.
}

\medskip

\begin{figure}[h]
\begin{center}
 \includegraphics[height=30mm]{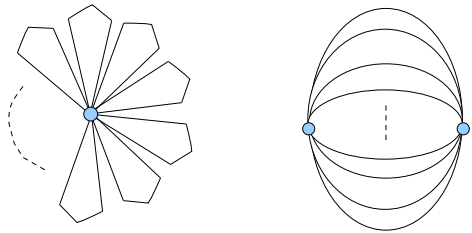}
\caption{The flower graph with a single vertex and $L$ petals, on the left, and the 2-vertex graph
with $L+1$ edges between its two vertices: both define the Hilbert space of $\SU(2)$-invariant spin
network states $L^2(\SU(2)^L/Ad\,\SU(2))$.
\label{flower}}
\end{center}
\end{figure}

A last remark that we would like to do in this section is on the role and relevance of the chosen
graph $\Gamma$. On the mathematical level, the phase space of gauge-invariant observables on a fixed
graph $\Gamma$, defined as $\C^{4E}//(\U(1)^E\times\SU(2)^V)$, turns out not to depend on the
particular combinatorics of the graph $\Gamma$ but only on the  number of edges and vertices. To
understand this, let us think in terms of the standard cylindrical functionals of loop quantum
gravity, that is the Hilbert space $L^2(\SU(2)^E/\SU(2)^V)$ of $\SU(2)$-invariant functions
$\vphi(\{g_e\})$. This space is isomorphic to $L^2(\SU(2)^{E-V})$ (assuming that $E>V$, else it is
isomorphic to $L^2(\SU(2)/Ad\SU(2))$ when $E=V$), which only depends on the number of (independent)
loops of the graph $L\equiv\,E-V+1$. Going further, the space $L^2(\SU(2)^E/\SU(2)^V)$ is actually
isomorphic to any space $L^2(\SU(2)^{E+n}/\SU(2)^{V+n})$ with $n\ge -V,\,n\in\Z$. This isomorphism
can be realized exactly through gauge fixing (or ``unfixing") of the $\SU(2)$-invariance at the
vertices of the graph $\Gamma$ (see e.g. \cite{noncompact} for a rigorous approach to gauge-fixing
spin network functionals). From this point of view, the Hilbert space of a fixed graph is always
isomorphic to the space of states on a flower graph $L^2(\SU(2)^L/Ad\,\SU(2))$ or on a 2-vertex
graph $L^2(\SU(2)^{L+1}/\SU(2)^2)$. Thus we can always write, both at the classical and quantum
levels, any gauge-invariant dynamics defined on $\Gamma$ on a corresponding flower or 2-vertex
graph.

The natural question is then why should we bother about using more complicated graphs involving more
vertices and a more complex combinatorial structure. Our point of view is that the combinatorial
structure of the graph $\Gamma$ provides us with an implicit vision of the space geometry. For
instance, through the implicit notion that a vertex of the graph represents a physical point of
space (or region of space) and that edges define directions around these points, the graph's
structure will define our notions of homogeneity and isotropy. Therefore, even though we can
translate any gauge-invariant observable or Hamiltonian operator from our potentially-complicated
graph $\Gamma$ to any graph with the same number of loops, the concepts of locality, homogeneity and
isotropy of our dynamics crucially depend on our original choice of $\Gamma$. From this viewpoint,
different graphs
will admit a different implementation of the symmetry reduction (to e.g. the homogeneous sector),
that in turn will lead to different mini-superspace models.

\section{Classical Phase Space on the 2-Vertex Graph}
\label{2v_classical}

\begin{figure}[h]
\begin{center}
\includegraphics[height=30mm]{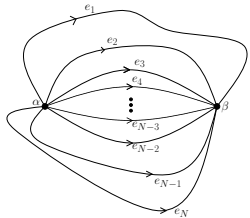}
\caption{The 2-vertex graph with $N$ edges linking the two vertices $\alpha$ and $\beta$.
\label{2v_fig}}
\end{center}
\end{figure}

Now that we have introduced and reviewed the spinorial formulation of the loop gravity phase space on an arbitrary fixed graph $\Gamma$, and discussed how to recover the standard observables -holonomies, fluxes, twisted geometry variables, $\U(N)$ observables- in this framework, we will now specialize to the case of the 2-vertex graph (see fig.\ref{2v_fig}). Indeed, there is serious evidence 
that truncating the loop quantum gravity dynamics to this very simple graph and implementing an
appropriate symmetry reduction leads to Friedmann-Robertson-Walker (FRW) homogeneous and isotropic
cosmologies \cite{LQGcosmo1,LQGcosmo2,SFcosmo,un3,un5}. In this section, we will study the
kinematics on the 2-vertex graph, clear up in this simpler context the geometrical meaning of the
variables defined in the previous section and explain how to reduce to the homogeneous and isotropic
sector following the ideas developed in \cite{un3,un5}. We will tackle the issue of the dynamics and
how the Hamiltonian generates the four-dimensional FRW metric in the next section.

\subsection{Kinematics on the 2-Vertex Graph}

Let $\alpha$ and $\beta$ denote the two vertices of the graph as on fig.\ref{2v_fig}.
Following Sec. \ref{sec1}, we denote by $z_i$ and $w_i$, with $i=1,...,N$,
the collection of spinors living on the $N$ edges respectively attached to the vertex $\alpha$ and to the vertex $\beta$.
These spinors are subject to the closure constraints,
\beq
\label{closure}
\sum_i |z_i\ra \la z_i|=\lambda(z)\id
&\quad\Longleftrightarrow \quad&
\sum_i \vec{V}(z_i)=0,\\
\sum_i |w_i\ra \la w_i|=\lambda(w)\id
&\quad\Longleftrightarrow \quad&
\sum_i \vec{V}(w_i)=0,\nn
\eeq
with $\lambda(z)\equiv\f12\sum_i \la z_i|z_i\ra=\f12\sum_i|\vec{V}(z_i)|$,
and to the matching constraints,
\be \label{gluing}
\la z_i|z_i\ra=\la w_i|w_i\ra \qquad \forall
i=1..N
\quad\Longleftrightarrow \quad
|\vV(z_i)|=|\vV(w_i)|\,.
\ee
In particular, the total boundary area is of course the same seen from the two vertices, $\lambda\equiv \lambda(z)=\lambda(w)$,.


The action principle summarizing the phase space structure can be equivalently written as a functional of the spinors $z_k,w_k$ or of the vectors $\vV_k\in\R^3$ and group elements $g_k\in\SU(2)$~:
\beq
\label{Skinz}
S^{(0)}[z_k,w_k]
&=&
\int dt\sum_{k=1}^N
-i\big(\la z_k|\partial_t z_k\ra+\la w_k|\partial_t w_k\ra\big)
+\la z_k|\Lambda_\alpha|z_k\ra
+\la w_k|\Lambda_\beta|w_k\ra
+\rho_k\big(\la z_k|z_k\ra-\la w_k|w_k\ra\big)\\
&=&
\int dt\sum_{k=1}^N \vV(z_k)\cdot \f i2\tr\vsigma g_k^{-1}\pp_tg_k
+\vec{\Lambda}_\alpha\cdot\sum_{k=1}^N\vV(z_k)
+\vec{\Lambda}_\beta\cdot\sum_{k=1}^N \vV(w_k)
+\vec{\rho_k}\cdot(\vV(w_k)+g_k\vartriangleright\vV(z_k)). \nn
\eeq
We can further re-express this action in terms of the twisted geometry variables or in terms of the $\SU(2)$-invariant $E^{\alpha,\beta}$ and $F^{\alpha,\beta}$ living at both vertices.

The interpretation of the 2-vertex graph as a discrete geometry is clearly of two polyhedra glued along their $N$ faces with matching areas. Nevertheless, the two polyhedra can still have different shapes and this can be interpreted as local degrees of freedom living at each vertex. We will see below that constraining the two polyhedra to have the same shape naturally projects us onto the homogeneous and isotropic sector of the phase space.

\subsection{Symmetry Reduction to a Isotropic Cosmological Setting}

In \cite{un3,un5}, a symmetry reduction of the 2-vertex graph model was proposed by
requiring a global $\U(N)$ invariance of the spinor networks. Indeed, this invariance leads to an
homogeneous and isotropic model parameterized by just one degree of freedom. Let us see how this reduction works. It is best understood when written in the $\U(N)$ observables which are $\SU(2)$-invariant. Reminding the definitions given above in section \ref{UN_def}, we have two sets of $\SU(2)$-invariant quantities defined at the vertices of the graph:
$$
\Ea_{ij}=\la z_i|z_j\ra,\quad
\Fa_{ij}=[ z_i|z_j\ra,\qquad
\Eb_{ij}=\la w_i|w_j\ra,\quad
\Fb_{ij}=[ w_i|w_j\ra.
$$
The matching constraints on the $N$ edges are simply expressed in terms of these observables:
$$
\forall i,\quad
\cM_i=\Ea_{ii}-\Eb_{ii}=0,
$$
and they generate an invariance under $\U(1)^N$ as said in the previous section. The point noticed in \cite{un3,un5} is that this symmetry can be naturally enlarged from $\U(1)^N$ to $\U(N)$. Let us indeed introduce:
\be
\forall i,j,\qquad
\cE_{ij}=\Ea_{ij}-\Eb_{ji},
\qquad
\cE_{ii}=\cM_i\,.
\ee
Considering the Poisson bracket of the $E$-observables, it is straightforward to check that these new $\SU(2)$-invariants form a $\U(N)$ algebra:
\be
\{\cE_{ij},\cE_{kl}\}
\,=\,
-i\,(\delta_{jk}\cE_{il}-\delta_{il}\cE_{kj})\,.
\ee
Let us further impose these constraints on the phase space of our spinor variables living on the 2-vertex graph:
\be
S^{(0)}_{homo}[z_k,w_k]
\,\equiv\,
\int dt\sum_{k=1}^N
-i\big(\la z_k|\partial_t z_k\ra+\la w_k|\partial_t w_k\ra\big)
+\la z_k|\Lambda_\alpha|z_k\ra
+\la w_k|\Lambda_\beta|w_k\ra
+\rho_{kl}\cE_{kl},
\ee
where the matching constraints $\cM_k$ are taken into account as the diagonal components (when $k=l$) of the $\U(N)$-constraints $\cE_{kl}$.

We insist on the fact that the new constraints $\cE_{kl}$ are first-class and implement the reduction to the isotropic sector through a symmetry reduction (by $\U(N)$, for a system who was originally only invariant under $\U(1)^N$), implying that the variation between edges are irrelevant for the dynamics and that only the global (boundary) area (between $\alpha$ and $\beta$) matters.

\medskip

This new set of constraints $\cE$ geometrically means that the polyedra dual to the two vertices $\alpha$ and $\beta$ are identical. Indeed requiring $\Ea_{ij}=\Eb_{ji}$ for all couple of (possibly identical) edges $i,j$ implies (by taking the norm squared) that all the scalar products between 3-vectors $\vV_i\cdot\vV_j$ at the two vertices are equal. This tells us that these $\U(N)$ constraints impose homogeneity: the intertwiner states at the two vertices are identical.

Coming back to the spinors, the constraints read $\la z_i|z_j\ra=[ w_i|w_j]$ for all $i,j$'s, which implies after a little algebra\footnotemark{} that the spinors $|z_i\ra$ are  equal to the spinors $|w_i]$ up to global $\U(2)$ transformation (it's an equality on scalar products) (see e.g. \cite{un5,spinor1}).
%
\footnotetext{
For example, we can translate the constraints on the $E$'s into constraints on the $F$'s using the fact that $|z\ra\la z|+|z][z|=\la z|z\ra\id$ for any spinor $z$:
$$
\forall i,j,k,\qquad
\bF_{ki}F_{kj}=E_{ij}E_{kk}-E_{ik}E_{kj}\,.
$$
}
Since both sets of spinors $z_i$ and $w_i$ are anyway defined up to $\SU(2)$ transformations generated by the closure constraints at the vertices $\alpha$ and $\beta$, this means that the spinors $|z_i\ra$ are exactly equal to the spinors $|w_i]$ up to a global phase:
\be
\forall k,l,\quad\cE_{kl}=0
\qquad
\Rightarrow
\qquad
\forall i\,
|w_i]=e^{i\phi}\,|z_i\ra\,.
\ee
This means that the unitary matrices $\Ua$ and $\Ub$ are also equal up to a phase, $\overline{\Ub}=e^{i\phi}\Ua$ (up to $\SU(2)\times\U(N-2)$ transformations). We can also translate this condition in the $\SU(2)$ group elements $g_k$ living on the edges. Just being careful that $e^{i\phi}\id\notin\SU(2)$, the holonomies read:
$$
g_k=\f{e^{i\phi}|z_k\ra\la z_k|+e^{-i\phi}|z_k][ z_k|}{\la z_k|z_k\ra},
$$
thus being the diagonal $\SU(2)$ matrix $(e^{i\phi},e^{-i\phi})$ in the $|z_k\ra,|z_k]$ orthonormal basis.

\medskip

Finally, putting back the solution $|w_i]=e^{i\phi}\,|z_i\ra$ in the action \eqref{UN_action} with $\U(N)$ constraints, the action greatly reduces and we simply get:
\be\label{reduced-action}
S^{(0)}_{homo}[z_k,w_k]
\,=\,
-2\int dt \,\,\lambda\pp_t\phi
\,=\,
S^{(0)}[\lambda,\phi]\,.
\ee
The only remaining degrees of freedom are thus the total area $A\equiv2\lambda$ and its
conjugate angle $\phi$. Thus we have reduced the 2-vertex graph to its homogeneous and isotropic
sector.

By isotropic, we mean that the only relevant degree of freedom is the total area and not the individual areas $A_k=\la z_k|z_k\ra=\la w_k|w_k\ra$. So the individual areas are not required to be the same on all edges. On the other hand, it means that the conjugate variable defined by the angle $\phi$ is the same for all edges. We can go further in understanding the $\U(N)$-constraints $\cE_{kl}$ leading to homogeneity. Indeed, they generate the following $\U(N)$-transformations on the spinors:
\be
z_k
\,\arr\,
(Uz)_k=\sum_l U_{kl}\,z_l,
\qquad
w_k
\,\arr\,
(\bU w)_k=\sum_l \overline{U_{kl}}\,w_l\,.
\ee
It is this huge gauge-invariance that kills the dependence on degrees of freedom living on each edge and effectively reduces the action to its isotropic sector.

We would also like to point out that the angle $\phi$ in the symmetry-reduced sector matches the angle $\xi$ parameterizing  twisted geometries, which provides it with a physical interpretation as related to the extrinsic geometry, i.e. related to the embedding of our spatial slice -the 2-vertex graph- into the 4d space-time.

\medskip

Therefore, to conclude this subsection, once we have accomplished the symmetry reduction by the $\U(N)$-constraints, we are left with homogeneous and isotropic 2-vertex spinor networks, which is described by the holonomies $e^{i\phi}$  along all the edges of the graph and by the
total boundary area $A=2\lambda$ around both vertices.

Our final comment is that the resulting homogeneous and isotropic sector is independent from the number of edges $N$ of the original 2-vertex graph. This is normal since imposing isotropy means getting rid of the dependence on the individual edges. On the other hand, the full theory and anisotropy should crucially depend on the allowed number of the edges $N$.

\subsection{On the Choice of Complex Variables}
\label{complex_strc}

As we have seen, in the isotropic sector, the phase space reduces to a one-dimensional system described by the total area $A=2\lambda$ and its conjugate angle $\phi$:
$$
\{A,\phi\}=1.
$$
Thinking of these as conjugate variables, it is usual to choose a complex structure defined by the variable $\zeta=(A-i\phi)/\sqrt{2}$:
\be
\{\zeta,\bar{\zeta}\}=i.
\ee
This is the standard choice done in the context of twisted geometries \cite{twisted1,twisted2} and
in the spinfoam cosmology approach of \cite{SFcosmo} which uses the twisted geometry phase space to
describe its boundary states. Another choice, better suited to the spinor approach used here and to
the coherent spin network states based on the spinorial framework \cite{spinor1}, is the complex
variable $z=\sqrt{2A}\,e^{-i\f\phi2}$:
\be
\{z,\bar{z}\}=i.
\ee
This is also a canonical choice of complex structure on this two-dimensional phase space. This is actually the choice that we will make when analyzing the dynamics of the 2-vertex graph phase space induced by the spinfoam amplitudes, as we will see in section \ref{spinfoam}.

The advantage of $z$ over $\zeta$ is that it reflects the original definition of the variables.
Indeed, $A$ is defined as a real positive number and $\phi$ is defined as an angle (modulo $2\pi$)
from the holonomy. Then $z=\sqrt{2A}\,e^{-i\f\phi2}$ is a more natural complex number in order to
define the unitary matrices $U^{\alpha,\beta}$. On the other hand, using the complex structure
defined by $\zeta=(A-i\phi)/\sqrt{2}$, and quantizing the system in terms of this variable, means
that one has to restrict by hand $A$ to be positive and $\phi$ to be bounded. In simpler terms,
$z\in\C$ parameterizes exactly our phase space taking into account the constraints
$A\in\R_+,\phi\in[0,2\pi]$ while $\zeta\in\C$ leads to some redundancies that have to be dealt with
in a nontrivial way at the quantum level. Finally, as we will see in section \ref{spinfoam}, the
spinfoam transition amplitudes in the isotropic and homogeneous sector will naturally be holomorphic
functions of $z$.

\subsection{Beyond the Isotropic Sector}

In the present work, we focus on the study of the homogeneous and isotropic sector and its dynamics both at the classical level through the definition of classical Hamiltonians as in the following section \ref{classical_dyn} and at the quantum level through the analysis of the spinfoam transition amplitudes in section \ref{spinfoam}. Nevertheless, from the perspective of cosmology, it is necessary to go beyond this simplistic model and study the departures from isotropy and homogeneity. In particular, we should study the dynamics of the inhomogeneities and their feed-back on the global homogeneous sector in order to compare to current measurements in cosmology (on the cosmic microwave background for instance). Extending the analysis of our 2-vertex model to the whole Hilbert space, beyond the $\U(N)$-invariant sector, would thus allow to test its relevance and validity as a quantum mini-superspace model for cosmology.

One goal would be to parameterize efficiently the reduced gauge-invariant phase space on the 2-vertex graph and understand how to project the phase space variables onto the spherical harmonics and define a multi-pole expansion of the corresponding degrees of freedom. For instance, the number of edges $N$ is irrelevant to the kinematics and dynamics of the homogeneous and isotropic sector since requiring the $\U(N)$-invariance kills any dependence on $N$. On the other hand, $N$ will play a non-trivial role out of the $\U(N)$-invariant sector as a cut-off in the number of degrees of freedom. It would then be interesting to understand how big should $N$ be to model our observed cosmology and, for example, allow anisotropy as seen in  the cosmic microwave background.

Actually, an analysis of the 2-vertex model beyond the isotropic sector has been carried out for
the case $N=4$ \cite{vmr}. In that work, the 2-vertex model is regarded as a triangulation of the
whole space, that is thus a three sphere. The reduced gauge-invariant phase space, which in this
$N=4$ case is formed by twelve degrees of freedom, has been then identified with the Bianchi IX
model (homogeneous and isotropic model with the spatial topology of $S^3$) plus perturbations. Since
the Bianchi IX model is described by six degrees of freedom, those perturbations account for the
remaining six degrees of freedom. Using the fact that $S^3$ is isomorphic to $SU(2)$, in \cite{vmr}
a expansion of the perturbations in terms of Wigner matrices is considered, and those six remaining
degrees of freedom are identified with the six components forming the diagonal part of the lowest
integer mode in that expansion, though the authors do not give a justification for such
identification.

We will not presently investigate in detail how to parameterize the full phase space out of the
$\U(N)$-invariant sector for general $N$ but merely discuss the possibilities from our perspective.
We postpone the full analysis to future work. The degrees of freedom of the 2-vertex graph can be
understood from the point of view of both the edges, as the holonomies defining the curvature
between the two vertices, or both the vertices, as the internal geometries of the two vertices and
the correlations between them. If one decides to focus on the holonomies, it seems natural to use
the holomorphic and anti-holomorphic components of the holonomies  around the loops of the graph
(expressed in terms of the $E$ and $F$ variables) as introduced in \cite{un5} and to look for a way
to combine them in order to get a finite Poisson algebra encoding the whole reduced phase space. The
alternative is to start with the unitary matrices $U^{\alpha,\beta}$ describing the geometry of each
vertex. Let us look a bit more into how to construct gauge invariant $\SU(2)$ observables from these
matrices.

First of all, as seen in section \ref{UN_def}, the observables must be invariant under the right action by $\SU(2)\times\U(N-2)$ at both vertices:
$$
U^\alpha\,\arr U^\alpha V^\alpha,
\quad
U^\beta\,\arr U^\beta V^\beta,
\qquad
V^{\alpha},V^\beta\,\in\SU(2)\times\U(N-2)\,.
$$
It is thus natural to make $U^{\alpha}$ and $U^\beta$ act on $\SU(2)\times\U(N-2)$ vectors. Following the previous work \cite{un1,un2}, we choose the irreducible representations of $\U(N)$ whose highest vector is invariant under $\SU(2)\times\U(N-2)$. These are labeled by an integer $J\in\N$ and correspond to Young tableaux with two horizontal lines of equal length $J$. Moreover they have been shown to be exactly the Hilbert spaces of $\SU(2)$-intertwiners  (at fixed total area $J$) resulting from the quantization of the spinorial phase space \cite{un1,un2,un5}. Let us call $|J,\Omega_N\ra$ the highest weight vector of these irreducible representations. As shown in \cite{un2}, acting with the matrix $U^\alpha\in\U(N)$ on $|J,\Omega_N\ra$ generates a coherent intertwiner state peaked on the classical phase space point with the spinors given by the first columns of the unitary matrix $U^\alpha$ according to the classical definition \eqref{zU}:
\be
U^\alpha\,|J,\Omega_N\ra=|J,\{z^\alpha_i\}_{i=1..N}\ra\,,
\ee
where the $z^\alpha_i$ satisfy the global normalization condition $\sum_i \la z^\alpha_i|z^\alpha_i\ra=1$.
The highest weight vector can be identified as the bivalent case, where all spinors vanish except the first two:
\be
|J,\Omega_N\ra= |J,\left\{\mat{c}{1\\0},\mat{c}{0\\1},\mat{c}{0\\0},..\right\}\ra
\ee
Then acting with the unitary matrix $U^\alpha$ on this state, we generate all coherent intertwiners. We similarly define the lowest weight vector by acting with the duality map on the spinors:
$$
|J,\varsigma\Omega_N\ra= |J,\left\{\mat{c}{0\\1},\mat{c}{-1\\0},\mat{c}{0\\0},..\right\}\ra
$$
More details on this construction can be found in \cite{un2,un4}. Then the scalar product between the states in $\alpha$ and $\beta$ is invariant under $\U(N)$ (acting as $U^\alpha\arr UU^\alpha,\,U^\beta\arr \bar{U}U^\beta$) and can be computed exactly:
\be
\la \varsigma\Omega_N,J | {}^tU^\beta U^\alpha|\Omega_N,J\ra
\,=\,
\la J,\{\varsigma z^\beta_i\}|J,\{z^\alpha_i\}\ra
\,=\,
\left(
\det_2 \sum_i|z^\alpha_i\ra [z^\beta_i|
\right)^J\,.
\ee
The first remark is that the $z^{\alpha,\beta}_i$ are just our spinors $z_i$ and $w_i$ up to the normalization factor $\sqrt{\lambda}$ defining the total area. Then the expression simplifies greatly assuming that the $z_i$'s and $w_i$'s satisfy the closure constraints and that we are in the isotropic sector with $|w_i]=e^{i\phi}|z_i\ra$ and we simply get:
\be
\la \Omega_N,J | {}^tU^\beta U^\alpha|\Omega_N,J\ra
\,=\,
(e^{-2i\phi})^J\,.
\ee
This scalar product can thus be considered as a definition of the conjugate angle $\phi$ in the full phase space.

The second remark is that the scalar product $\la \Omega_N,J | {}^tU^\beta U^\alpha|\Omega_N,J\ra$ is exactly the evaluation at the identity of the 2-vertex spin network state with coherent intertwiners attached to the vertices $\alpha$ and $\beta$. It is then straightforward to obtain an overcomplete basis of gauge-invariant observables by considering the evaluation of this spin network state on arbitrary group elements $g_i\in\SU(2)$ living on every edge of the graph. This amounts to inserting these group elements in the scalar product expression between the two $\U(N)$ matrices. Such evaluations are by definition both $\SU(2)$-invariant at the vertices and $\U(1)$-invariant along the edges. These evaluations and their Poisson algebra might not be easy to compute, but this procedure still hints towards a parametrization of the gauge-invariant phase space in terms of $\U(N)$ representations. From this viewpoint, it might be interesting to investigate the relation between the Peter-Weyl expansion in $\U(N)$ representation and a multi-pole expansion of the observables in our phase space. We would then expect to recover the continuum limit (i.e no cut-off in $j$ in the spherical harmonics expansion) as $N$ is sent to infinity.

A detailed analysis of the full phase space is postponed to future investigation of anisotropy in the 2-vertex model for cosmology.

\section{(Effective) Classical dynamics on the 2-Vertex Graph}
\label{classical_dyn}
\label{dynamics}

As said before, the above kinematical setting, namely the U(N)-invariant spinor network defined on
the 2-vertex graph, seems suitable to model effective dynamics for Friedmann-Robertson-Walker
(FRW) like cosmologies. In this section we will face this issue from the canonical point of view, by adding an appropriate Hamiltonian to the action, as it was done in \cite{un5}. We will analyze
whether the resulting model can be understood as an effective FRW model, that introduces corrections coming from the discrete theory.

\subsection{Hamiltonian for FRW Cosmology}

Before describing the effective dynamics that arises naturally from the two-vertex model, let us
first review the dynamics of FRW models as dictated by general relativity, so that we can make the
link between both theories.

The FRW model represents homogeneous and isotropic spacetimes.
Their 4-metric is given by the general form
\be\label{frw-metric}
ds^2\,=\,-{\cal N}^2(t)dt^2+a^2(t)d\mathcal{S}^2,\qquad d\mathcal{S}^2\equiv{}^oq_{ab}dx^adx^b.
\ee
Here ${\cal N}$ is the lapse function and $a(t)$ is the scale factor, that we consider with
dimensions of length so that the coordinates $x^a$ are dimensionless. The indices $a$ and $b$ go
from 1 to 3 and denote spatial indices.
Finally, ${}^oq_{ab}$ stands for the 3-metric of a 3-dimensional space of uniform curvature. There
exist three such spaces: Euclidean space, spherical space, or hyperbolic space.
In (dimensionless) polar coordinates $(r,\theta,\phi)$ we simply have
\be\label{Sigma}
d\mathcal{S}^2\,=\,\frac{r^2}{1-kr^2}+r^2(d\theta^2+\sin^2\theta d\phi^2),
\ee
where $k$ is a constant representing the curvature of the space: $k=1$ for the sphere, $k=-1$ for
the hyperboloid, and $k=0$ for the Euclidean space.

In order to get the dynamics of a generic FRW model, the easiest way to proceed is to take the
Einstein-Hilbert action for general relativity with cosmological constant $\Lambda$,
\be\label{S-gr}
S_{GR}\,=\,\frac1{16\pi G}\int dt \int_\Sigma d^3x \sqrt{\det g}(R[g]+\Lambda),
\ee
and particularize it to the FRW metric given in Eqs. \eqref{frw-metric}-\eqref{Sigma}. After
employing a 3+1 ADM decomposition to express it in the Hamiltonian formalism, the result
for $S_{GR}[g_{FRW}]$ is
\be\label{S-frw}
S_{FRW}\,=\,\int dt\int_\Sigma d^3x\sqrt{{}^o q} \big[ \pi_a \pp_t a -{\cal N}C\big], \qquad
C_{grav}\,=\,-\frac{2\pi G}{3}
\frac{\pi_a^2}{a}-\frac{3k}{8\pi G}a+\frac{\Lambda}{8\pi G}a^3.
\ee
In the above expressions $G$ denotes the Newton constant, ${}^o q$ denotes the determinant of the
fiducial metric ${}^oq_{ab}$, and $\Sigma$ stands for a 3-dimensional spatial slice of the
space-time. In presence of homogeneity, since all the functions entering the action are homogeneous,
the spatial integral over the slices $\Sigma$ diverges for the non-compact spatial topologies
corresponding to $k=0,-1$.
Nonetheless, one usually avoids that divergence by restricting the above spatial integral just to a
finite cell $\mathcal{V}$ with finite fiducial volume
\be
V_o\,\equiv\,\int_\mathcal{V} d^3x\sqrt{{}^o q}.
\ee
Indeed, because of the homogeneity, the dynamics of the whole spacetime is the same in every
point and then a finite region is enough to capture the dynamics.

Note that the lapse ${\cal N}$ is a Lagrange multiplier which imposes the Hamiltonian constraint
$C_{grav}=0$.
Actually, since this model is homogeneous, the spatial diffeomorphism symmetry of general relativity
trivializes and there is only left the invariance under time reparameterizations, which is
implemented by the Hamiltonian constraint.

Remarkably, the FRW cosmologies, being a solution of the equations of motion obtained by varying the
reduced action \eqref{S-frw}, are also a particular solution of the Einstein equations (that are
obtained by varying the full action \eqref{S-gr}). In other words, the FRW model is a symmetry
reduction of general relativity, more concretely, its homogeneous and isotropic sector.

\begin{figure}[h]
\begin{center}
 \includegraphics[height=55mm]{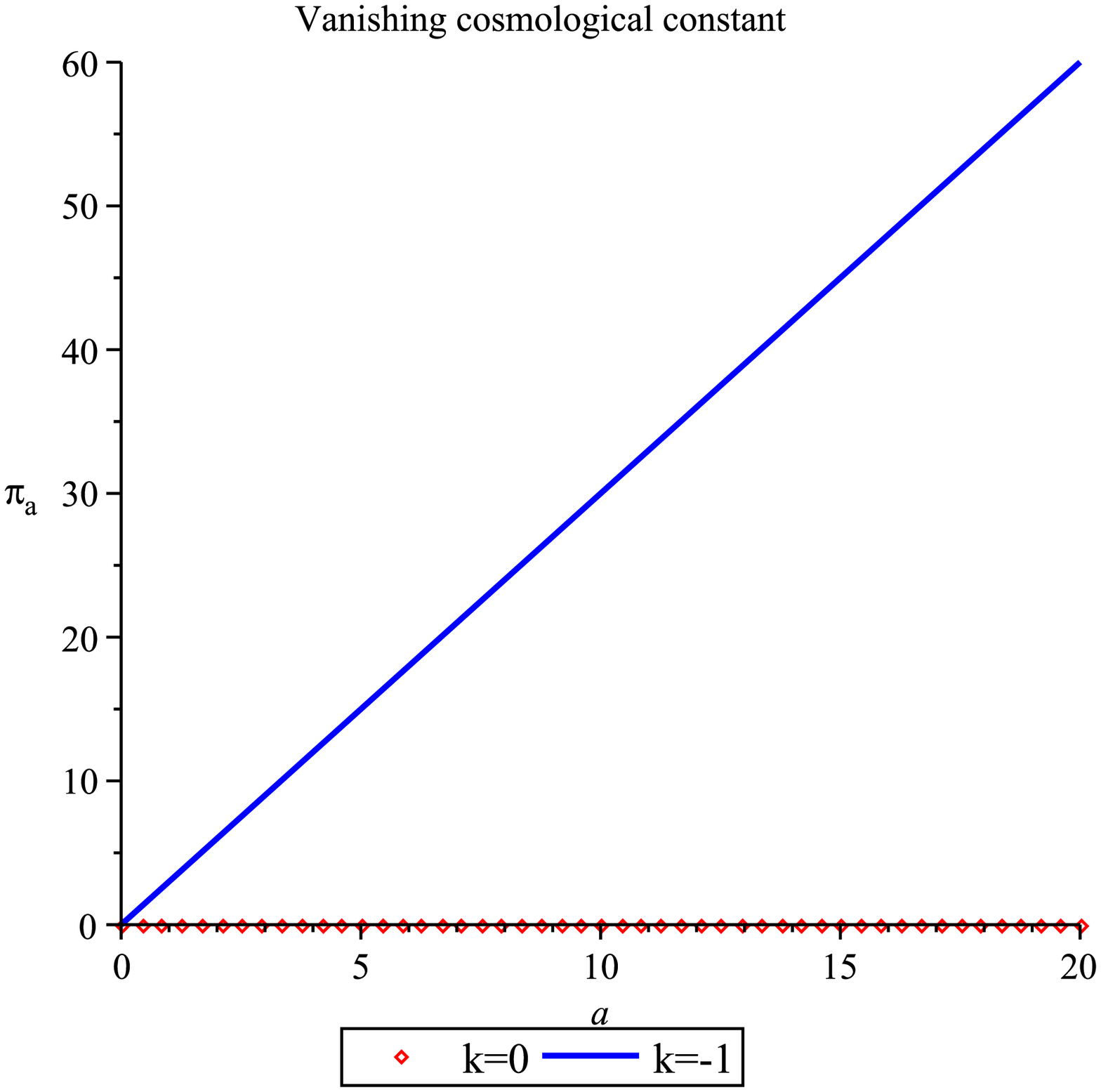}
\includegraphics[height=55mm]{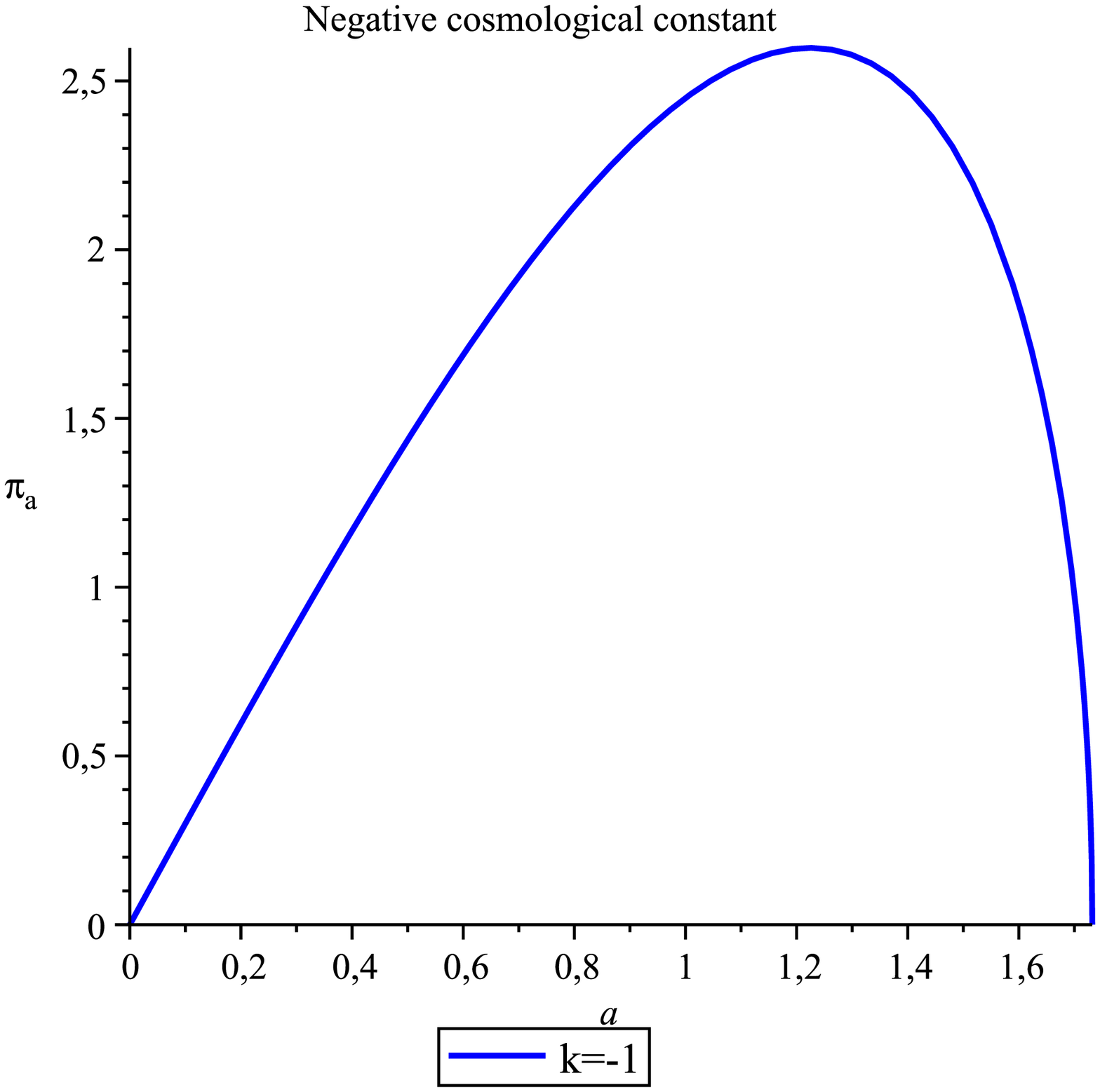}
\includegraphics[height=55mm]{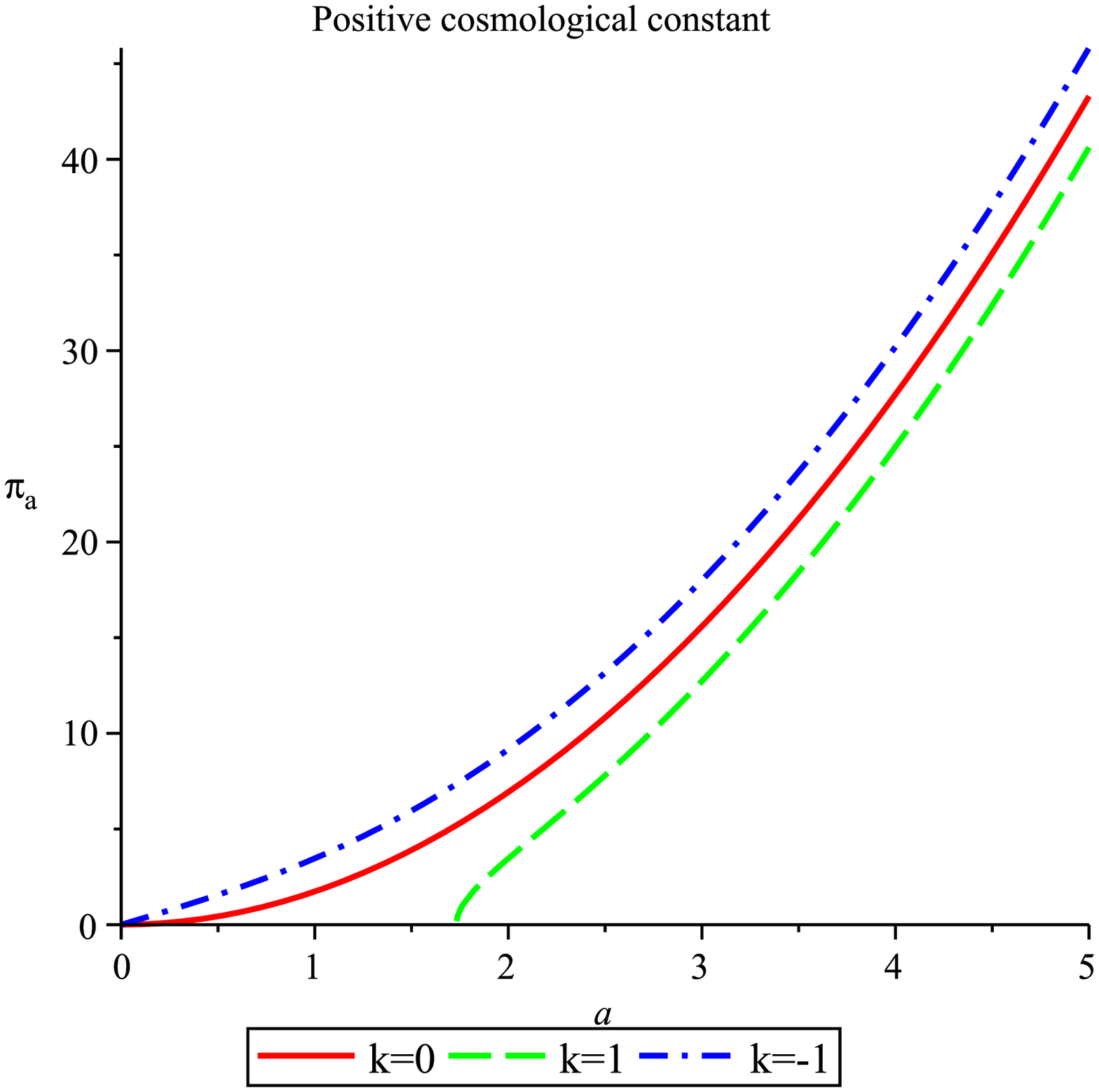}
\caption{Trajectories $\pi_a(a)$ in the different vacuum FRW models. In these plots we have used
the conventions $4\pi G=1$, and $\Lambda=\pm1$ for the cases with non-vanishing cosmological
constant.
\label{frw-traj}}
\end{center}
\end{figure}

In the vacuum case under study it is very easy to obtain the trajectories in phase space,
$\pi_a(a)$, using the Hamiltonian constraint $C_{grav}=0$. They are given by
\be
\pi_a(a)\pm\frac{a}{4\pi G}\sqrt{3\Lambda a^2-9k}.
\ee
In fig. \ref{frw-traj} we show them just for positive $\pi_a$.

This vacuum case is not very interesting as long as there are no degrees of freedom. In fact, the
geometry is totally fixed by the Hamiltonian constraint $C=0$. As a consequence many trajectories
are rather boring, like e.g that of the flat case ($k=0$) with vanishing cosmological
constant. Other trajectories are not even real, as happens for the cases
$\{\Lambda=0,k=1\}$, $\{\Lambda<0,k=0\}$, $\{\Lambda<0,k=1\}$. In order to have more interesting
phenomenology one should add matter to the model. Actually, already the simplest form of matter,
a minimally coupled massless scalar, provides a quite non-trivial evolution, as we explain below.
However in this paper we choose to focus on the vacuum case, as a first step to start with in our
derivation of effective dynamics for cosmology using the 2-vertex graph.
In the following sections, we will derive this effective dynamics and discuss if it can be mapped in some regime to the cases occurring in the FRW model. We leave for the future the study of
non-vacuum effective models.

\subsection{A Remark on the Coupling to Scalar Field and Deparametrization}
\label{scalar}

Although it is not the most physically relevant type of matter, it is nevertheless interesting to couple a free massless scalar field to the FRW cosmology, because the coupled system now has one physical degree of freedom and that we can choose the scalar field as an internal clock allowing to deparameterize the gravitational evolution.

Let us thus introduce a homogeneous scalar field, described by a single canonical pair, $\{\vphi,p_\vphi\}=1$, where $\vphi$ is the massless scalar and $p_\vphi$ its momentum. This modifies the Hamiltonian constraint, by adding the contribution of the new matter field:
\be
C=C_{grav}+\f{p_\vphi^2}{2a^3}
=-\frac{2\pi G}{3} \frac{\pi_a^2}{a}+\f{p_\vphi^2}{2a^3}\,,
\ee
where we simplified the gravitational contribution by considering the flat case with vanishing cosmological constant, $k=0,\,\Lambda=0$, which will be the most relevant in the rest of the paper. Obviously $p_\vphi$ is a constant of motion since it Poisson commutes with the Hamiltonian constraint. We can then deparameterize the system taking $\vphi$ as the internal time and $p_\vphi$ as the physical Hamiltonian which generates evolution with respect to the time $\vphi$. Solving the Hamiltonian constraint,
\be
p_\vphi=\pm \sqrt{\f{4\pi G}{3}} \,\pi_a a\,=\,\pm H\,,
\ee
we obtain two branches, which are the time reversal of each other. Considering the positive branch, we can look at the equation of motion:
\be
\pp_\vphi a = \{a,H\}=\sqrt{\f{4\pi G}{3}} \,a
\quad\Rightarrow\quad
\f{\pp_\vphi a}{a}=\sqrt{\f{4\pi G}{3}}
\quad\Rightarrow\quad
a(\vphi)=a(\vphi_0)e^{\sqrt{\f{4\pi G}{3}}(\vphi-\vphi_0)},
\ee
with a constant expansion rate with respect to the internal time.

One can easily go back to the proper time $t$ (defined by taking the lapse $\cN=1$) by computing the evolution of the internal time $\vphi$ by the Hamiltonian constraint $C$:
\be
\f{d\vphi}{dt}=\{\vphi,C\}=\f{p_\vphi}{a^3},\qquad
d\vphi\, =\,\f{p_\vphi}{a^3} \,dt\,.
\ee
From this, we can compute the evolution of the scale factor and recover the standard Friedman equation:
\be
\pp_t a = \pp_\vphi a\,\f{d\vphi}{dt}=\sqrt{\f{4\pi G}{3}}\f{p_\vphi^2}{a^2}
\quad\Rightarrow\quad
\left(\f{\pp_t a}{a}\right)^2=\f{8\pi G}{3}\rho
\qquad\textrm{with}\quad
\rho\equiv=\f12\f{p_\vphi^2}{a^6},
\ee
in terms of the matter density $\rho$. We would recover the same result from computing the Hamiltonian flow of $C$ on the variables $a$ and $\pi_a$.

The case of the massless scalar field is interesting because it is the simplest matter field to couple to the FRW cosmology: it allows to introduce one physical degree of freedom in the system and to explore regimes where $C_{grav}$ does not vanish (as in the vacuum case). It is also particularly relevant to our context because it will be straightforward to introduce on the 2-vertex graph, as we will explain below in section \ref{vidotto}.

\subsection{Cosmological Hamiltonian on the 2-Vertex Graph}

Let us now add an appropriate Hamiltonian to our kinematical action \eqref{reduced-action}.
We require this Hamiltonian to be, first $SU(2)$-invariant, so that the gauge
invariance of the theory is preserved, and second $U(N)$-invariant, so that it leads to  homogeneous and isotropic dynamics.
In this way, the resulting dynamics will be consistent with the kinematical setting, and
even more, it may be regarded as generating the reduced (homogeneous and isotropic) sector of the
full theory (as the FRW Hamiltonian does for general relativity).

In order to construct such an ansatz for the Hamiltonian, the simplest $SU(2)$ invariants on a given graph are the holonomies along its loops, or more generally the generalized holonomies constructed as a product of $E$ and $F$ observables as defined in \cite{un5}.
In the case of the 2-vertex graph, we consider the elementary loops made of two edges. These generalized holonomy observables are then simply $E^\alpha_{ij}E^\beta_{ij}$, $F^\alpha_{ij}F^\beta_{ij}$, and $\bF^\alpha_{ij}\bF^\beta_{ij}$, for the pair of edges $i,j$.
Now, the symmetry reduction to the homogeneous and isotropic sector implemented by the
$U(N)$-invariance, reduces the above $SU(2)$ invariants to the $U(N)$-invariant terms $\tr
E^\alpha=\tr E^\beta$, $\tr {}^tE^\alpha E^\beta\propto(\tr E^\alpha)^2$,
$\tr F^\alpha F^\beta$, and $\tr\bF^\alpha\bF^\beta$, where we look at the $E$'s and $F$'s as $N\times N$ matrices indexed by the edges.  As proved in \cite{un3,un5}, these are the only $\U(N)$ invariant polynomial in the spinor variables and of lowest order (beside the trivial quadratic invariant $\tr E^\alpha \propto \lambda$).

Made up of these  terms, we will consider the following ansatz for
an action with non-trivial dynamics, on the $U(N)$-invariant two-vertex spinor network
\begin{align}
\label{UN_action}
S_{homo}[z_k,w_k]\,&=\,S^{(0)}_{homo}[z_k,w_k]-\int dt\, \tilde{\cal N}
\,(H_{homo}[z_k,w_k]-H_o),\nonumber\\
H_{homo}[z_k,w_k]\,&\equiv\, \gamma^o\tr {}^tE^\alpha E^\beta+\gamma^+\tr F^\alpha F^\beta+
\gamma^-\tr\bF^\alpha\bF^\beta+\frac{\gamma^1}{4}[\tr {}^tE^\alpha]^{3}
\end{align}
Here $\gamma^o$, $\gamma^+$, $\gamma^-$, and $\gamma^1$ are some real coupling constants.
The above ansatz was actually introduced in \cite{un5}, but we have added an additional term in $[\tr {}^tE^\alpha]^{3}$ with coupling constant $\gamma^1$. This term corresponds to a cosmological constant term, as we explain below.
Then $\tilde{\cal N}$ is a Lagrange multiplier imposing the Hamiltonian constraint
$H_{homo}[z_k,w_k]-H_o=0$. The  real constant $H_o$ accounts for the fact that
the energy of the fundamental state in the quantum theory could be nonzero
(similarly to the energy of the fundamental state of the harmonic oscillator or of the hydrogen atom is not null).

Using the expression of the matrices $E^v$ and $F^v$ in terms of $\lambda_v$ and $U^v$ and that remembering that the $U(N)$-invariance implies $\overline{U^\beta}=e^{i\phi}{U}^\alpha$ and $\lambda_\alpha=\lambda_\beta=\lambda$, the above action reduces to a single degree of freedom in the homogeneous and isotropic sector:
\be
S[\lambda,\phi]\,=\,-2\int dt \lambda\pp_t\phi
-\int dt \,\tilde{\cal N}(H[\lambda,\phi]-H_o),\qquad H(\lambda,\phi)\,=\,
2\lambda^2(\gamma^o-\gamma^+e^{-2i\phi}-\gamma^-e^{2i\phi}+\gamma^1\lambda)
\label{2vertex_cosmo_H}
\ee
We will further choose $\gamma^+=\gamma^-\equiv\gamma/2\in\R$ so that the Hamiltonian is real \footnotemark
and given by
\be
H(\lambda,\phi)\,=\,2\lambda^2[\gamma^o-\gamma\cos(2\phi)+\gamma^1\lambda].
\label{2vertex_cosmo_Hbis}
\ee
\footnotetext{
In order to get a real Hamiltonian, we just need to require that $\gamma^+=\overline{\gamma^-}$. We can nevertheless take $\gamma^+=\gamma^-\in\R$ since we can re-absorb any phase in a constant off-shift for the angle $\phi$.
}

\medskip

This Hamiltonian constraint is actually the (gravitational part of the) effective action for the FRW cosmology in loop quantum cosmology (LQC) in its older version \cite{aps1}, with an exact matching at least in the flat case with vanishing cosmological constant. Indeed this similarity between the two-vertex model and the effective dynamics of LQC  was already pointed out in \cite{un5}. This already establishes a link between our 2-vertex graph Hamiltonian and FRW cosmology. The interested reader will find details on the LQC effective dynamics in \cite{aps1,aps3,tom,vand,luc,apsv,skl}.

We will furthermore show in the following section III.D that this Hamiltonian constraint is also recovered directly from a discretization of the loop quantum gravity Hamiltonian on the 2-vertex graph and matches an earlier proposal by Rovelli and Vidotto  \cite{LQGcosmo1}.

In the loop quantum cosmology context, one would include matter in the model, at least a massless scalar field, and then the main prediction is a big bounce replacing the big bang singularity due to the $\cos$ term (the ``holonomy correction" in LQC). The problem of such dynamics is that the matter density at the bounce depends on the initial conditions at late times and could be classical and not in the deep quantum regime as would be expected. This issue was addressed in the LQC framework by moving on to an improved dynamics scheme \cite{aps3}. We do not go yet in this direction but we will comment shortly on the relevance of this scheme for our approach at the end of this section.

\medskip

We can now compare $S[\lambda,\phi]$ with the standard FRW action given in Eq.
\eqref{S-frw}. In order to do that, let us express $S_{FRW}$ in our variables $(\lambda,\phi)$.
Since $\lambda$ has the interpretation of an area, we can take the following
canonical transformation to relate the variables $(a,\pi_a)$, with Poisson bracket
$\{a,\pi_a\}=1/V_o$, with our variables $(\lambda,\phi)$, with Poisson bracket
$\{\lambda,\phi\}=1/2$:
\be\label{canonical-aA}
a\,=\,\frac{l}{V_o^{1/3}}\sqrt{\lambda},\qquad\pi_a=\frac{4}{lV_o^{2/3}}\sqrt{\lambda}\phi,
\ee
where $l$ serves as a unity of length defining the relation between our dimensionless area $\lambda$
with the dimensionless scale factor $a/l$.
This identification is natural due to the geometrical interpretation of $\lambda$ as the boundary area between the two vertices $\alpha$ and $\beta$ and the physical role of the scale factor $a$, both defining the unique length unique in the homogeneous and isotropic setting of FRW cosmology.
As before, $V_o$ represents the (dimensionless) volume of
a finite cell $\mathcal{V}$, measured with respect to the fiducial metric ${}^oq_{ab}$ of the
3-dimensional spaces with constant curvature.
Namely, we regard the graph as dual (not to the whole space but just) to a finite region
$\mathcal{V}$, which it is enough for determining the dynamics of the space because of the
homogeneity.

\begin{figure}[h]
\begin{center}
 \includegraphics[height=55mm]{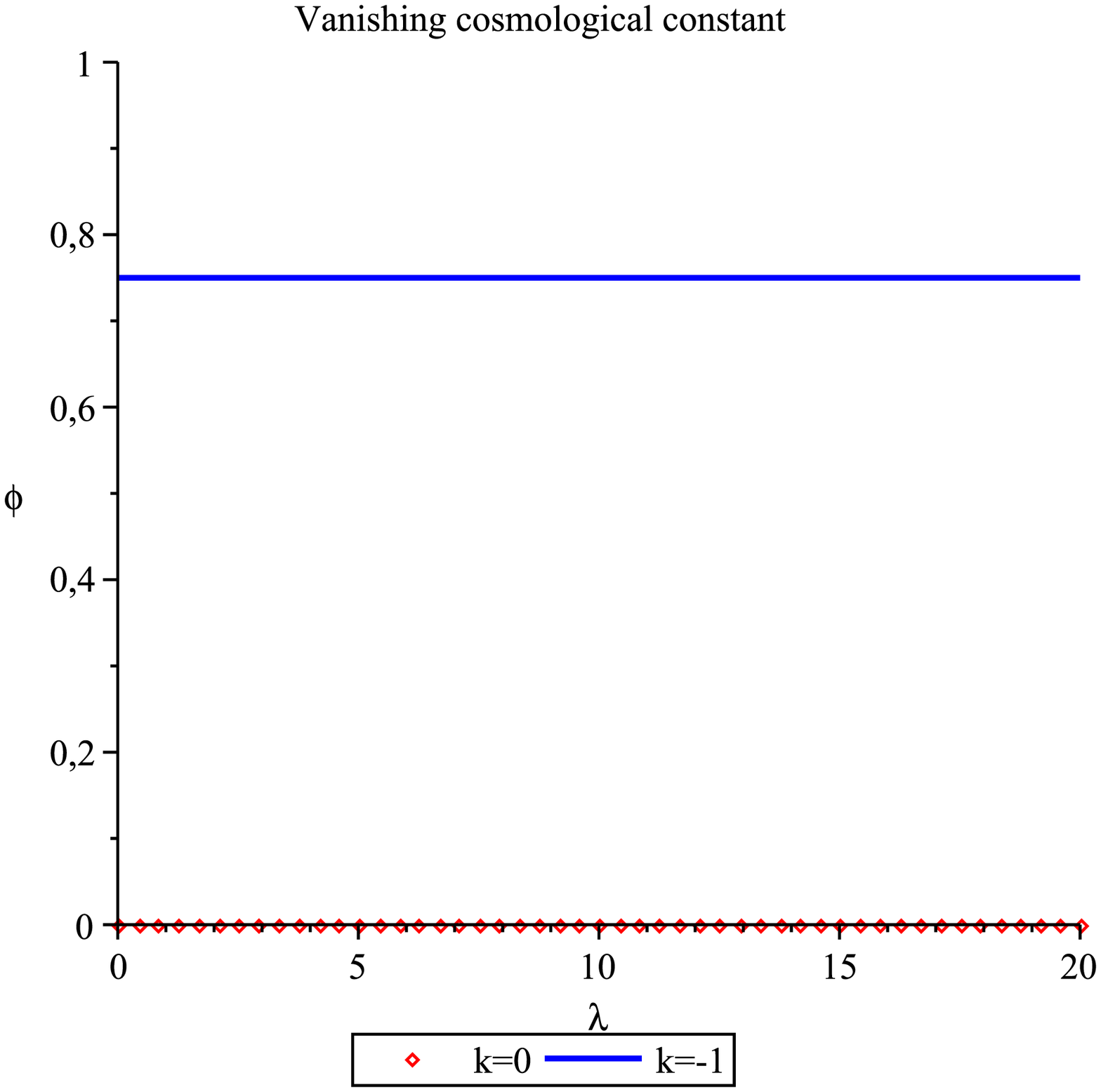}
\includegraphics[height=55mm]{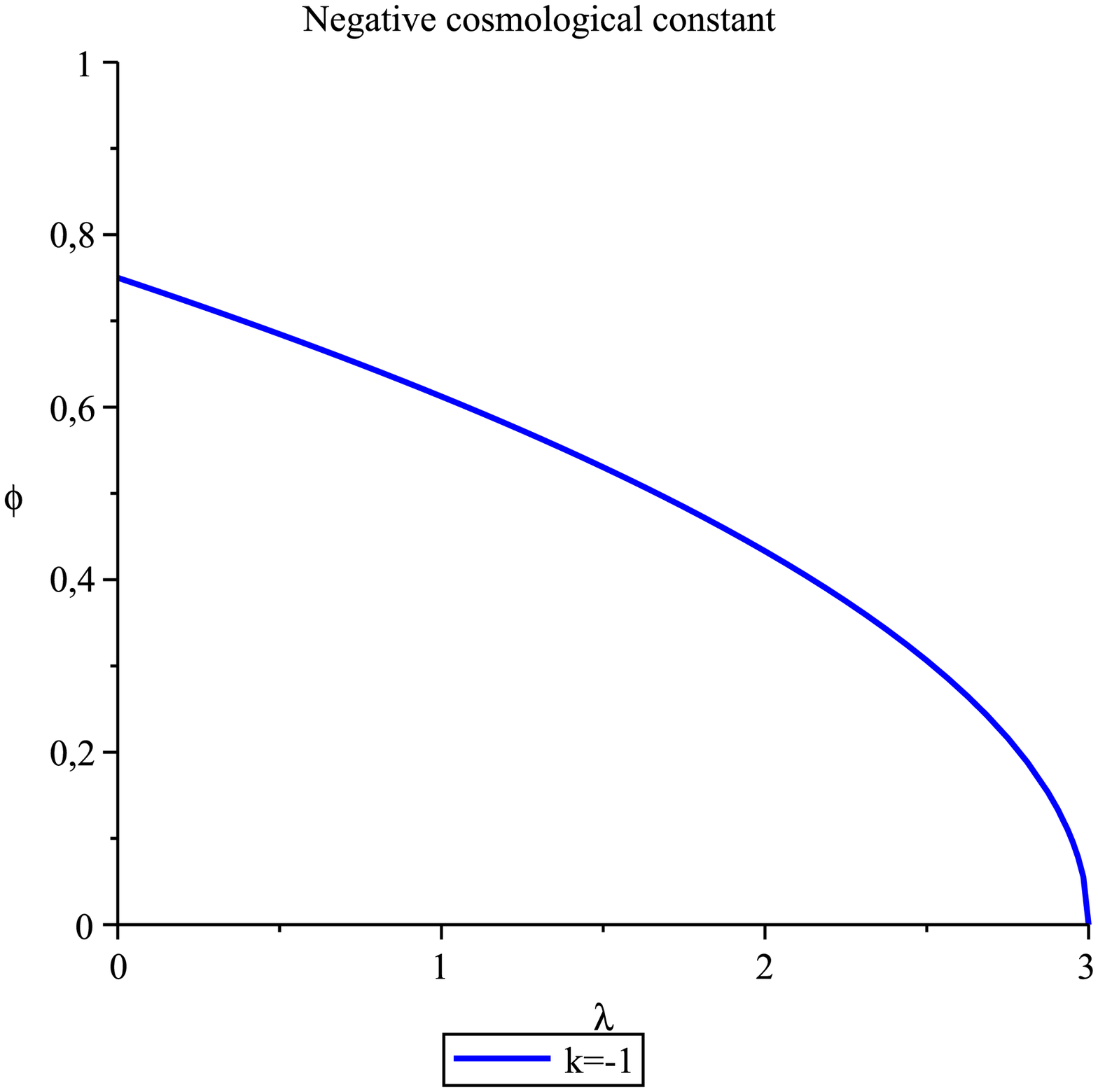}
\includegraphics[height=55mm]{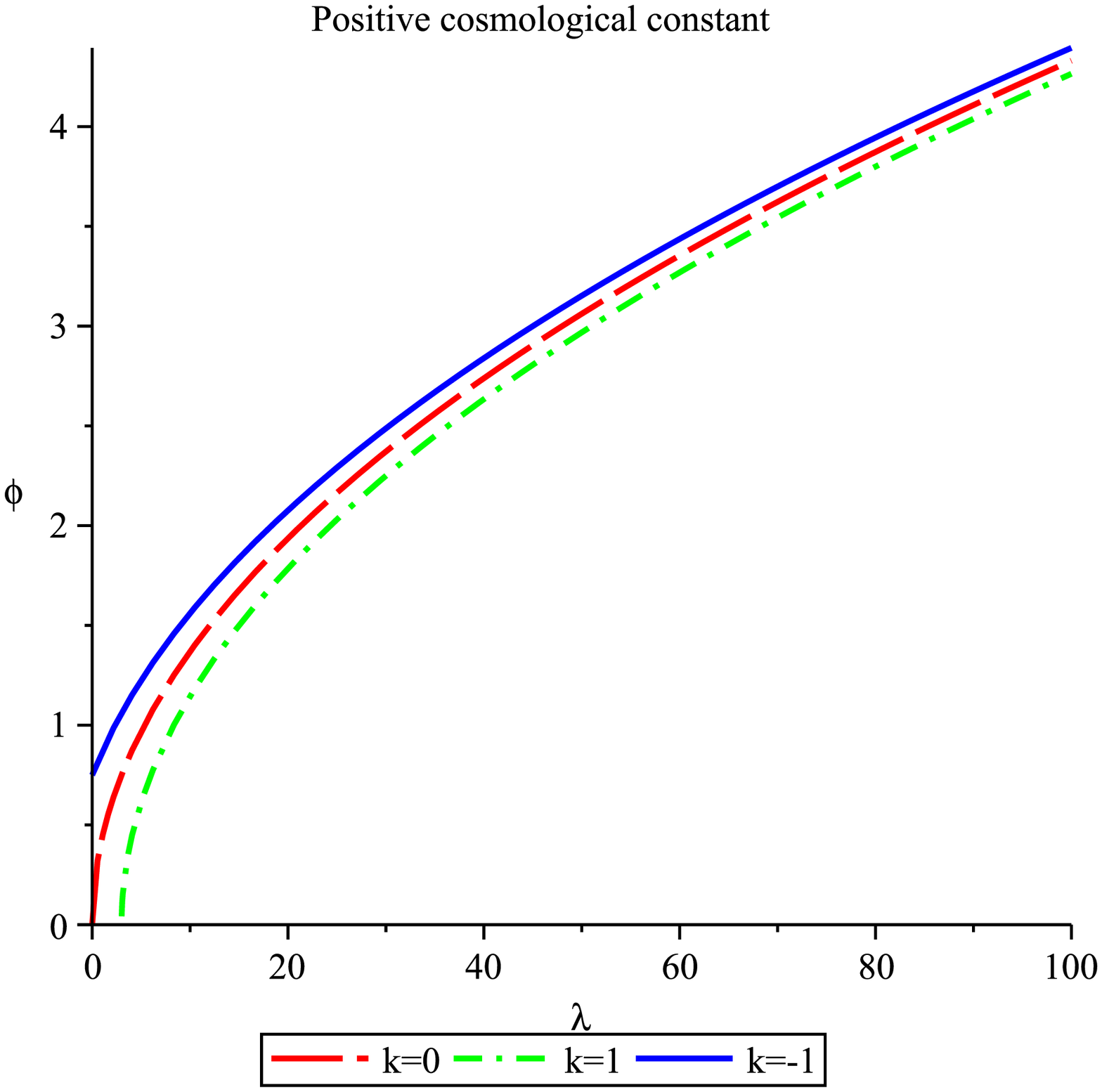}
\caption{Trajectories $\phi(\lambda)$ in the different vacuum FRW models. In these plots we have
used the conventions $V_o=1$, $4\pi G=1$, and $\Lambda=\pm1$ for the cases with non-vanishing
cosmological constant.
\label{frw-traj-phi}}
\end{center}
\end{figure}

Employing the above canonical transformation we obtain the Hamiltonian of the FRW model
in our variables. It has the form
\be\label{hamFRW-A}
H_{FRW}\,\equiv\,V_oNC_{grav}\,=\,-\frac{32\pi G}{3l^3}
\lambda^2\phi^2-\frac{3Kl}{8\pi G}\lambda^2+\frac{\Lambda l^3}{8\pi
G}\lambda^{3},
\ee
where we have chosen as lapse $N=\lambda^{3/2}\propto a^3$ (to ensure the correct  matching of the scaling of the  $\lambda^2\phi^2$ term in $H_{FRW}$ with the $\lambda \cos 2\phi$ term in $H$), and we have defined $K\equiv kV_o^{2/3}$.
Then the trajectories in phase space in terms of these variables are now given by
\be
\phi(\lambda)=\pm\frac{l^2}{4(4\pi G)}\sqrt{3\Lambda l^2\lambda-9K}.
\ee
The  positive and real trajectories (shown before in fig. \ref{frw-traj}) have now the graphic
shown in fig. \ref{frw-traj-phi}.

\bigskip

Now in the limit $\phi\rightarrow0$ we can do the approximation $\cos(2\phi)\approx1-2\phi^2$
and identify our Hamiltonian $H$ with that of the FRW model. Indeed, upon that approximation both
Hamiltonians agree by doing the identifications
\be\label{ident}
\gamma^o\,=\,-\frac{8\pi G}{3l^3}-\frac{3Kl}{16\pi G},\qquad
\gamma\,=\,-\frac{8\pi G}{3l^3},\qquad
\gamma^1\,=\,\frac{\Lambda l^3}{16\pi G},\qquad
H_o\,=\,0.
\ee
As we said before, the term with coupling constant $\gamma^1$ represents a
cosmological constant term. On the other hand, the other two terms, with coupling constants
$\gamma^o$ and $\gamma$ account for the curvature term and for the other term, the only term
remaining when there is neither cosmological constant nor curvature.
The energy off-set $H_o$ naturally vanishes in this identification but that's mainly because we are considering the classical limit where the scale factor (and thus $\lambda$) is large: in that case, the $H_o$ becomes simply negligible and actually setting it to 0 or not will not affect at all the classical behavior. We will thus keep $H_o$ arbitrary for the sake of completeness when analyzing the classical trajectories of our Hamiltonian on the 2-vertex graph. Moreover, a non-vanishing $H_o$ allows to explore the regime where $C_{grav}$ does not vanish, which will become useful as soon as we couple matter to the system.

Outside the regime in which the above approximation is valid, the Hamiltonian of the two-vertex
model with the above identifications reads
\be\label{H-phi}
H(\lambda,\phi)=2\lambda^2\left[-\left(\frac{8\pi G}{3l^3}+\frac{3Kl}{16\pi G}\right)+\frac{8\pi
G}{3l^3}\cos(2\phi)+\frac{\Lambda l^3}{16\pi G}\lambda\right].
\ee
Alternatively, we can express it in the variables $\{a,\pi_a\}$ commonly employed in cosmology, in
which case is given by
\be\label{H-a}
H(a,\pi_a)=2V_o^{4/3}\frac{a^4}{l^4}\left[-\left(\frac{8\pi G}{3l^3}+\frac{3Kl}{16\pi
G}\right)+\frac{8\pi
G}{3l^3}\cos\left(\frac{V_o^{1/3}}{2}\frac{l^2\pi_a}{a}\right)+\frac{\Lambda
l^3}{16\pi G}V_o^{2/3}\frac{a^2}{l^2}\right].
\ee
We would like to emphasize that here we are pushing forward the identifications of Eq.
\eqref{ident} further from the regime in which $\phi\rightarrow0$, where they have been obtained.

Our goal is to analyze whether the Hamiltonian of the two-vertex model $H(\lambda,\phi)$ can be
regarded as an effective Hamiltonian for the FRW models. Upon this interpretation, this Hamiltonian
introduces corrections to the results predicted by general relativity. On the one hand we will have
corrections coming from the gauge invariance of loop gravity, and they are negligible whenever
the angle $\phi$ approaches a vanishing value. On the other hand we are considering that the
Hamiltonian of the two-vertex model  is fixed to be equal to some constant value $H_o$ that does not
necessarily vanish, unlike in general relativity. This fact will also induce corrections.

\begin{figure}[h]
\begin{center}
 \includegraphics[height=55mm]{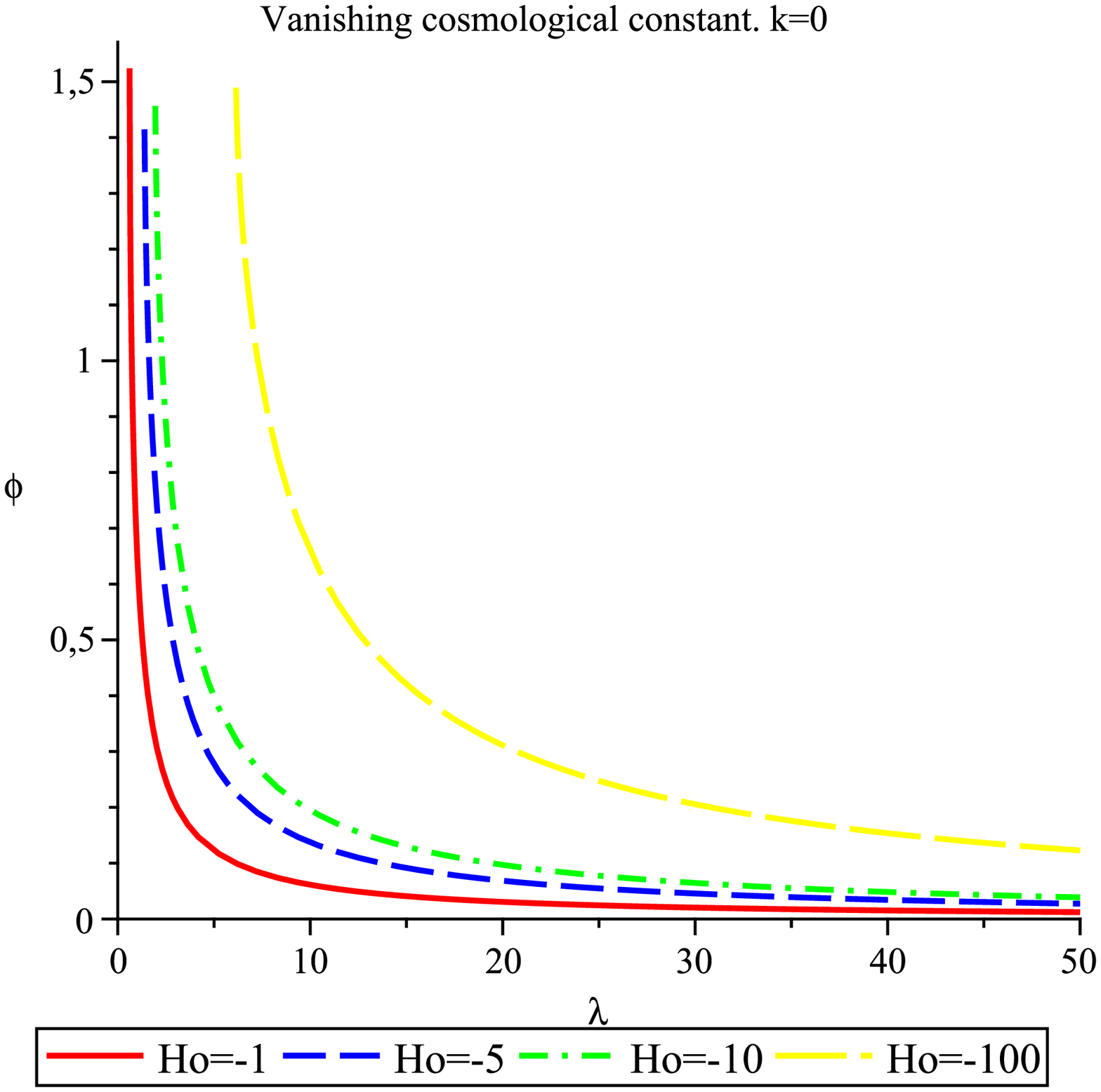}
\includegraphics[height=55mm]{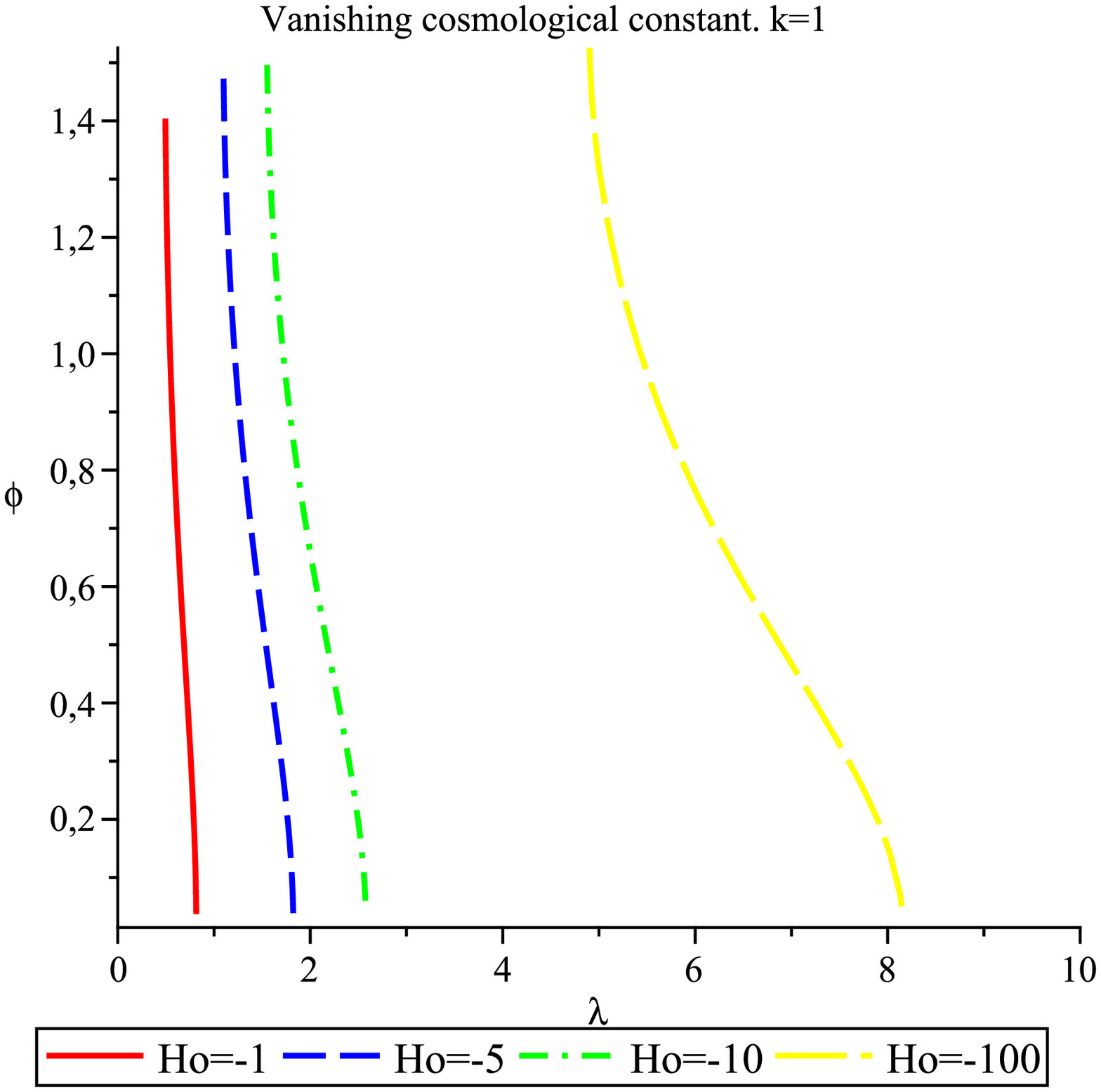}
\includegraphics[height=55mm]{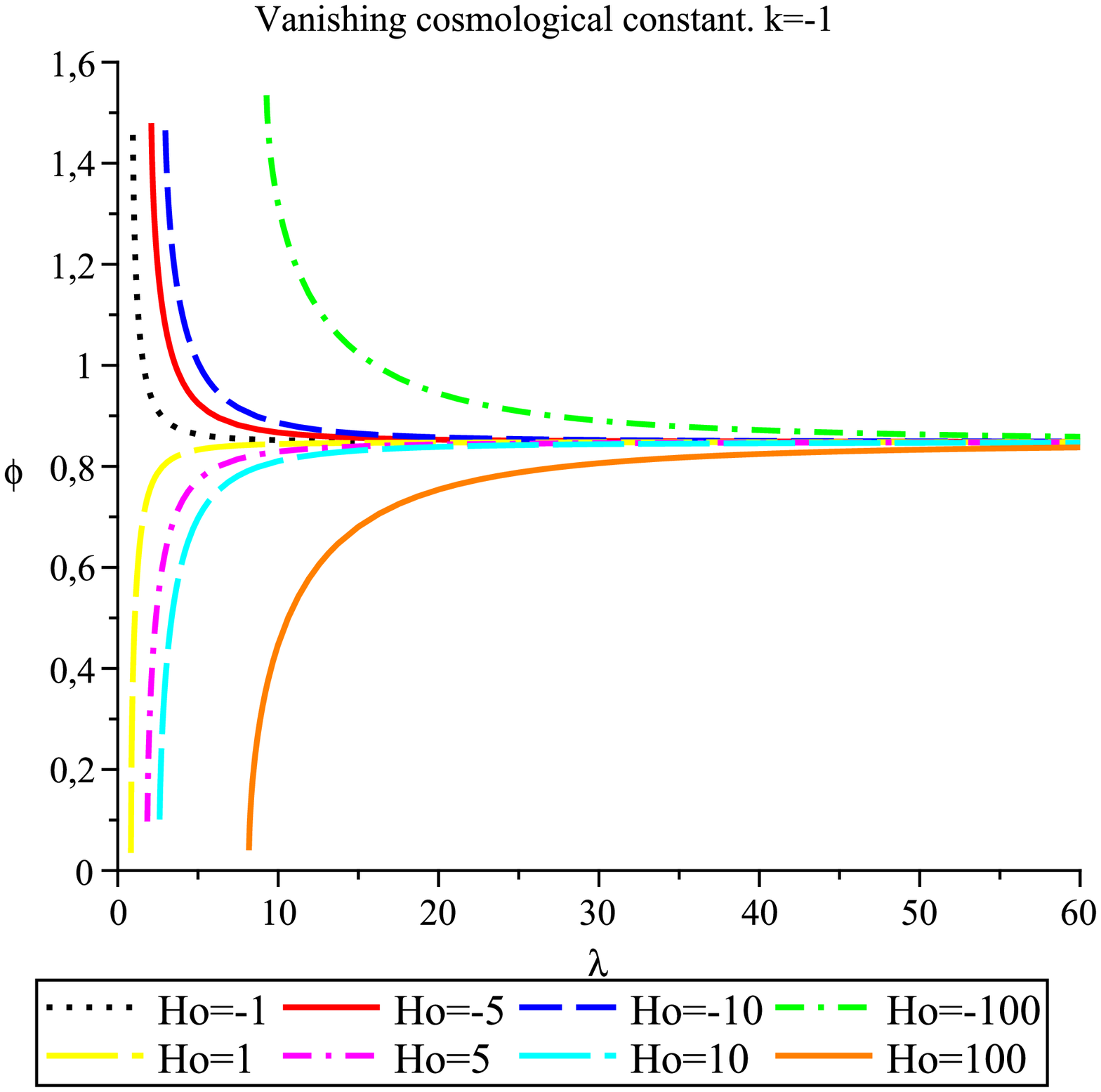}
\includegraphics[height=55mm]{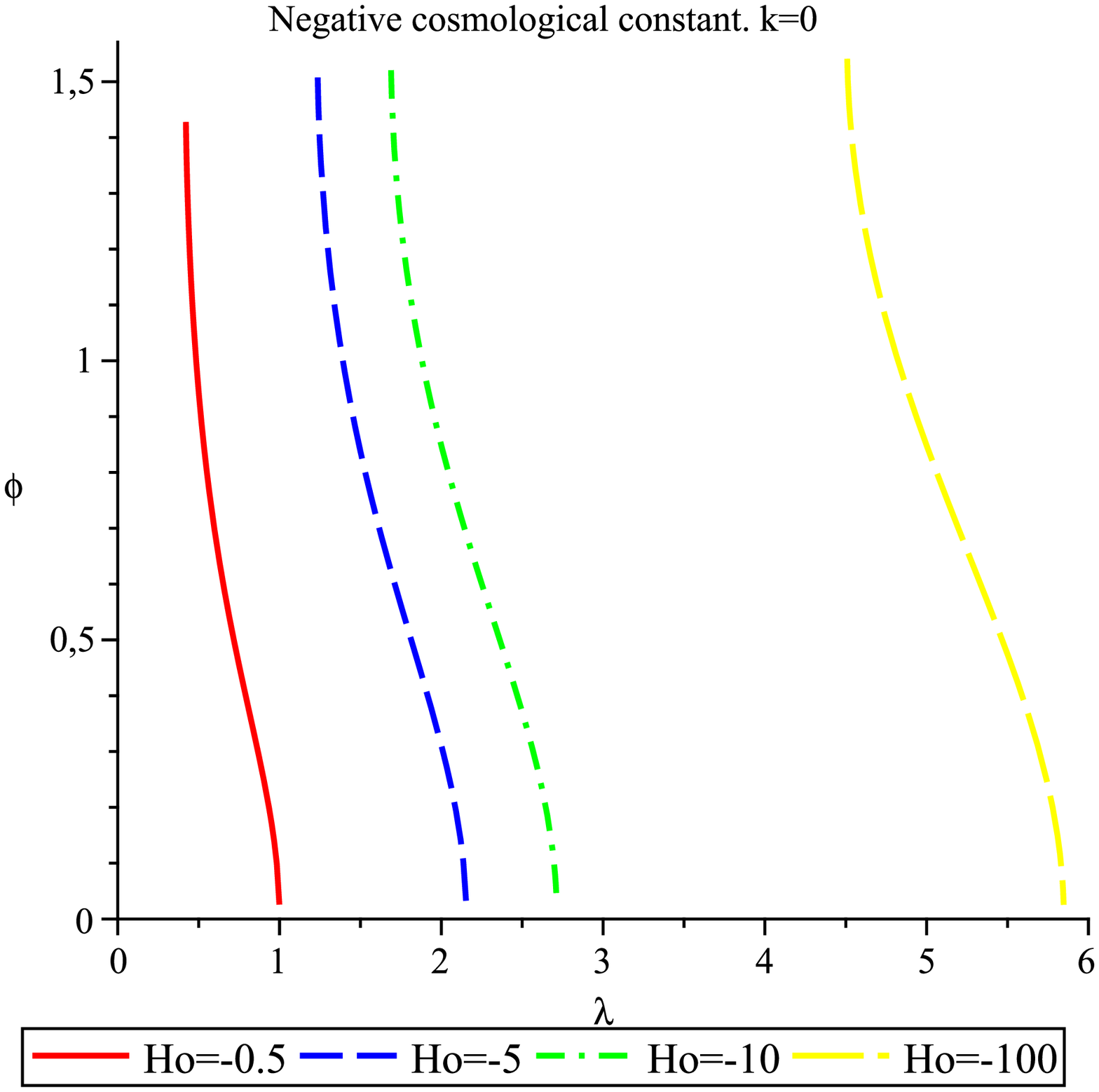}
 \includegraphics[height=55mm]{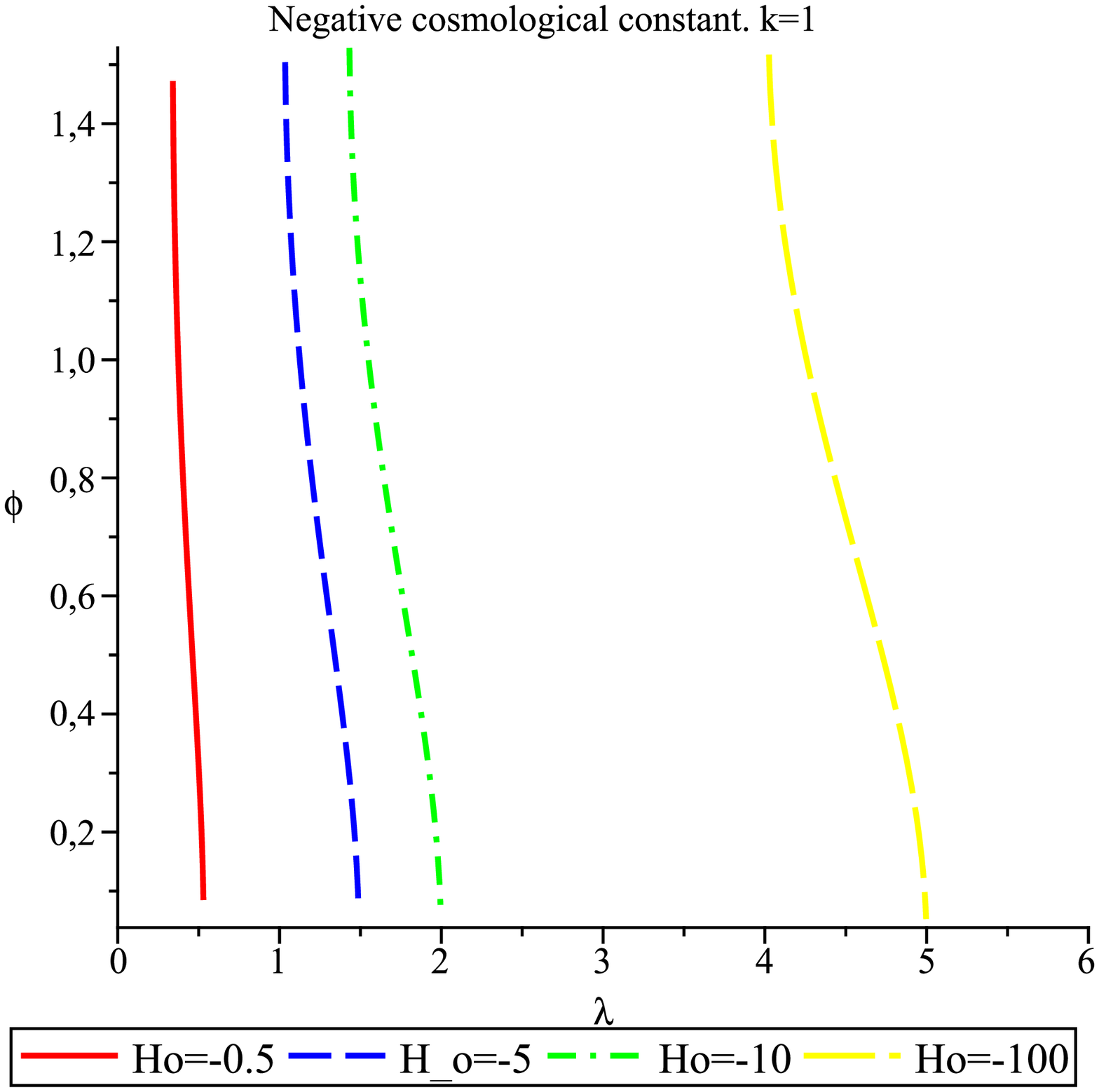}
 \includegraphics[height=55mm]{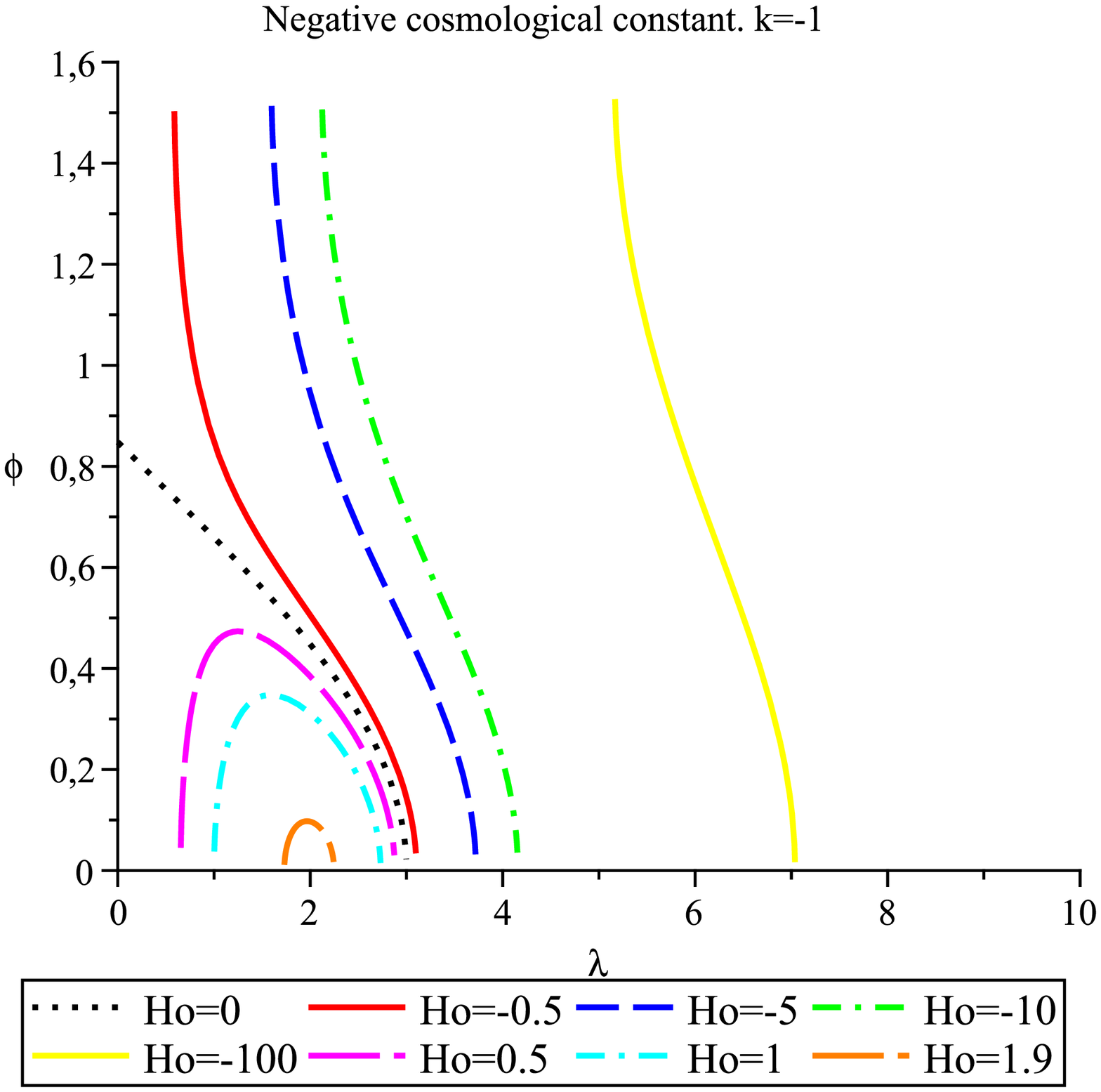}
 \includegraphics[height=55mm]{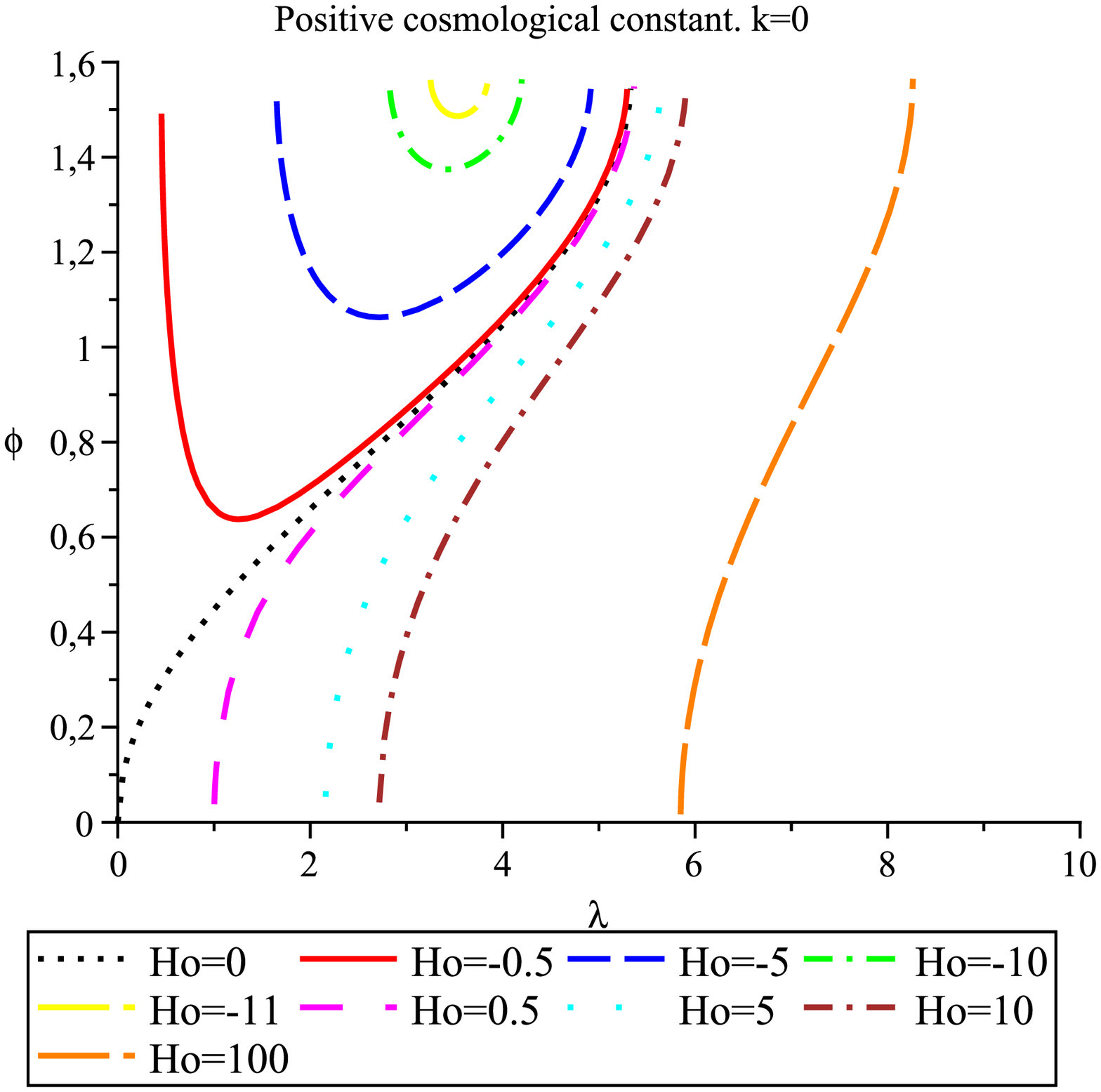}
\includegraphics[height=55mm]{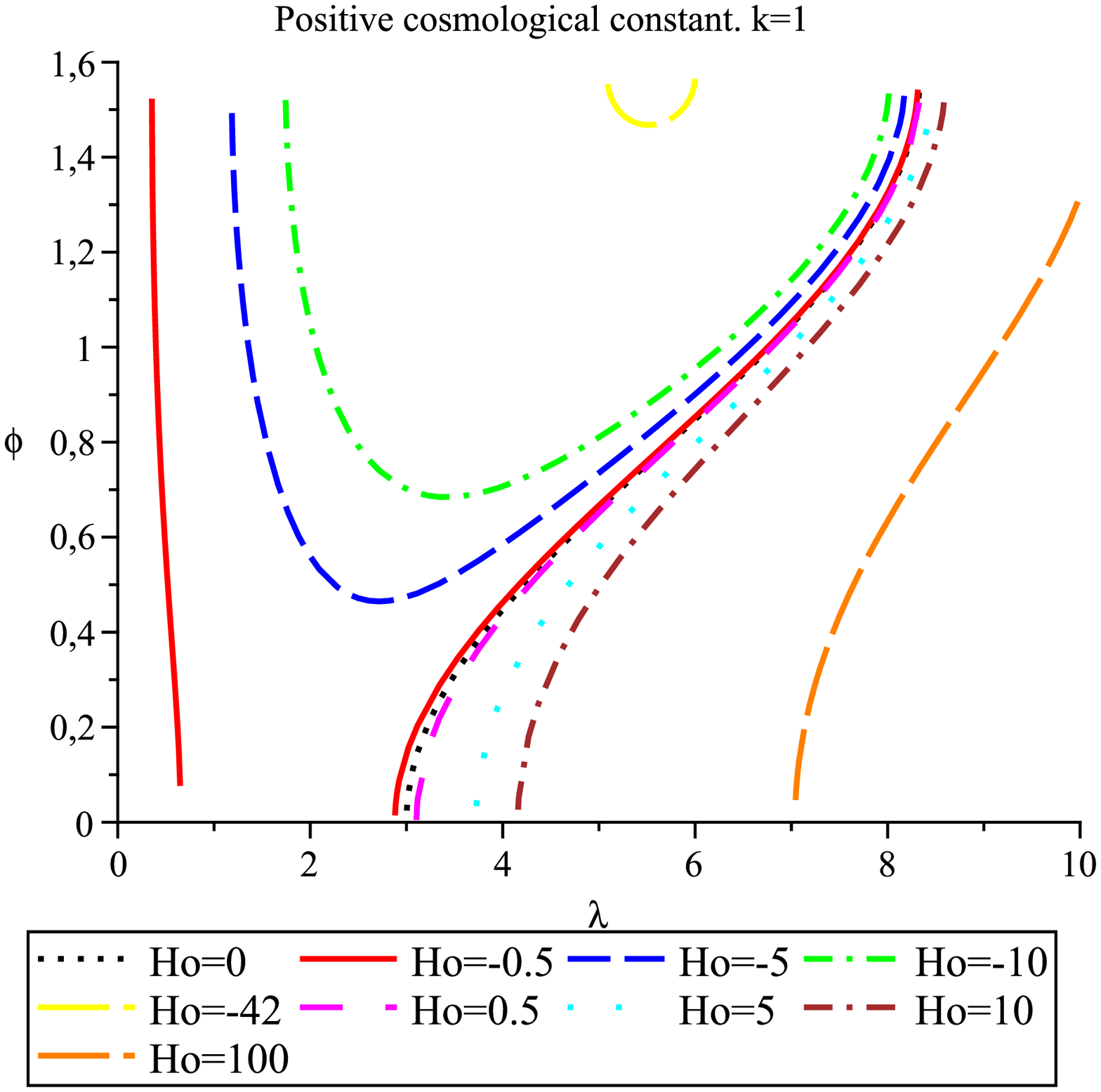}
\includegraphics[height=55mm]{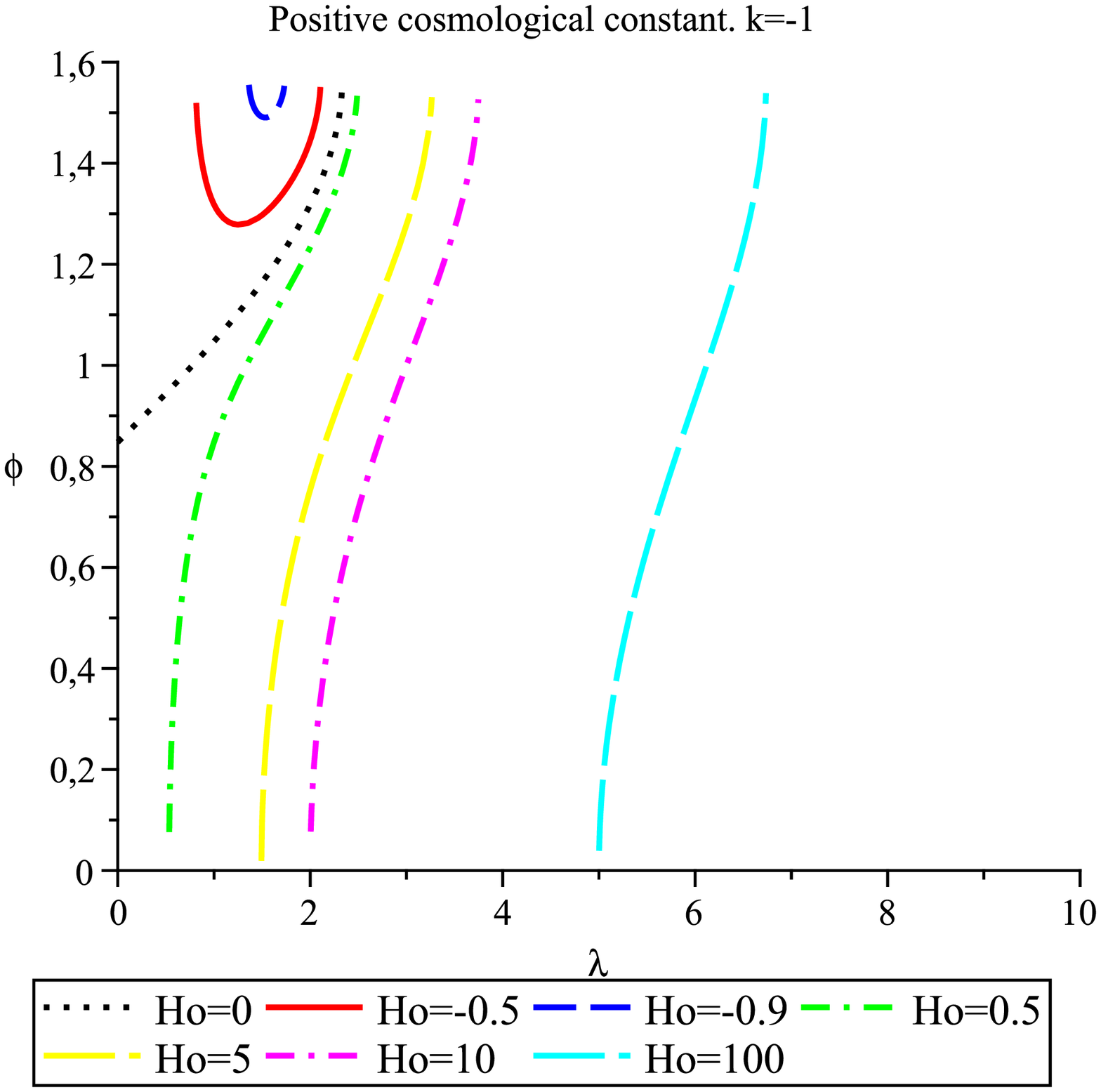}
\caption{Trajectories $\phi(\lambda)$ in the different 2-vertex models. In these plots we have used
the conventions $l=1$, $V_o=1$, $4\pi G=1$, and $\Lambda=\pm1$ for the cases with non-vanishing
cosmological
constant.
\label{eff-traj-phi}}
\end{center}
\end{figure}

Before doing the comparison between our two-vertex model and the FRW models, we first need
to study the phase space trajectories resulting in the two-vertex model. Using the constraint
$H(\lambda,\phi)=H_o$ we easily get
\be
\phi(\lambda)=\frac1{2} \arccos\left[1+\left(\frac{l}{4\pi
G}\right)^2\left(\frac{9K}{8}-\frac{3\Lambda l^4}{8}\lambda\right)+\frac{l^3}{4\pi
G}\frac{3H_o}{4\lambda^2}\right]
\ee
These trajectories are drawn in fig. \ref{eff-traj-phi}. The graphics only show the trajectory for
$\phi\in[0,\pi/2]$. In the range $\phi\in[\pi/2,\pi]$ the trajectory is given by the specular image
of that in the previous range with respect to the edge $\phi=\pi/2$, and for bigger range it is
periodic with period $\pi$. We show them for different values of $H_o$. This constant is usually
bounded, either from below or from above, as shown in table \ref{table}. We see that indeed it can
not be zero for many cases.

\begin{table}[!hbt]
\begin{center}
\begin{tabular}{|l|c|c|c|}
\hline
 & $k=0$ & $k=1$ & $k=-1$ \\
\hline
$\Lambda=0$ & $H_o<0$ & $H_o<0$ & $H_o\neq0$\\
\hline
$\Lambda=-1$ & $H_o<0$ & $H_o<0$ & $H_o\lesssim 1.9$\\
\hline
$\Lambda=1$ & $H_o\gtrsim -11$ & $H_o\gtrsim -42$ & $H_o\gtrsim -0.9$\\
\hline
\end{tabular}
\caption{Ranges of $H_o$. \label{table}}
\end{center}
\end{table}

Let us now compare the trajectories of the FRW model in the six cases with real solution (fig.
\ref{frw-traj-phi}) with the analog cases in the two-vertex model:
\begin{itemize}
 \item $\Lambda=0$, $k=0$: The trajectory of the two-vertex model agree with that of the FRW model
for increasing values of $\lambda$. Moreover $\lambda$ is bounded from below. The bound depends on
the particualr value of $H_o$ and it is always bigger than zero.

\item $\Lambda=0$, $k=-1$: For increasing value of $\lambda$ the trajectory tends to a constant
value (independent of $H_o$), as in the FRW model. This value is
\[\phi=\frac1{2} \arccos\left[1+\left(\frac{l}{4\pi
G}\right)^2\frac{9K}{8}\right]\]
and agrees with that of the FRW model, $\phi=3l^2V_o^{1/3}/16\pi G$, for an appropriate value of the
length parameter $l$. As in the previous case, $\lambda$ is strictly
positive and its minimum depends on the particular value of $H_o$.

\item $\Lambda<0$, $k=-1$: Only for $H_o\geq 0$, $\phi$ is bounded from above, as happens in the FRW
model. For $H_o=0$ the trajectory is quite similar to that of the FRW model, approaching it as
$\lambda$ tends to its maximum value. More interestingly, for values of $H_o$ slightly bigger than
zero, the trajectory still agrees with that of FRW for the maximum value of $\lambda$, but it
deviates from the FRW trajectory as $\lambda$ decreases, in such a way that $\lambda$ is bounded
also from below and it never vanishes.

\item $\Lambda>0$, $k=0,\pm1$: In these three cases, in the two-vertex model $\lambda$ is bounded
both from below and from above, while in the FRW model $\lambda$ is not bounded from above.
\end{itemize}

This analysis points out the limitations of our present two-vertex model to
provide an effective cosmological model. Actually, it is not suitable to model the FRW models with positive cosmological constant, since the area of the cell under study can not increase arbitrarily, unlike in the FRW model. However, the other FRW models admit their analog in the two-vertex model. In these cases, there exists an effective two-vertex model that introduces corrections to the FRW trajectories as the area decreases, in such a way that the area turns out to be strictly positive. The corrections are then unimportant in the classical regime of large areas, as desirable, since in this regime any effective theory should agree with general relativity. The fact that in the effective model the area has a positive bound resembles the results obtained in LQC, where the scale factor never vanishes and bounces instead of collapsing at the big bang singularity (see e.g \cite{abl,aps1,aps3}).

As said before, this analysis is just a starting point in the derivation of effective cosmological
models from loop gravity formulated on a fixed graph.
This is an approach to be improved. The main issue is that the semiclassical limit fails to be the correct one since the large area $\lambda \gg 1$ limit does not necessarily corresponds to $\phi\rightarrow 0$, in which case one can not approximate the cosine $\cos2\phi$ as $1-2\phi^2$. Then our na\"\i ve identification of the classical FRW Hamiltonian and our 2-vertex Hamiltonian totally fails.
Already before the introduction of any kind of matter, we see that our 2-vertex graph model does not provide any effective model for homogeneous and isotropic cosmologies with positive cosmological constant. The problem does not lead in the fact that the area can display a positive bound, feature which in turn is a consequence of the dependence of the Hamiltonian in the variable $\phi$ through the cosine (which is a bounded function), but rather in the fact that the semiclassical limit fails to be the correct one, since the large area limit does not necessarily corresponds to $\phi\rightarrow 0$. Therefore, if we want to obtain a successful effective model for all possible homogeneous and isotropic cosmologies, we need to improve our approach.
In the present canonical framework different possibilities seem to be at hand:
\begin{itemize}
\item We could redefine the canonical transformation \eqref{canonical-aA} doing
\be
\pi_a\rightarrow \tilde\pi_a=\pi_a+f(\lambda)
\ee
such that the canonical commutation relation $\{a,\tilde\pi_a\}=1/V_o$ is preserved. Such a
modification would change the phase space trajectories, and since the function $f(\lambda)$ is
arbitrary we could try to choose it conveniently such that the resulting model indeed succeeds in providing a correct effective FRW model.

We could go further and drop the implicit assumption that our variables $(\lambda,\phi)$ are
canonically related with the common ones $(a,\pi_a)$. Indeed a deeper understanding of our model makes us think that our variables, coming from a discrete theory (that in
principle is based on a quantum theory), may not be canonically related with the classical ones $(a,\pi_a)$, but only approximately recovered in the $\phi\rightarrow 0$ regime. The physical meaning of the coupling
constants $\gamma^0$, $\gamma$, and $\gamma^1$ would then be different for the one assumed in our
previous analysis. In consequence, this idea could allow us to drastically change our two-vertex
model to find a successful link between it and the classical FRW cosmologies.

\item Bringing to our framework the ideas of LQC, other possibility is to modify the physical
meaning of the angular variable defined out of the holonomies of the loop formalism. This can be
done by rescaling the variable $\phi$ by a function of the area,
\be
\phi\rightarrow\tilde\phi=\frac{\phi}{f(\lambda)},
\ee
as it is done in the {\it improved
dynamics} of LQC \cite{aps3}, where one chooses $f(\lambda)\propto \sqrt{\lambda}$, or in the {\it
lattice refinement} approach \cite{bck}, where a more general rescaling is considered, but still of
potential form $f(\lambda)\propto \lambda^a$. In this way the new angular variable $\tilde\phi$ is
no longer canonically conjugate to the area but to some function of it, for instance the volume in
the case of the improved dynamics of LQC. Note that after the rescaling the large area limit
corresponds to the limit $\tilde\phi\rightarrow 0$, as desired.

Such a rescaling does not seem neither natural nor simple to implement on the 2-vertex graph. In order to reproduce an improved dynamics setting \`a la LQC taking as canonical variables the volume and its conjugate variable instead of the area and its conjugate holonomy, it seems more likely that we should move to a different more complicated graph and possibly allow graph changing dynamics. This means revising the definition of the homogeneous and isotropic sector accordingly with the new class of graphs considered. This is out of the scope of the present study and will be investigated in future work.


\item Finally we have as so far only studied the 2-vertex model dynamics in vacuo without any matter field. Coupling matter will render useless the off-shift $H_o$ that we have introduced by hand and will affect the trajectory. This possibility of ``correcting" the trajectory and the behavior of $\phi$ at large scale factor using the matter contribution to the Hamiltonian constraint is particularly physically relevant since it is necessary step towards building a realistic cosmological model. However this demands understanding how to couple matter consistently to the geometrical data in LQG (and in particular understand if this requires graph changing or can be achieved on a fixed graph).

\end{itemize}

In summary, just because of the limitations of the two-vertex model, pointed out from the
analysis of the phase space trajectories, we cannot rule out this model in our aim of modeling
effective FRW cosmologies with it. The model formulated in the 2-vertex graph is not as simple as
it seems, in the sense that we may not yet understand completely the exact physical meaning of the
variables $(\lambda,\phi)$ that describe it, as suggested by the first two points listed above.

More precisely, there is no freedom in the construction of the dynamics on the 2-vertex graph once we have the homogeneous and isotropic sector through the $\U(N)$ symmetry. The ambiguity lies in the physical interpretation of the $(\lambda,\phi)$ variables. We can try to change their physical meaning in order to improve the matching on the 2-vertex Hamiltonian with the one from FRW cosmologies, but we would then lose their natural geometrical interpretation as area and curvature.

We will postpone the investigation of the possibilities pointed out above for future work, when we will also be able to take into account the coupling with matter, in order to study true dynamical cosmological models.
Instead, in the next section, we will rather look for the derivation of effective FRW
models from loop gravity formulated on the 2-vertex graph following the ideas of spin foam
cosmology \cite{SFcosmo}.

\subsection{Discretizing the Loop Gravity Hamiltonian Constraint: the Rovelli-Vidotto proposal}
\label{vidotto}

Up to now, we have simply constructed the canonical Hamiltonian on the 2-vertex graph out of all possible (lowest order) operators compatible with the $\SU(2)$ gauge invariance and our isotropy requirement. There is a priori no relation to gravity or cosmology. The link to FRW cosmology is established a posteriori (up to the limitations underlined above in the previous section)in the large scale factor regime.

It would be interesting to see if we could derive our Hamiltonian from loop quantum gravity. Actually, a discretization of the loop quantum gravity Hamiltonian constraint operator on the 2-vertex graph was already proposed by Rovelli and Vidotto in \cite{LQGcosmo1}. They furthermore couple the system to a massless scalar field. One discretizes the scalar field. Since there are only two vertices, the scalar field will be discretizes on those two space points. We introduce the two canonical pairs, $\{\vphi_\alpha,p_\alpha\}=\{\vphi_\beta,p_\beta\}=1$. Then we have two Hamiltonian constraints, one for each vertex $\alpha$ and $\beta$. The gravitational part of the Hamiltonian constraint is constructed as a discretization of its classical counterpart and consists in two triad insertions times a holonomy operator. As constructed in \cite{LQGcosmo1}, this leads to:
\be
C^\alpha=C^\alpha_{grav}+\f{p^2_\alpha}{2}=
\sum_{i,j}\tr \tV_i\tV_j g_j^{-1}g_i +\f{p^2_\alpha}{2},
\qquad
C^\beta=C^\beta_{grav}+\f{p^2_\beta}{2}=
\sum_{i,j}\tr \tW_i\tW_j g_jg_i^{-1} +\f{p^2_\beta}{2},
\ee
where the group elements $g_i\in\SU(2)$ are the holonomies living on the edges $i$ of the graph, while the 2$\times$2 matrices corresponding to the triad insertions around $\alpha$ and $\beta$ are defined as $\tV_i\equiv \vV(z_i)\cdot\vsigma$ and $\tW_i\equiv \vV(w_i)\cdot\vsigma$.

\medskip

We will show here that this construction matches exactly the Hamiltonian as we have defined from the simple requirement of $\SU(2)$ gauge invariance and $\U(N)$ invariance. More precisely, it corresponds to a special (trivial) choice of coupling constants in our ansatz. It thus justifies our ansatz as coming from an implementation of the loop quantum gravity dynamics.

First, as already noticed by Rovelli and Vidotto in \cite{LQGcosmo1}, the gravitational part $C_{grav}$ is the same at both vertices, due to the $\SU(2)$ gauge invariance. Indeed using that $\vV(w_i)=-g_i\vV(z_i)$, we get that the 2$\times$2 matrices at the two vertices are related by conjugation, $\tW_i=-g_i\tV_ig_i^{-1}$. This leads to the trivial identity:
$$
\tr \tV_i\tV_j g_j^{-1}g_i
=\tr g_i\tV_i\tV_j g_j^{-1}
=\tr \tW_ig_ig_j^{-1}\tW_j
=\tr \tW_j\tW_i g_ig_j^{-1},
$$
thus implying that $C^\alpha_{grav}=C^\beta_{grav}$. In turn, this implies that we have the (Hamiltonian) constraint $p_\alpha^2-p_\beta^2=0$ obtained as $C^\alpha=C^\beta$. It means that the dynamics of the scalar field is homogeneous and we can choose a homogeneous scalar field $\vphi_\alpha=\vphi_\beta$ without loss of generality. In our context, when the $\U(N)$ symmetry imposes both isotropy and homogeneity (same states around both vertices up to global $\SU(2)$ rotation) of the geometrical sector, it is natural to also get homogeneity of the scalar field.

Beside this trivial constraint, we still have the constraint relating the scalar field density to the geometry:
$$
C=C_{grav}+\f{p^2}2,
$$
where we have dropped the index $\alpha$ or $\beta$ since it is irrelevant.
Now we have computed $C_{grav}$ in terms of the spinor variables. This is straightforward using the definition of the 3-vector $V(z_i)$ and of the holonomies $g_i$ in terms of $z_i$ and $w_i$:
\beq
\tr \tV_i\tV_j g_j^{-1}g_i
&=&
\f{\la z_i|z_i\ra\la z_j|z_j\ra}{\sqrt{\la z_i|z_i\ra\la z_j|z_j\ra\la w_i|w_i\ra\la w_j|w_j\ra}}\,
\left(
\Ea_{ij}\Eb_{ij}+\Ea_{ji}\Eb_{ji}-\Fa_{ij}\Fb_{ij}-\overline{\Fa_{ij}}\overline{\Fb_{ij}}
\right)\nn\\
&\simeq&
\Ea_{ij}\Eb_{ij}+\Ea_{ji}\Eb_{ji}-\Fa_{ij}\Fb_{ij}-\overline{\Fa_{ij}}\overline{\Fb_{ij}}\,,
\eeq
where the last equality $\simeq$ is weak in the sense that it only holds assuming the matching constraints $\la z_i|z_i\ra=\la w_i|w_i\ra$. Summing over all pairs of edges, we finally get:
\be
C=C_{grav}+\f{p^2}2
\simeq
\f{p^2}2
+\sum_{ij}\Ea_{ij}\Eb_{ij}+\Ea_{ji}\Eb_{ji}-\Fa_{ij}\Fb_{ij}-\overline{\Fa_{ij}}\overline{\Fb_{ij}}
\,.
\ee

\medskip

Let us first comment on the gravitational part. This is exactly our ansatz \eqref{UN_action} for $\gamma^o=2$, $\gamma^+=\gamma^-=-1$ and $\gamma^1=0$. In particular, this means that the discretized LQG Hamiltonian (as defined in \cite{LQGcosmo1}) is invariant under $\U(N)$ and defines an isotropic and homogeneous cosmological dynamic in our context. Moreover it legitimizes our ansatz, showing its clear relation with the standard loop quantum gravity framework, and our requirement of $\U(N)$ symmetry. Finally, we can evaluate $C_{grav}$ in terms of the boundary area $\lambda$ and its conjugate curvature $\phi$. For this special choice of coupling constants, we get:
\be
C_{grav}= 2\lambda^2\,\left[
1-\cos(2\phi)
\right],
\ee
thus corresponding to the flat case with vanishing cosmological constant. In particular, in vacuum without the scalar field, it implies that the angle $\phi$ vanishes.

Second, looking at the coupling to the scalar field, we notice that the gravitational part goes in $\lambda^2\propto a^4$ and thus provides the proper relative scaling of the matter density with the scale factor $a$ as we expect from the classical FRW cosmology, as reviewed in section \ref{scalar}.

In conclusion of the canonical analysis of the 2-vertex graph model, we have explained how the requirement of a $\U(N)$ symmetry reduces the classical phase space to its isotropic and homogeneous sector. And we have accordingly introduced the more general $\U(N)$-invariant Hamiltonian, explained its relation to FRW cosmology in the large scale factor regime and showed that the usual LQG Hamiltonian constraint operator is a special case of our more general ansatz.

\section{Spinfoam Dynamics}
\label{spinfoam}
\label{sec4}

The spinfoam framework defines a path integral formalism for quantum gravity, which allows to compute well-defined transition amplitudes for spin network states (see e.g. \cite{SFreview} for a review). We propose here to use it to define the transition amplitudes between coherent spin network states peaked on the classical spinor network data and derive from this an effective classical dynamics for the spinor networks taking into account the spinfoam quantum gravity effects.

We will not review the spinfoam framework, and we will assume that the reader is familiar with the various spinfoam constructions and methods. We will only introduce the necessary concepts for our derivation and refer to the known literature on the subject for the details.

Other perspectives on the developing topic of spinfoam cosmology can be found in \cite{SFcosmo,SFcosmoLambda} and \cite{frank}.

\subsection{The Spinfoam Cosmology Setting}

\medskip
\begin{center}
{\it 1. Transition Amplitudes from Spinfoams: the General Framework}
\end{center}
\smallskip

Given a boundary graph $\Gamma$ and a spin network state $\Psi_\Gamma$ on that boundary, a spinfoam model defines possible bulk structure as 2-complexes $\Delta$ whose boundary is $\Gamma$ and builds a spinfoam probability amplitude $\cA^{(\Gamma)}_{\Delta}[\Psi_\Gamma]$ for each of these admissible 2-complexes. The boundary graph $\Gamma$ and the spin network state define the three-dimensional state of geometry and metric on the boundary, while the 2-complex $\Delta$ and the spinfoam amplitude $\cA_\Delta$ defines the bulk space-time structure.

More precisely, the spinfoam amplitude are defined as local state-sums. One associates algebraic
data to the edges and faces of $\Delta$, which usually have an interpretation in terms of discrete
space-time geometry. Then all the dynamics is assumed to take place at the vertices $\sigma$ of the
2-complex and a local amplitude $\cA_\sigma$ is defined as a function of the algebraic data living
on the edges and faces meeting at the vertex $\sigma$. The spin network state $\psi_\Gamma$ is
understood as defining the probability amplitude of the algebraic data on the boundary of $\Delta$:
the data associated to edges (resp. faces) of $\Delta$ meeting the boundary will be associated to
the vertices (resp. the links) of the boundary graph $\Gamma$. Finally, the spinfoam amplitude
associated to $\Delta$ is defined as the sum over all possible algebraic data of the product of the
vertex amplitudes in the bulk and the spin network state on the boundary, which roughly read as:
\be
\cA^{(\Gamma)}_{\Delta}[\Psi_\Gamma]
\,=\,
\sum_{\{j_f,i_e\}}
\prod_{\sigma\in\Delta} \cA_\sigma[j_{f\ni \sigma},i_{e\ni\sigma}]
\,\psi_\Gamma(j_{f\in\pp\Delta},i_{e\in\pp\Delta}),
\ee
where we have implicitly defined the algebraic data as representations $j_f$ of a certain Lie group (usually $\SU(2)$ or the Lorentz group) on the faces and intertwiners $i_e$ between these representations on the edges of $\Delta$.

Now considering a fixed graph $\Gamma$, one can use this formalism to define transition amplitudes between two spin network states living on that graph. Indeed, introducing the disconnected boundary $\Gamma \cup \Gamma$, one consider a 2-complex $\Delta$ whose boundary is $\Gamma \cup \Gamma$ and which basically interpolates between an initial copy of $\Gamma$ and the final copy of $\Gamma$ and interprets the spinfoam amplitude as a transition amplitude between a initial spin network state $\psi^i_\Gamma$ and a final spin network state $\psi^f_\Gamma$.

Going further we will use coherent spin networks peaked on spinor networks as boundary states. These
coherent spin network states should be considered as coherent wave-packets around the classical
spinor networks. Such states have been defined in \cite{un5} and are based on the coherent
intertwiner techniques developed earlier in \cite{LS,un2,un4}. Let us label such coherent spin
network states by their corresponding classical phase space data $|\{z^v_e\}_{v,e\in\Gamma}\ra$.
Then spinfoam models define transition amplitudes between such coherent spin network states, which
can be interpreted as defining an effective dynamics for the corresponding spinor networks. This
dynamics should correspond to first order at large scales to a classical Hamiltonian dynamics plus
extra quantum gravity corrections coming from the specific chosen spinfoam model for quantum
gravity.

Our strategy will be to fix the boundary $\Gamma$ and the bulk 2-complex $\Delta$ defining the  space-time structure interpolating between the initial space slice and the final space slice, and to study the dynamics resulting from the associated spinfoam amplitudes. Just the same way, that a fixed simple graph  should be enough to describe simple spatial geometrical structures and metrics, a simple 2-complex should be enough to describe simple space-time structure defining homogeneous dynamics such as the FRW 4-metric. One goal is then to identify which bulk structure $\Delta$ corresponds to which mini-superspace model of cosmology.

A more involved strategy would be to sum over all 2-complexes, as we will comment below from the perspective of group field theories for spinfoam models. This would correspond to summing over all possible geometrical and topological degrees of freedom for the bulk 4-metric. Here, we are aiming to describing symmetry-reduced physical situations which are defined through a finite number of parameters and degrees of freedom. It thus makes sense to truncate such a sum over 2-complexes and try to identify the minimal 2-complex faithfully representing the relevant physical context and the degrees of freedom of the considered mini-superspace model(s).

\medskip
\begin{center}
{\it 2. Recursion Relations to Differential Equations: to Einstein equations?}
\end{center}
\smallskip

From the procedure described above, we will obtain transition functions $\cA^{\Gamma}_{\Delta}[{}^iz^e_v,{}^fz^e_v]$ depending on the fixed graph $\Gamma$, the bulk 2-complex $\Delta$ and the classical initial and final spinor data on the graph $\Gamma$. These transition amplitudes will satisfy some differential equations in the spinor variables $z$'s. These differential equations encode both the symmetries and the dynamics of the spinfoam amplitudes. Indeed one must keep in mind that symmetries and dynamics are intimately intertwined in quantum gravity, since the dynamics and evolution are defined by the Hamiltonian constraints generating space-time diffeomorphisms. In the context of spinfoam models, this relation is also present.

More precisely, focusing on the spinfoam model for topological BF theory and restricting ourselves to the gauge group $\SU(2)$ for the sake of simplicity, one has recursion relations for invariants of $\SU(2)$ representations (such as the 6j-symbol and more generally 3nj-symbols). These recursion relations are understood to be related to the topological invariance of the spinfoam amplitudes (see e.g. \cite{SFrecursion_simone}). They have also been recently shown to be a quantization of the Hamiltonian constraint encoding the evolution and projecting on the Hilbert space of physical states \cite{SFrecursion_valentin}. Furthermore, decomposing explicitly the coherent intertwiners and coherent spin network functionals in representations of $\SU(2)$, one can interpret these coherent spin networks as generating functionals for the $\SU(2)$ invariants and corresponding spinfoam amplitudes \cite{SFrecursion_final}. Then the recursion relations for the the $\SU(2)$ invariants and spinfoam amplitudes expressed in terms of $\SU(2)$ spins get translated into differential equations satisfied by the spinfoam transition amplitudes for coherent spin networks on the boundary \cite{SFrecursion_final}. These differential equations expresses the invariance of the spinfoam amplitudes under certain deformations of the boundary geometry. Therefore they should translate symmetries of the considered spinfoam model. But, since some of these deformations induces diffeomorphisms in the time direction, they should also encode the dynamics of the theory.

We will not show this mechanism in details here. We will restrict ourselves to the 2-vertex graph $\Gamma$ and to the simplest 2-complex $\Delta$ and compute the spinfoam transition amplitudes between homogeneous and isotropic spinor data. We will show in this restricted setting that the spinfoam amplitudes does satisfy a differential equation, which can be interpreted as a Hamiltonian constraint for homogeneous and isotropic cosmology and from which one can infer the classical 4-metric encoded by the spinfoam amplitude.
The goal here is to derive the FRW equation and metric from this procedure. Applying the same method to generic spinfoam amplitudes, the hope is to derive more general differential equations describing the discrete diffeomorphism invariance at the level of 2-complexes and encoding general relativity's Einstein equations for 4-metrics.

\medskip
\begin{center}
{\it 3. Spinfoam Transition Amplitudes: a Leading Order Calculation}
\end{center}
\smallskip

\begin{figure}[h]
\begin{center}
\psfrag{Delta}{\Large{$\Delta_1$}}
\psfrag{initial}{\Large{$\Gamma$}}
\psfrag{final}{\Large{$\Gamma$}}
 \psfrag{state-fin}{\Large{$\psi^f$}}
 \psfrag{state-in}{\Large{$\psi^i$}}
 \psfrag{v}{\Large{$\sigma$}}
\psfrag{Transition}{ }
\includegraphics[width=6cm]{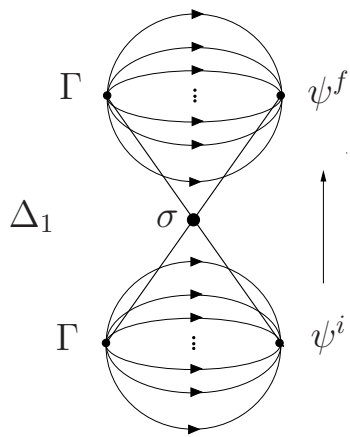}
\caption{Simplest 2-complex $\Delta_1$ interpolating between the initial state $\psi^i$ and the
final state $\psi^f$. Its disconnected boundary $\Gamma\cup\Gamma$ coincides
with the boundary graph of its single vertex $\sigma$.
\label{transition}}
\end{center}
\end{figure}

In the present paper, we will study states on the 2-vertex graph $\Gamma$ with an arbitrary number
of edges and choose the simplest compatible 2-complex $\Delta_1$, made of a single vertex in the
bulk, following the spinfoam cosmology approach  introduced in \cite{SFcosmo}. This 2-complex
$\Delta_1$ represents single four-dimensional cell interpolating between the initial 2-vertex graph
and the final 2-vertex graph, as illustrated in fig. \ref{transition}

As explained above, the spinfoam ansatz in this simplest case of a single spinfoam vertex is that the spinfoam amplitude is given by the evaluation of the boundary spin network. Therefore, we choose our spin network states on the initial and final 2-vertex graphs and the spinfoam amplitude is the evaluation of these two spin networks. Since the boundary is disjoint, we simply get the decoupled product of the evaluations:
\be
\cA^{\Gamma\cup \Gamma}_{\Delta_1}[\psi^i,\psi^f]
\,=\,
\psi^i_\Gamma(\id)\psi^f_\Gamma(\id)
\,.
\ee
In future work, we would like to study the spinfoam amplitudes beyond this leading order 2-complex with a single vertex. Adding vertices and/or using a non-trivial bulk topology, we will then obtain various more complicated transition amplitudes beyond this decoupled first order calculation.

In the next subsection, we will compute explicitly this first order spinfoam amplitude based on $\Delta_1$ for both $\SU(2)$ BF theory and a spinfoam model for quantum gravity taking into account the simplicity constraints using coherent spin networks on the boundary.

\medskip
\begin{center}
{\it 4. The Group Field Theory Point of View and the Issue of Renormalization}
\end{center}
\smallskip


Here, we have taken the point of view of fixing both the boundary graph $\Gamma$ on which our spin networks live {\it and} the bulk spinfoam 2-complex $\Delta$. Our goal is to compute the corresponding spinfoam amplitudes describing the evolution and dynamics of the spin networks for this fixed choice of bulk structure and interpret as a mini-superspace model (for cosmology).

An alternative would be to fix the structure of the boundary but sum over all ``admissible" bulks. In order to do this, we need to define the list of admissible 2-complexes and to fix their relative weights in the sum. This is done automatically by the group field theory formalism which provides us with a non-perturbative definition of the sum over spinfoam histories for fixed boundaries (see e.g. \cite{gftreview_daniele, gftreview_razvan}).

More precisely, the standard group field theory (GFT) partition function is expanded in Feynman diagrams, which are understood as the spinfoam 2-complexes. These are interpreted as the dual 2-skeleton of (pseudo-)triangulations made of 4-simplices glued together (along their boundary tetrahedra), with each 4-simplex dual to a spinfoam vertex. The resulting spinfoam amplitude can be written as a sum over all 2-complexes dual 4d triangulations with fixed boundary with a statistical weight:
\be
\cA^\Gamma[\psi_\Gamma]
\,=\,
\sum_{\Delta|\Gamma=\pp\Delta} \f1{w[\Delta]} \gamma^{\#\sigma_\Delta} \cA^\Gamma_\Delta[\psi_\Gamma],
\ee
where the factor $w[\Delta]$ is the symmetry factor coming from the Feynman diagram expansion, $\gamma$ is the GFT coupling constant and $\#\sigma_\Delta$ counts the number of spinfoam vertices (or 4-simplices) of $\Delta$.

\smallskip

Let us revisit our setting with the fixed 2-complex $\Delta_1$ for the 2-vertex graph boundary. It consists in a single spinfoam vertex and we have chosen the corresponding spinfoam ansatz for such a configuration given by the straightforward evaluation of the boundary spin network. However, from the GFT point of view, $\Delta_1$ is not dual to a 4d triangulation made of a single 4-simplex. On the contrary, we would need several 4-simplices to get a bulk topologically equivalent to $\Delta_1$. Thus, {\it if} we define the spinfoam amplitudes from the group field theory framework, it seems that a priori we should {\it not} assume the simplest spinfoam ansatz for the amplitude for the 2-complex $\Delta_1$. On the other hand, we should compute the full sum over 4d triangulations compatible with the 2-vertex graph on the boundary (more exactly whose boundary is the union of both initial and final 2-vertex graphs). Actually, we do not yet know how to control and compute such a sum, which is likely to be divergent, despite the recent progress on this issue. We would need to study the coarse-graining of spinfoam amplitudes and the renormalization of the group field theory. We would need to extract the relevant interaction terms which should be in the effective group field theory at large scale and study the running of the corresponding coupling constants (with the scale of the boundary geometry).

One possibility is that the leading order term after renormalization is exactly given by the simplest spinfoam ansatz for the 2-complex $\Delta_1$. This is what happens for the topological BF theory if we assume a trivial bulk topology (which seems to dominate the GFT partition function in colored group field theory models \cite{gftreview_razvan}) and we can expect this to remain true for spinfoam quantum  gravity in a low curvature regime. Nevertheless, one should keep in mind the limits of our present point of view of fixing the bulk structure and not following the group field theory prescription.

\smallskip

One ingredient from the GFT perspective which we could keep is the zeroth order term of the
expansion. This is the identity map between the initial and final boundary coming from the trivial
contribution of the 2-complex directly interpolating between the initial and final graph without any
vertex. Therefore, we could define a truncated transition  amplitude as the sum of the identity term
plus the first order contribution coming from the one-vertex 2-complex $\Delta_1$:
\be
\cA^{\Gamma\cup \Gamma}_{\Delta_1}[\psi^i,\psi^f]
\,\equiv\,
\cA^{\Gamma\cup \Gamma}_{\Delta_0}[\psi^i,\psi^f]+
\gamma\cA^{\Gamma\cup \Gamma}_{\Delta_1}[\psi^i,\psi^f]
\,=\,
\la \psi^i_\Gamma|\psi^f_\Gamma\ra+
\gamma\,\psi^i_\Gamma(\id)\psi^f_\Gamma(\id)\,.
\ee
This new zeroth order, with a trivial propagation, gives a totally coupled term between the initial
and final boundaries while our first order is totally decoupled. In order to focus on the
simplest possibility, we will nonetheless not consider this
zeroth order term but just the first order term
when computing the transition amplitude in next sections.

\subsection{Spinfoam Amplitude and Dynamics for BF Spinfoam}

Let us start applying our program to the spinfoam model for topological BF theory with gauge group $\SU(2)$. We have already described in great details in section \ref{2v_classical} the classical phase space for $\SU(2)$ spin networks on the 2-vertex graph and its reduction to the homogeneous and isotropic sector through a symmetry reduction by $\U(N)$. We have two sets of $N$ spinors, $z_i$ and $w_i$, attached to the $N$ edges around respectively the vertex $\alpha$ and the vertex $\beta$. Both sets of spinors satisfy the closure constraint. Moreover they satisfy the matching constraint along each edge, $\la z_i|z_i\ra=\la w_i|w_i\ra$. Then the homogeneous and isotropic sector is defined as assuming that the spinors $w_i$ are equal to the dual of the spinors $z_i$ up to a global phase, $|w_i]=e^{i\phi}\,|z_i\ra$.
%

At the quantum level, the components of the spinors are quantized as harmonic oscillators and we recover the Hilbert space of spin networks on the considered graph (see \cite{un1,un2,un4,un5,spinor1} for more details). At the end of the day, we define coherent spin network states by attaching two coherent intertwiners, $|\{z_i\}\rangle$ and $||\{\varsigma w_i\}\rangle$, to both the source vertex $\alpha$ and the target vertex $\beta$ respectively. These coherent intertwiners are defined as diagonalizing the annihilation operators $\hat{F}_{ij}$ and are labeled by the classical phase space points \cite{un4,spinor1}. One can furthermore show that they transform covariantly under the $\U(N)$-action on the spinors and that they provide a decomposition of the identify of the space of intertwiners. Finally, they are good semi-classical states, minimally spread about the corresponding phase space point.
Moreover we understand very well their decomposition in terms of the Livine-Speziale intertwiners $|\{j_i,z_i\}$ (labeled by spins $j_i\in\N/2$ and spinor variables) introduced in \cite{LS} or in terms of the $\U(N)$ coherent intertwiners $|J,\{z_i\}\ra$ (labeled by the total area $J=\sum_i j_i$ and spinor variables) introduced in \cite{un2}. For instance, it will be useful for computational purposes to remind the formula established in \cite{un4,spinor1}:
\be
|\{z_i\}\ra=\sum_{J\in\N}\f1{\sqrt{J!(J+1)!}}\,|J,\{z_i\}\ra\,,
\ee
since the scalar product between the $\U(N)$ coherent intertwiners $|J,\{z_i\}\ra$ are explicitly known and are homogeneous of degree $J$ in the spinor variables \cite{un2}.

\smallskip

\begin{figure}[h]
\begin{center}
 \psfrag{Delta}{ }
\psfrag{initial}{\large{$\{z_i\}$}}
\psfrag{final}{\large{$\{\tilde z_i\}$}}
 \psfrag{state-fin}{\hspace*{-0.3cm}\large{$\{\tilde w_i\}$}}
 \psfrag{state-in}{\hspace*{-0.3cm}\large{$\{ w_i\}$}}
 \psfrag{v}{\Large{$\sigma$}}
\psfrag{Transition}{ }
\includegraphics[width=6cm]{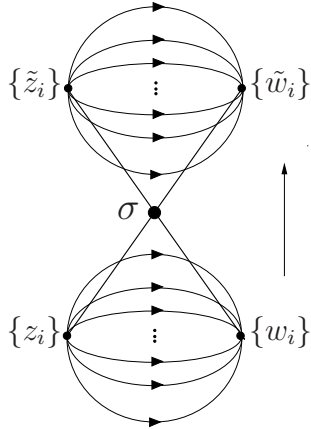}
\caption{Boundary data on the 2-complex $\Delta_1$ defining initial and final boundary coherent spin
networks. Focusing e.g. in the initial 2-vertex graph, we attach to the vertex $\alpha$ a coherent
intertwiner labeled by the spinors $\{z_i\}$ and to the vertex $\beta$ another coherent
intertwiner labeled by the spinors $\{w_i\}$, both consistently matched to form a coherent spin
network $\psi_{\{z_i,w_i\}}$.
\label{amplitude}}
\end{center}
\end{figure}

Now,  we have our initial 2-vertex graph and our final 2-vertex graph, with the bulk in between
defined by the 2-complex $\Delta_1$ with a single vertex, as shown in figure \ref{amplitude}.
Our initial spin network state is the coherent spin network labeled with spinors $z_i$ and $w_i$:
$$
\psi_{\{z_i,w_i\}}(g_i)=\langle\{\varsigma w_i\}|\otimes_i g_i |\{z_i\}\ra,
$$
where the spin network is a gauge-invariant functional of the $\SU(2)$ group elements $g_i$  on the edges.
Our final spin network state is the coherent spin network labeled with spinors $\tz_i$ and $\tw_i$. And the spinfoam amplitude associated to $\Delta_1$ is the evaluation of the boundary spin network, which is the decoupled product of the evaluations of the initial spin network and of the final spin network:
\be
\cA_{\Delta_1}[z_i,w_i,\tz_i,\tw_i]
\,=\,
\psi_{\{z_i,w_i\}}(\id)
\psi_{\{\tz_i,\tw_i\}}(\id)
\,=\,
\la \{\varsigma w_i\}|\{z_i\}\ra\,\la \{\varsigma \tw_i\}|\{\tz_i\}\ra,
\ee
where the evaluation of a spin network state on the 2-vertex graph at the identity is simply given by the scalar product of the intertwiners living at the two vertices.

Let us focus on a single scalar product, $\cW(z_i,w_i)=\psi_{\{z_i,w_i\}}(\id)=\la \{\varsigma w_i\}|\{z_i\}\ra$. We know how to compute it exactly \cite{un2,un4,spinor1}:
\beq
\cW(z_i,w_i)&=&\la \{\varsigma w_i\}|\{z_i\}\ra
\,=\,
\sum_J\f1{J!(J+1)!}\la J,\{\varsigma w_i\}|J,\{z_i\}\ra \nn\\
&=&
\sum_J\f1{J!(J+1)!}\,\left(\det \sum_i|z_i\ra [w_i|\right)^J\nn\\
&=&
\sum_J\f1{J!(J+1)!}\,\left(\f12\sum_{i,j}[z_i|z_j\ra [w_i|w_j\ra\right)^J
\eeq

We can further simplify this expression in the homogeneous and  isotropic case by plugging in the relation between the two sets of spinors $|w_i]=e^{i\phi}|z_i\ra$ and taking into account the closure constraint $\sum_i |z_i\ra\la z_i| = \lambda\id$. We then obtain a formula depending only on the two conjugate variables $\lambda$ and $\phi$:
\be
\cW(z_i,w_i)=\sum_J \f1{J!(J+1)!}\, (\lambda e^{-i\phi})^{2J}\,.
\ee
As mentioned earlier in section \ref{complex_strc}, we define the complex variable:
\be
z=2\sqrt{\lambda}\,e^{-i\f\phi2},
\qquad
\{\lambda,\phi\}=\f12
\,\Rightarrow\,
\{z,\bz\}=i\,.
\ee
Then the spinfoam amplitude is simply a Bessel function in $z^2$:
\be
\cW(z_i,w_i)\equiv \cW(z)
\,=\,
\sum_J \f1{J!(J+1)!}\,\left(\f z2\right)^{4J}
\,=\,
\f4{z^2}\,I_1\left(\f {z^2} 2\right),
\ee
where $I_1$ is the first modified Bessel function (of the first kind). It is known that Bessel
functions satisfy a second order differential equations. Here, one can deduce directly from the
expansion of $\cW(z)$ in $J$ that it satisfies:
\be
\hat{\cC}\,\cW=0,
\qquad\textrm{with}\quad
\hat{\cC}
\,=\,
z^2\pp_z^2+5z\pp_z-z^4
\,=\,
\pp_z^2z^2+z\pp_z-z^4-2
\,=\,
z^{-3}\pp_z z^5\pp_z-z^4
\,,
\ee
which basically translates the obvious recursion relation on the series' coefficients, $(J+1)!(J+2)!=(J+1)(J+2)\,J!(J+1)!$.

\smallskip

This constraint $\hcC$ satisfied by the spinfoam amplitude expresses an invariance of the amplitude under certain deformations of the boundary. From a canonical point of view, it should be interpreted as the Hamiltonian constraint. Indeed, it has already been pointed out that the Hamiltonian constraint, as well as the space diffeomorphism constraints, get translated into recursion relations and differential equations satisfied by the amplitudes in the spinfoam framework \cite{SFrecursion_simone,SFrecursion_valentin,SFrecursion_final}. We will show below that this is indeed the case and that $\hcC$ is indeed the quantization  of the Hamiltonian constraint (for BF theory).

Considering $\hcC$ as the quantization of a classical (Hamiltonian) constraint, let us point out that the two relevant terms in the differential operators are $z^2\pp_z^2$ and $z^4$, while the first derivative term $z\pp_z$ and the constant term come from ordering ambiguities (for instance, changing the order of the operators in $z^2\pp_z^2$ clearly affects those terms).

From this perspective, we also expect this Hamiltonian constraint to express the FRW equation in our present cosmological setting. Therefore, we look below at the large scale behavior of the spinfoam amplitude $\cW(z)$ and of the quantum constraint operator $\hcC$ and show their relation to FRW cosmology and to the classical effective dynamics on the 2-vertex graph.

\subsection{Asymptotic Behavior and FRW Equation}

It is well-known that the Bessel function is approximate at large argument by a Gaussian, similarly to a Poisson distribution. Let us apply this to our spinfoam amplitude $\cW(z)$. More precisely, let us perform a stationary point approximation of the sum over $J$. To this purpose, we use the Stirling formula to approximate the factorials for large $J$:
\be
\f1{J!(J+1)!}\left(\f z2\right)^{4J}
\underset{J\gg1}{\sim}
\f1{2\pi J^2}\,e^{4J\ln(\f z 2)-2J(\ln J-1)}.
\ee
Then we look for the stationary point(s) of the exponent $\vphi(J)\equiv 4J\ln(\f z 2)-2J(\ln J-1)$ and compute the second derivative:
$$
\pp_J \vphi =0
\,\Leftrightarrow\,
J=\f{z^2}4,
\qquad
\pp_J^2\vphi_{|J=\f{z^2}4}
=\f{-2}{J}=\f{-8}{z^2}\,.
$$
We get a unique stationary point $J=z^2/4$, which justifies that our large $J$ approximation holds when $z$ itself is large (in modulus).
Then approximating the sum over $J$ by a Gaussian integral around this stationary point, we get:
\be
\label{Gauusianapprox}
\cW(z)
\underset{z\gg1}{\sim}
\f4{z^3\sqrt{\pi}}\,e^{\f{z^2}2}
\equiv W(z)\,.
\ee
We have checked this formula numerically.
What is interesting is that $W(z)$ is simply a Gaussian distribution in $z$ with a pre-factor. Under this form, it is obvious that it satisfies a 2nd order differential equation, which is the same as $\cW$ up to a constant shift:
\be
(\hcC+3)\,W(z)=0.
\ee
The fact that it is exactly the same differential operator up to a simple constant shift means that the large scale behavior is exactly the same at the classical level and that the deviation from the exact amplitude is at small scales in the quantum regime.

We see that in the asymptotic regime, both the spinfoam amplitude and the constraint operator are very similar to the one derived previously in the earlier work on spinfoam cosmology \cite{SFcosmo}, except that we do not use the exact same complex variable ($z\propto \sqrt{\lambda}e^{-i\f\phi2}$ here instead of $\zeta\sim \lambda-i\phi$) and that we easily control the level of approximation that we do. Indeed, we know both the exact spinfoam amplitude and constraint operator and their large scale approximations. Despite these minor differences, the physical interpretation will be exactly the same.

\smallskip

Let us focus on the constraint operator $\hcC=z^2\pp_z^2+5z\pp_z-z^4$. Considering the canonical Poisson bracket, $\{z,\bz\}=i$, the differential operator $\pp_z$ is the quantization of the classical variable $\bz$ and thus our differential operator turns out to be the quantization of the classical Hamiltonian constraint:
\be
\cC= z^2(\bz^2-z^2) + (5z\bz).
\ee
Replacing $z$ by its definition $2\sqrt{\lambda}e^{-i\f\phi2}$, we see that the first term goes in $\lambda^4$ and clearly dominates the second term which goes in $\lambda^2$ and can be neglected. This is consistent with the fact that the second term comes from the differential operator $z\pp_z$, which comes from ordering ambiguity in the leading order operator $z^2\pp_z^2$. Therefore, we neglect the term in $z\bz$ in the definition of our classical Hamiltonian and we factor out the pre-factor $z^2$, which leaves us will a renormalized classical Hamiltonian derived from the spinfoam amplitude:
\be
\label{spinfoam_H}
\tcC \,\equiv\,\bz^2-z^2
\,=\,
8i\lambda\sin\phi\,.
\ee
This is exactly the Hamiltonian for FRW cosmology in the simplest case, with vanishing curvature $k=0$ and cosmological constant $\Lambda=0$, where we don't have any evolution with $\phi=0$ and constant $\lambda$, as we see by comparing with the explicit expressions of section \ref{dynamics}. More precisely, we have derived the effective dynamics on the 2-vertex graph with the Hamiltonian in $\lambda \sin\phi$ instead of the usual $\lambda\phi$ of classical FRW (see section \ref{dynamics} for the comparison between the effective and standard Hamiltonians).

Going into more details, it is actually to understand how this Hamiltonian constraint comes truly from the dynamics of BF theory, as we explain below. Since BF theory is a theory of flat connections, this is the reason that we obtain  only the flat cosmology case with $k=0$. Let us point out that the terms in $z\bz$ that we neglected are not with the correct scaling in $\lambda$ and  do not generate any non-trivial term in $k$ or $\Lambda$.

\subsection{Recovering the Hamiltonian Constraint}

For now, we have made the link between the spinfoam (transition) amplitude for the quantum dynamics of coherent spin networks on the 2-vertex graph and the classical dynamics defined in the earlier section \ref{classical_dyn}. We can actually make this link stronger and show that the relation holds at the quantum level and not only in the classical regime at large scale $\lambda\arr\infty$. More precisely, we show the relation between the differential equation satisfied by the spin network evaluation derived in the covariant spinfoam context and the quantization of the classical Hamiltonian on the 2-vertex graph defined from a canonical point of view and worked out in \cite{un3}.

Considering the evaluation of the coherent spin network state $\cW(z_i,w_i)=\psi_{\{z_i,w_i\}}(\id)$, we remind the method introduced in \cite{SFrecursion_simone} and refined in \cite{SFrecursion_valentin,SFrecursion_final} to derive recursion relation or differential equations on the evaluation by acting with holonomy operators on the spin network state. Indeed, let us consider the holonomy operator around the loop formed by the two edges $j$ and $k$. It acts by multiplication on the spin network state $\psi_{\{z_i,w_i\}}(g_i)$. On the other hand, it is compute its action on the labels $\{z_i,w_i\}$ of the spin network state using the recoupling theory of representations on $\SU(2)$ as discussed in \cite{SFrecursion_simone,SFrecursion_valentin,SFrecursion_final}. In short, we get:
\beq
&&\widehat{\chi(g_jg_k^{-1})}\,\psi_{\{z_i,w_i\}}(g_i)
\,=\,
\chi(g_jg_k^{-1})\,\psi_{\{z_i,w_i\}}(g_i)
\,=\,
\hcD^{(j,k)}_{\{z_i\}}\,\psi_{\{z_i,w_i\}}(g_i)\nn\\
&
\Rightarrow
&
\hcD^{(j,k)}_{\{z_i\}}\,\cW(z_i,w_i)
\,=\,
\left.\chi(g_jg_k^{-1})\,\psi_{\{z_i,w_i\}}(g_i)\,\right|_{g_i=\id}
\,=\,
2\,\cW(z_i,w_i)\,,
\eeq
where the holonomy is taken in the fundamental spin-$\f12$ representation (for the sake of simplicity) and $\hcD^{(j,k)}_{\{z_i\}}$ is a to-be-determined differential operator in the spinor variables. Then by evaluating the action of the holonomy operator on the identity, we obtain a differential equation on the spin network evaluation $\cW(z_i,w_i)$.

We can of course consider operators more complicated than a single holonomy operators. However, as soon as it does not act anymore as a multiplication operator in the group elements $g_i$, one has to be careful with the operator ordering (e.g. \cite{SFrecursion_final}), but the method still works.
Here, we will apply it using a $\U(N)$-invariant combination of renormalized holonomy operator. We consider the classical $\U(N)$-invariant observable:
\beq
\cQ
&\equiv&
\sum_{j,k}\sqrt{\la z_j|z_j\ra\la z_k|z_k\ra\la w_j|w_j\ra\la w_k|w_k \ra}\,
\,\left[\chi(g_jg_k^{-1})-2\right]
\nn\\
&=&
\sum_{j,k}\left(2\la z_j|z_k\ra\la w_j|w_k\ra+[ z_j|z_k\ra [w_j|w_k\ra+\la z_j|z_k]\la w_j|w_k]\right)
-2\left(\sum_{j}\la z_j|z_j\ra\right)^2\nn\\
&=&
\sum_{j,k}\left(2\Ea_{jk}\Eb_{jk}+\Fa_{jk}\Fb_{jk}+\overline{\Fa_{jk}}\overline{\Fb_{jk}}\right)
-2\left(\sum_{j}\Ea_{jj}\right)^2
\,,
\eeq
where we used the expression of the group elements $g_j$ and $g_k$ in terms of the spinor variables in order to compute the holonomy $\chi(g_jg_k^{-1})$ and where we assumed the matching conditions,
$\la z_i|z_i\ra\la w_i|w_i\ra$ for all edges $i$, to simplify the constant term. It is clear that this observable vanishes where the group elements are fixed to the identity $g_i=\id$:
$$
\left.\cQ\right|_{g_i=\id}=0\,.
$$
Our strategy will be to compute the action of this operator on the spin network state $\psi_{\{z_i,w_i\}}(g_i)$ at the quantum level and deduce the differential equation satisfied by the evaluation $\cW(z_i,w_i)$ by taking $g_i=\id$ at the end.

We chose this particular observable $\cQ$, constructed from the holonomies $\chi(g_jg_k^{-1})$, polynomial (of the lowest possible order) in the spinor variables (which explains the norm pre-factors), vanishing on the identity (reason for the $-2$ terms) and invariant under $\U(N)$ (which implies summing over all pairs of edges $j,k$). In our 2-vertex graph context, this determines more or less entirely the observable $\cQ$. Indeed, as proved in \cite{un3,un5}, there are only three $\U(N)$-invariant polynomial terms of order 4 in the spinor variables, $\sum EE$, $\sum FF$ and $\sum \overline{FF}$, which explains the structure of our observable $\cQ$. Finally, only a specific combination of those will vanish on the identity.

Let us first see how is the value of $\cQ$  in the homogeneous and isotropic sector, with $|w_i]=e^{i\phi}\,|z_i\ra$ for all $i$'s. An easy calculation gives:
\be
\cQ=4\lambda^2(1+\cos 2\phi)-8\lambda^2
=-8\lambda^2\sin^2\phi\,.
\ee
We recognize both our classical Hamiltonian \eqref{2vertex_cosmo_H} on the 2-vertex graph for the special values $\gamma^o=\gamma$ and $\gamma^1=0$ and the square of the constraint $\tcC$ derived in \eqref{spinfoam_H} from the spinfoam amplitude. This makes the link between the canonical formalism with a Hamiltonian and the covariant perspective with the spinfoam amplitude satisfying  certain constraint.

\medskip

Let us now work out the quantized version $\hcQ$ and check that its action on $\psi_{\{z_i,w_i\}}(g_i)$ vanishes on trivial holonomies $g_i=\id$. This will also provide us with the exact differential equation generically satisfied by the evaluation $\cW(z_i,w_i)$, even without assuming the homogeneous and isotropic ansatz. We follow the quantization procedure for the spinor variables: holomorphic coordinates $z_i$ and $w_i$ are quantized as multiplication operators while the anti-holomorphic variables $\bz_i$ and $\bw_i$ respectively become the differential operators $\pp_{z_i}$ and $\pp_{w_i}$. This leads to the following quantization for the $E$ and $F$ observables \cite{un5,spinor1}:
\beq
{\hEa_{jk}}&=& z^0_k\pp_{z^0_j}+z^1_k\pp_{z^1_j}, \nn\\
{\hFa_{jk}}&=& (z^0_jz^1_k-z^1_jz^0_k),\nn\\
{\hFa_{jk}}{}^\dagger&=& (\pp_{z^0_j}\pp_{z^1_k}-\pp_{z^1_j}\pp_{z^0_k}),
\eeq
and similarly for the operators acting at the vertex $\beta$ as differential operators in the $w_i$'s. We can then compute the action of these operators on $\cW(z_i,w_i)=\sum_J (\det X)^J/J!(J+1)!$ where $X$ is the following 2$\times$2 matrix:
$$
X=\sum_i |z_i\ra[w_i|,
\qquad
\det X=\f12\sum_{jk}\Fa_{jk}\Fb_{jk}\,.
$$
Following the natural notation as in \cite{un3}, we will denote the functional $(\det X)^J$ as the quantum state $|J\ra$. Then the three components of the $\hcQ$ operator, ${\hEa_{jk}}{\hEb_{jk}}$, ${\hFa_{jk}}{\hFb_{jk}}$ and ${\hFa_{jk}}{}^\dagger{\hFb_{jk}}{}^\dagger$, respectively act as a number of quanta operator, a creation operator and an annihilation operator in the $|J\ra$ basis. More precisely, after a lengthy but straightforward calculation, we get:
\beq
\sum_{j,k}{\hEa_{jk}}{\hEb_{jk}}\,|J\ra
&=&
2J(J+N-2)\,|J\ra\\
\sum_{j,k}{\hFa_{jk}}{\hFb_{jk}}\,|J\ra
&=&
2\,|J+1\ra\nn\\
\sum_{j,k}{\hFa_{jk}}{}^\dagger{\hFb_{jk}}{}^\dagger\,|J\ra
&=&
2J(J+1)(N+J-1)(N+J-2)\,|J-1\ra\nn\,.
\eeq
It is worth pointing out that the same expressions were computed earlier in a more elegant way in
\cite{un3} by working out the commutation relations between these operators in the $\U(N)$-invariant
space.
Putting these results together, we define the quantized version of the $\cQ$ observable:
\be
\hcQ\,\equiv\,
\sum_{j,k}
\left(
2{\hEa_{jk}}{\hEb_{jk}}
+{\hFa_{jk}}{\hFb_{jk}}
+{\hFa_{jk}}{}^\dagger{\hFb_{jk}}{}^\dagger
\right)
-2(\hat{E}+N-1)^2-2(N-1),
\ee
where $\hat{E}\equiv\sum_i{\hEa}_ii$ is shown to act as $\hat{E}\,|J\ra=2J\,|J\ra$. The last term $(\hat{E}+N-1)^2+(N-1)$ corresponds to the quantization of the classical term $(\sum_{j}\Ea_{jj})^2$ up to sub-leading contributions interpreted as ordering ambiguities. This specific choice gives the expected result:
\be
\hcQ\,\cW(z_i,w_i)
\,=\,
\hcQ\,\sum_J\f{(\det X)^J}{J!(J+1)!}
\,=\,
0.
\ee
This is a fourth order differential equation on the spinfoam amplitude $\cW(z_i,w_i)$, which correspond to the quantization of the classical Hamiltonian on the 2-vertex graph.

It is further possible to write all the differential operators in terms of derivative of $\det X$, and thus in terms of the single complex variable $z$, then the operator $\hcQ$ should give the square of the differential equation $\hcC^2$ (up to ordering terms) satisfied by the spinfoam amplitude.

\subsection{How to Depart from Flat Cosmology?}

Working with the spinfoam model for topological BF theory with gauge group $\SU(2)$, we have derived the differential equation satisfied by the spinfoam amplitude for fixed boundary (the 2-vertex graph $\Gamma$) and fixed bulk (the 2-complex $\Delta_1$) and shown its relation to the classical Hamiltonian for the dynamics on the 2-vertex graph.
The Hamiltonian that we have obtained corresponds to the flat FRW cosmology (in vacuum). The natural question is whether it is possible or not to obtain the models of effective dynamics of section \ref{classical_dyn} for non-vanishing curvature $k\ne0$ and cosmological constant $\Lambda\ne 0$ with this spinfoam cosmology framework.

From the previous analysis, we understand that the flatness of the Hamiltonian $k=0=\Lambda$, or equivalently $\gamma^o=\gamma$ and $\gamma^1=0$ in the cosmological Hamiltonian ansatz \eqref{2vertex_cosmo_H}, comes from the definition of the spinfoam amplitude as the evaluation of the boundary spin network at the identity. This comes from working with the spinfoam path integral for topological BF theory, which projects onto trivial connections and holonomies.
Therefore,  in order to obtain non-flat FRW cosmology, it is clear that we have to depart from BF theory and the trivial spinfoam ansatz defining the amplitude as the evaluation of the boundary spin network state at the identity. For instance, it is natural to expect that evaluating the boundary spin network on non-trivial holonomies will produce curvature and lead to FRW cosmology models with non-vanishing curvature and cosmological constant.

In order to derive such a new spinfoam amplitude,  we see two non-exclusive possibilities.
First, we can change the spinfoam model. Either, we can attempt to modify the spinfoam amplitudes by hand and introduce curvature and cosmological constant terms in the BF amplitude. Or we can use a spinfoam model built for quantum gravity, such as the EPRL-FK spinfoam amplitudes \cite{EPR,EPRL,FK}. However, if we use such a model, the standard spinfoam amplitude ansatz is still to evaluate the boundary spin network around each vertex at the identity. Therefore, we expect no difference from topological BF theory if our bulk 2-complex contains a single vertex as considered here with our simplest 2-complex $\Delta_1$. We illustrate this by working out in the next subsection the spinfoam transition amplitude for the 2-vertex graph still with the single-vertex 2-complex $\Delta_1$ for the Dupuis-Livine variant of the EPRL model based on the holomorphic simplicity constraints \cite{spinor1,proceeding}.
Then the second alternative is to change the bulk 2-complex, and even the boundary graph, in order to allow for more intricate space-time structure with non-trivial topology and geometry. By allowing more than one vertex in the bulk 2-complex, we expect to have curvature generated through the gluing of the dual 4-cells.

\smallskip

Here, we will show how to modify the spinfoam amplitude and depart from the mere evaluation of the boundary spin network state in order to obtain non-flat FRW cosmology. We will use the realization of the Hamiltonian constraint as a differential equation satisfied by the spinfoam amplitude. Indeed, we have derived the differential equation satisfied by the evaluation of the boundary spin network state at the identity. We can now modify this differential equation to take into account non-vanishing curvature and cosmological constant and investigate how the spinfoam amplitude changes.

Calling $e\equiv2\sum_{j,k}{\hEa_{jk}}{\hEb_{jk}}-2(\hat{E}+N-1)^2-2(N-1)$,
$f\equiv{\hFa_{jk}}{\hFb_{jk}}$ and $f^\dag\equiv{\hFa_{jk}}{}^\dagger{\hFb_{jk}}{}^\dagger$ the operators introduced above and defined as differential operators in the spinor variables $z_i$ and $w_i$, the differential equation satisfied by the coherent spin network evaluation was simply defined by the operator $\hcQ=e+f+f^\dag$:
\be
\hcQ\,\la \{\varsigma w_i\}|\{z_i\} \ra
=
\hcQ\,\sum_J\f1{J!(J+1)!}\,(\det X)^J
=
0\,.
\ee
Following our analysis of the classical dynamics on the 2-vertex graph in section \ref{classical_dyn}, we consider the modified differential operator $\hcQ^{(\ka)}=e+\ka(f+f^\dag)$ for $\ka\in\R$. The corresponding classical constraint (in the large $\lambda$ regime) in the homogeneous and isotropic sector is:
\be
\cQ^{(\ka)}=-2\,2\lambda^2(1-\ka\cos 2\phi),
\ee
which corresponds to our effective FRW Hamiltonian \eqref{2vertex_cosmo_Hbis} with parameters $\gamma^o=1$, $\gamma=\ka$ and $\gamma^1=0$ (up to a global $(-2)$-factor), thus leading to modified FRW cosmology with non-vanishing curvature $k\ne 0$. We can easily take into account a non-vanishing cosmological constant by adding a $\hat{E}^3$ term in our constraint. But for the sake of simplicity, we will simply describe here the case $\gamma^1=0$.

Let us then solve the constraint $\hcQ^{(\ka)}=0$ at the quantum level. Noting as before $|J\ra\equiv(\det X)^J$ , we apply the modified operator $\hcQ^{(\ka)}$ to a linear combination of the basis vectors $|J\ra$ and we obtain a 2nd order recursion relation on the coefficients:
\be
\hcQ^{(\ka)}\,\sum_J\f{\alpha_J^{(\ka)}}{J!(J+1)!}\,|J\ra
\,=\,0
\,\Leftrightarrow\,
\left|
\begin{array}{l}
(J+N)(J+N-1)\alpha_{J+1}^{(\ka)}
-\f1\ka [2J(J+N)+N(N-1)]\alpha_{J}^{(\ka)}
+J(J+1)\alpha_{J-1}^{(\ka)}=0\\
\ka(N+1)\alpha_1=(N-1)\alpha_0
\end{array}
\right.
\,.
\ee
The spectral properties of similar operators were studied in \cite{un3} when analyzing the Hamiltonian dynamics of the quantum 2-vertex model. Applying the same techniques, we look at the asymptotics of the recursion relation at large $J$ and solve it at second order (in $1/J$). We then get:
\be
\alpha_J\underset{J\gg 1}{\propto} \f1{J^{N-1}}\cos\om J,
\qquad
\cos\om=\f1\ka\,.
\ee
We have checked the accuracy of this asymptotics numerically using Maple 15. For $\ka=1$, the oscillation frequency $\om$ vanishes and we recover the flat case which we have already described. For $|\ka|<1$, the frequency $\om$ becomes purely imaginary and we have an exponential solution instead of the oscillating behavior.

We can modify slightly the action of the operators $e,f,f^\dag$ on the states $|J\ra$ in order to obtain an exact analytical expression for the physical state. For instance, we can take:
\be
\begin{array}{lcl}
\te\,|J\ra &\equiv& -4J^2\,|J\ra \\
\tf\,|J\ra &\equiv& 2\,|J+1\ra \\
\tf^\dag\,|J\ra &\equiv& 2J^2(J-1)^2\,|J-1\ra \\
\end{array}
\ee
and solve the modified equation $\htcQ^{(\ka)}\equiv \te+\ka(\tf+\tf^\dag)=0$. Then we get the states:
\be
\htcQ^{(\ka)}\,\sum_J\f{\cos\om J}{J!^2}\,|J\ra
=
\htcQ^{(\ka)}\,\sum_J\f{\cos\om J}{J!^2}\,(\det X)^J
=0.
\ee
This new operator $\htcQ^{(\ka)}$ has the same classical limit than our original  operator $\hcQ^{(\ka)}$ but can be interpreted as differing in operator ordering.
It is simple to define these modified operators $\te,\tf,\tf^\dag$ as differential operators in the single complex variable $\det X$, or equivalently $z$, but we lose the direct translation as differential operators in the original spinor variables which allow to define these operators as acting on the whole phase space and not only in the homogeneous and isotropic sector. Moreover, this hides the relation of the operators $\te,\tf,\tf^\dag,\htcQ^{(\ka)}$ with the holonomy operators and the interpretation of the solution state as a coherent spin  network  evaluation.

Even if we do not have an explicit closed formula for the coefficients $\alpha_J^{(\ka)}$, this gives the spinfoam amplitude than we should define in order to get non-flat FRW cosmology. The next step is to understand how this amplitude can be defined as the evaluation of the coherent spin network on non-trivial holonomies or with an operator insertion, i.e of the type $\la \{\varsigma w_i\}|\cO^{(\ka)}|\{z_i\} \ra$. Already the asymptotics of the coefficients $\alpha_J^{(\ka)}$ provide us with some clues about how to realize this. Indeed, the series based on the asymptotics can be easily realized as the evaluation of the coherent spin network on some non-trivial holonomies:
\beq
\sum_J \f{\cos\om J}{J!(J+1)!}\,(\det X)^J
&=&
\sum_J \f1{J!(J+1)!}\,\f12\left[(e^{i\om}\det X)^J+(e^{-i\om}\det X)^J\right]\nn\\
&=&
\f12\left[
\la \{\varsigma w_i\}|\otimes_ie^{i\om} |\{z_i\} \ra
+
\la \{\varsigma w_i\}|\otimes_ie^{-i\om} |\{z_i\} \ra
\right]\,
\eeq
where we act with the same $\U(1)$ group element $e^{i\om}$, resp. $e^{-i\om}$, on all the legs of the 2-vertex graph.
This corresponds to the evaluation of the coherent spin network states on $\SU(2)$ group elements defined as the 2$\times$2 diagonal matrix $[e^{i\om}, e^{-i\om}]$ (resp. $[e^{-i\om}, e^{+i\om}]$) in the $(|z_i\ra,|z_i])$ basis on each edge of the graph.
As we have shown above, this functional satisfies a differential equation, whose classical counterpart at large scale, is $\cQ^{(\ka)}$, and thus corresponds to our Hamiltonian \eqref{2vertex_cosmo_Hbis} for the modified FRW cosmology with non-vanishing curvature.

It seems possible to follow the same strategy to take into account a cosmological constant. We expect to find a spinfoam amplitude similar the one postulated in \cite{SFcosmoLambda} or/and related to the $q$-deformation of the spinfoam amplitudes. Then the goal will be to interpret it as some particular evaluation or operator insertion of the boundary coherent spin network state. We postpone such a detailed analysis to future investigation.

\subsection{Cosmological Dynamics with Holomorphic Simplicity Constraints}

Let us apply our method to analyze the spinfoam amplitude for the Dupuis-Livine (DL) spinfoam model for Riemannian quantum gravity based on holomorphic simplicity constraints. We will not review the details and the definition of this model, but we will refer the interested reader to \cite{spinor1}  for the full presentation or to \cite{proceeding} for a quick summary of the model's basic features. We will only point out here that the resulting spinfoam amplitudes are very similar  to the ones of the EPRL-FK spinfoam models \cite{EPRL,FK}, except that the diagonal simplicity constraints are not strongly imposed and that the wave-packets for the boundary states are thus slightly enlarged Gaussian distributions compared to the coherent states used for the EPRL-FK  models. This allows for an easier analysis and understanding of the symmetries, constraints and amplitudes of the model.

The DL model is based on the gauge group $\Spin(4)=\SU(2)_L\times\SU(2)_R$ made of two copies of $\SU(2)$. The classical phase space on a fixed graph is therefore simply two copies of the $\SU(2)$ spinor network phase space and the boundary spin networks at the quantum level are simply tensor products of spin network states for the left copy of $\SU(2)$ and the right copy. The the holomorphic simplicity constraints are easily imposed at both the classical level and the quantum level. The resulting structure is coherent spin network states labeled by two sets of spinors living on the graph, the $z^{L,R}$'s respectively from the left and right sectors:
\be
\Psi_{\{z^L{}^v_e,z^R{}^v_e\}}(g^L_e,g^R_e)
\,=\,
\psi_{\{z^L{}^v_e\}}(g^L_e)\,\psi_{\{z^R{}^v_e\}}(g^R_e)\,,
\ee
where $(g^L,g^R)\in\SU(2)_L\times\SU(2)_R$ represents a $\Spin(4)$ group element. Then the holomorphic simplicity constraints simply impose that all the right spinors are equal to their left counterpart up to a fixed factor, $\forall v,e,\,\,z^R{}^v_e=\rho\,z^L{}^v_e$ where $\rho>0$ is related to the Immirzi parameter \cite{spinor1}. Therefore, removing the useless subscript $L$ or $R$ for the spinor variables, the DL spinfoam model is based on the coherent spin network states for $\Spin(4)$ defined as:
\be
\Psi_{\{z^v_e\}}(g^L_e,g^R_e)
\,=\,
\psi_{\{z^v_e\}}(g^L_e)\,\psi_{\{\rho z^v_e\}}(g^R_e)\,.
\ee

Repeating our derivation of the spinfoam amplitude for the 2-vertex graph and $\Delta_1$ 2-complex in the bulk, we obtain once more a decoupled spinfoam amplitude which is the product of the initial and final spin network evaluations:
$$
\cA_{\Delta_1}^{DL}[z_i,w_i,\tz_i,\tw_i]
\,=\,
\cW^{DL}(z_i,w_i)
\cW^{DL}(\tz_i,\tw_i),
$$
\be
\textrm{with}\quad
\cW^{DL}(z_i,w_i)=
\Psi_{\{z_i,w_i\}}(\id)
=\psi_{\{z_i,w_i\}}(\id)\psi_{\{\rho z_i,\rho w_i\}}(\id)
=\la \{\varsigma w_i\}|\{z_i\}\ra\la \{\varsigma \rho w_i\}|\{\rho z_i\}\ra
\,.
\ee
Since these scalar products are known analytically, we can compute exactly these amplitudes. Nevertheless, it is simpler to have a look at the asymptotic behavior. Using the stationary point approximation derived earlier \eqref{Gauusianapprox}, we get for the homogeneous and isotropic sector:
\be
\cW^{DL}(z)\sim
\f4{z^3\sqrt{\pi}}\,e^{\f{z^2}2}\,\f4{(\rho z)^3\sqrt{\pi}}\,e^{\f{(\rho z)^2}2}
=
\f{4^2}{z^6\rho^3\pi}\,e^{(1+\rho^2)\f{z^2}2}\,.
\ee
This Gaussian wave-packet is once more solution to a 2nd order differential equation whose leading order is of the type $\pp_z^2-z^2$ (up to some global power of $z$). Considering the large scale regime, we obtain  a classical constraint identical to the BF theory case corresponding to the flat FRW cosmology with no dependence (at leading order) on the Immirzi parameter $\rho$.

This confirms that we should study more complicated bulk configurations or change the spinfoam amplitude ansatz for the simple single-vertex bulk $\Delta_1$ in order to obtain some effective dynamics for more generic cosmology, with possibly curvature and cosmological constant.


\section{Conclusion and Outlook}

We have discussed the general procedure of truncating loop quantum gravity and spinfoams to a fixed finite graph and studying the dynamics of the geometry on that fixed graph at both classical and quantum levels.
We distinguish our procedure from the standard approach of loop quantum cosmology, where one implements the symmetry reduction to the relevant homogeneous and isotropic 4-metrics directly at the level of the classical phase space in the continuum and then quantizes the resulting reduced cosmological phase space. Instead, here we start with the finite phase space of loop gravity on a fixed finite graph and investigate if there exists a similar symmetry reduction to a homogeneous and isotropic sector which could then be interpreted as a cosmological sector. 

The goal is to define mini-superspace models from loop quantum gravity that can be interpreted as cosmological models or as toy models to investigate and test the spinfoam dynamics.
Following this logic, we have studied in detail the kinematics and dynamics of loop gravity on a
2-vertex graph. We have defined the dynamics from both the canonical point of view and using the
covariant spinfoam amplitudes and showed the consistence between the two approaches. We have further defined the reduction of a homogeneous and isotropic sector at the kinematical level and showed that the dynamics of this sector can be understood as some modified effective Friedmann-Robertson-Walker (FRW) cosmology. This confirms and improves the earlier results about the 2-vertex model derived in \cite{LQGcosmo1,SFcosmo,un3,un5}.
We show in particular that the Rovelli-Vidotto approach of \cite{LQGcosmo1} of discretizing the loop gravity Hamiltonian constraint on the 2-vertex graph and the strategy presented in \cite{un3,un5} aiming at implementing isotropy by a $\U(N)$-symmetry reduction on the 2-vertex phase space actually converge to the same proposal for a Hamiltonian constraint for the 2-vertex model. 
These results show that the present approach might be very promising to derive mini-superspace models from spinfoam models and to study the quantum gravity dynamics and fluctuations in these restricted settings.

Concerning the interpretation of the 2-vertex model as a cosmological mini-superspace model, the
limitations of our approach are clear. First, we absolutely need to include matter in our analysis,
both at the canonical level (following the earlier work \cite{LQGcosmo1}) and in the spinfoam
calculations, in order to obtain a realistic FRW model. Second, we have identified and explained
some discrepancies in the large scale behavior of our 2-vertex dynamics with the standard FRW
dynamics when we turn on the curvature $k$ or the cosmological constant $\Lambda$. This shows the
limits of our current approach.

Moreover, concerning the derivation of the 2-vertex cosmology from spinfoam amplitudes, we also see
two main issues. On the one hand, we have discussed the effects that curvature and
cosmological constant cause on the spinfoam transition amplitude, but we still need to understand
how to encode them in the coherent spin networks living in the boundary of the two-complex.
On the other hand, we need to study the
transition amplitudes for more complicated bulk structures, beyond our calculations for a simple one
4-cell bulk. This involves studying more the spinfoam amplitudes and their
coarse-graining/renormalization in order to understand how the transition amplitudes for more
complex bulks look like and to determine if the one 4-cell bulk is actually the dominating
contribution. Both issues are intertwined since we expect that considering more complicated bulks
would (at least) induce curvature. Actually, for the simplest 2-complex that we have considered,
the boundary of the spinfoam vertex coincides with the boundary of the 2-complex, and then the
transition amplitude is just the evaluation of the boundary spin network on the identity (according
with the spinfoam ansatz), meaning that we can only describe a flat model (BF theory) in this simple
setting.
Nevertheless, besides these questions to solve, we have shown how to derive
differential equations satisfied by spinfoam amplitudes and how they are equivalent to the
Hamiltonian constraint, in the context of the 2-vertex graph. We hope that this method will be
generalizable to more complicated graphs and bulk structures.

Beyond these issues to solve, we would like to point out three promising directions of work. First,
we would like to investigate the freedom in modifying the one 4-cell  spinfoam amplitude and study
the resulting dynamics and corresponding Hamiltonian. Second, we would like to apply our methods to
more complicated graphs, for example the 3+N-vertex graph proposed in \cite{un3} or possibly graphs
with an infinite number of vertices, in order to test the validity of truncating loop (quantum)
gravity to a fixed graph beyond the simplest 2-vertex graphs and to try to generate more realistic
cosmological models, beyond the  homogeneous and isotropic FRW cosmology. For instance, we hope to
identify some simple family of graphs with a number of vertices that can be sent to infinity in
order to model midi-superspace models for cosmology with inhomogeneities.
Finally, it could be interesting to develop some simplified group field theory for our 2-vertex
graph model, which would admit specific boundary graph and would sum over a restricted set of
2-complexes, and compare it to the group field theory for the EPRL-FK spinfoam model \cite{EPRL_GFT}
and to the spinfoam amplitudes for loop quantum cosmology \cite{GFTcosmo1,GFTcosmo2,GFTcosmo3}.

\section*{Acknowledgments}

EL is partially supported by the ANR ``Programme Blanc" grants LQG-09, and MMB by the
Spanish MICINN Projects No. FIS2008-06078-C03-03 and No. FIS2011-30145-C03-02.
The present work was also supported by the ESF Quantum Geometry and Quantum Gravity Network through
the travel grants 3595 and 3770 for EL and 3048 and 3939 for MBM.

We thank the Perimeter Institute for Theoretical Physics (Waterloo, Ontario, Canada) for its hospitality during the last stages of our work.

\appendix

\section{Spinors and Notations}

In this appendix, we introduce spinors and the related useful notations, following the previous
works \cite{un3,un4,twisted2}.
Considering a spinor $z$,
$$
|z\ra=\mat{c}{z^0\\z^1}, \qquad
\la z|=\mat{cc}{\bar{z}^0 &\bar{z}^1},
$$
we associate to it a geometrical 3-vector $\vec{V}(z)$, defined from the projection of the $2\times 2$ matrix $|z\ra\la z|$ onto Pauli matrices $\sigma_a$ (taken Hermitian and normalized so that $(\sigma_a)^2=\id$):
\be \label{vecV}
|z\ra \la z| = \f12 \left( {\la z|z\ra}\id  + \vec{V}(z)\cdot\vec{\sigma}\right).
\ee
The norm of this vector is obviously $|\vec{V}(z)| = \la z|z\ra= |z^0|^2+|z^1|^2$  and its components are given explicitly as:
\be
V^z=|z^0|^2-|z^1|^2,\qquad
V^x=2\,\Re\,(\bar{z}^0z^1),\qquad
V^y=2\,\Im\,(\bar{z}^0z^1).
\ee
The spinor $z$ is entirely determined by the corresponding 3-vector $\vec{V}(z)$ up to a global phase. We can give the reverse map:
\be
z^0=e^{i\phi}\,\sqrt{\f{|\vec{V}|+V^z}{2}},\quad
z^1=e^{i(\phi-\theta)}\,\sqrt{\f{|\vec{V}|-V^z}{2}},\quad
\tan\theta=\f{V^y}{V^x},
\ee
where $e^{i\phi}$ is an arbitrary phase.

Following \cite{un3}, we also introduce the map duality $\varsigma$ acting on spinors:
\be
\varsigma\mat{c}{z^0\\ z^1}
\,=\,
\mat{c}{-\bar{z}^1\\\bar{z}^0},
\qquad \varsigma^{2}=-1.
\ee
This is an anti-unitary map, $\la \varsigma z| \varsigma w\ra= \la w| z\ra=\overline{\la z| w\ra}$, and we will write the related state as
$$
|z]\equiv \varsigma  | z\ra,\qquad
[z| w]\,=\,\overline{\la z| w\ra}.
$$
This map $\varsigma$ maps the 3-vector $\vec{V}(z)$ onto its opposite:
\be
|z][  z| = \f12 \left({\la z|z\ra}\id - \vec{V}(z)\cdot\vec{\sigma}\right).
\ee

Finally considering the setting necessary to describe intertwiners with $N$ legs, we consider $N$ spinors $z_i$ and their corresponding 3-vectors $\vV(z_i)$.
Typically, we can require that the $N$ spinors satisfy a closure
condition, i.e  that the sum of the corresponding 3-vectors
vanishes, $\sum_i \vec{V}(z_i)=0$. Coming back to the definition of
the 3-vectors $\vV(z_i)$, the closure condition is easily translated
in terms of $2\times 2$ matrices:
\be
\sum_i |z_i\ra \la z_i|=A(z)\id,
\qquad\textrm{with}\quad
A(z)\equiv\f12\sum_i \la z_i|z_i\ra=\f12\sum_i|\vec{V}(z_i)|.
\ee
This further translates into quadratic constraints on the spinors:
\be
\sum_i z^0_i\,\bar{z}^1_i=0,\quad
\sum_i \left|z^0_i\right|^2=\sum_i \left|z^1_i\right|^2=A(z).
\ee
In simple terms, it means that the two components of the spinors, $z^0_i$ and $z^1_i$, are orthogonal $N$-vectors of equal norm.

\section{From Spinor Networks to Twisted Geometry}
\label{twisted_app}

Another way to parameterize the $\U(1)$-invariant phase space is to use the twisted geometry variables developed by Freidel and Speziale \cite{twisted1,twisted2}, which are particularly relevant to building semi-classical coherent spin network states for spinfoam models.

The starting point of this approach is to define a spinor $z$ through the unique group element which maps the ``origin" spinor $\Omega\equiv\,\mat{c}{1\\ 0}$ onto the normalized spinor $z/\sqrt{\la z|z\ra}$. Then we decompose that group element as the product of a rotation along the $z$-axis and a rotation with axis in the $(0xy)$ plane:
\be
|z\ra
\,=\,
\sqrt{\la z|z\ra}\,n(Z)\,e^{i\phi\sigma_3}\,|\Omega\ra
\,=\,
e^{i\phi}\sqrt{\la z|z\ra}\,n(Z)\,|\Omega\ra
\,=\,
\f{e^{i\phi}\sqrt{\la z|z\ra}}{\sqrt{1+|Z|^2}}\,
\mat{c}{1 \\ Z}
\qquad
n(Z)=\f{1}{\sqrt{1+|Z|^2}}\,\mat{cc}{1 & -\bZ \\ Z & 1},
\ee
where $n(Z)\in\SU(2)$ defines a rotation with axis in the $(0xy)$ plane\footnotemark.
\footnotetext{The $\SU(2)$ group element $n(Z)$ is a rotation $e^{\alpha\,\hat{m}}$ with $\hat{m}\cdot \hat{e}_z=0$ and  $|\hat{m}|^2=1$. Indeed we can expand this group element as:
$$
e^{\theta\,\hat{m}}=\cos\theta\id+i\sin\theta\mat{cc}{1 & \overline{m} \\ m & 1},
\quad m=m_x+im_y,\quad |m|^2=1,
$$
which matches $n(Z)$ for $Z=i\tan\theta\,m$.
}
This means that we have parameterized the spinor $z$ with four real numbers $\phi,\la z|z\ra\in\R^2$ and $Z\in\C$, so there shouldn't be any (continuous) redundancy in this parametrization. We can furthermore easily check that the map is one-to-one, since we can invert this definition:
$$
Z=\f{z_1}{z_0},
\qquad
e^{i\phi}=\f{z_0}{|z_0|},
\qquad
\la z|z\ra=z_0^2+z_1^2
$$
The complex variable $Z$ defines the direction of the 3-vector $\vV$. Indeed, we have:
$$
|z\ra\la z|
\,=\,
\la z|z\ra\,\, n(Z)\,|\Omega\ra\la\Omega|\,n(Z)^{-1},
$$
which is straightforward to translate into 3-vectors:
\beq
|z\ra\la z| &=& \f12\left(|V|\id+\vV\cdot\vsigma\right), \nn\\
|\Omega\ra\la \Omega| &=&
\f12\left(\id+\sigma_3\right)
=\f12\left(\id+\hat{e}_z\cdot\vsigma\right), \nn
\eeq
Thus we simply have $\vV/|\vV|=n(Z)\vartriangleright \hat{e}_z$. Finally, it is also useful to
write explicitly the dual spinor $|z]$ in terms of the same variables. Using that $\eps h\eps^{-1}
=\overline{h}$ for all $\SU(2)$ group elements $h$ and in particular for $h=n(Z)$, we easily
get\footnotemark:
\be
\f{|z]}{\sqrt{\la z|z\ra}}
=\f{\eps|\bz\ra}{\sqrt{\la z|z\ra}}
\,=\,
e^{-i\phi}n(Z)\eps\,|\Omega\ra
\,=\,
n(Z)\eps \,e^{-i\phi\sigma_3}\,|\Omega\ra
\,=\,
n(Z)\,e^{+i\phi\sigma_3}\,|\tOmega\ra,
\quad
\textrm{with}
\quad
\tOmega\equiv|\Omega]=\eps\Omega=\mat{c}{0 \\ 1}.
\ee
\footnotetext{
%
We can actually re-absorb the $\eps$ factor in the  $n(Z)$ group element itself by switching from $Z$ to $-1/\bZ$. After a few algebraic manipulations, we get:
$$
n\left(-\f1\bZ\right)
\,=\,
n(Z) \eps^{-1} e^{i\theta\sigma_3},
\qquad
\textrm{with}
\quad
\theta\,=\,\textrm{Arg}\,Z\,.
$$
Thus, we can re-write the dual spinor as:
$$
|z]
\,=\,
-\sqrt{\la z|z\ra}\,e^{-i(\phi+\theta)}\,n\left(-\f1\bZ\right)\,|\Omega\ra
\,=\,
-\sqrt{\la z|z\ra}\,n\left(-\f1\bZ\right)\,e^{-i(\phi+\theta)\sigma_3}\,|\Omega\ra\,.
$$
}

\medskip

Applying these definitions to all the spinors $z_e^{s,t}$, we can write all the functions over the $\C^{4E}$ phase space in terms of the new variables $\phi^{s,t}_e$ and $Z^{s,t}_e$.
For instance, we can express the holonomy variables $g_e$ in terms of these twisted geometry variables. The group element $g_e$ sends $|z^s_e\ra$ back to the origin $\Omega$ and then maps it to $|z^t_e]$:
\be
g_e
\,=\,
n(Z^t_e)\,\eps\,e^{-i\phi^t_e\sigma_3}\,e^{-i\phi^s_e\sigma_3}\,n(Z^s_e)^{-1}
\,=\,
n(Z^t_e)\,\eps\,e^{-i\xi_e\sigma_3}\,n(Z^s_e)^{-1},
\qquad\textrm{with}\quad
\xi_e\equiv\,(\phi^t_e+\phi^s_e)
\,,
\ee
where the variables $Z^{s,t}_e$ and $\xi_e$ are all $\U(1)$-invariant.

Finally, we can write the whole action in terms of the twisted geometry variables. First, we compute the kinematical term for a single spinor, $\la z|\pp_t z\ra$ , in terms of the new variables:
$$
\la z|\pp_t z\ra
\,=\,
i\la z| z\ra\pp_t \phi
\,+\,
\la z| z\ra\,\la\Omega|n(Z)^{-1}\pp_tn(Z)|\Omega\ra
\,+\,
\f12\pp_t\la z| z\ra\,.
$$
We further compute explicitly the derivative term in the variable complex $Z\in\C$~\footnotemark:
\be
\la\Omega|n(Z)^{-1}\pp_tn(Z)|\Omega\ra
\,=\,
\f{\bZ\pp_t Z - Z\pp_t \bZ}{2(1+|Z|^2)}
\,=\,
i\,\f{|Z|^2}{(1+|Z|^2)}\,\pp_t\theta,
\qquad
\textrm{with}
\quad
\theta\,=\,\textrm{Arg}\,Z\,.
\ee
\footnotetext{
This hints that there might be a more convenient choice of complex variable $\cz=\cR e^{i\theta}$ instead of $Z=Re^{i\theta}$ such that:
$$
\cR=\f{R}{\sqrt{1+R^2}}\in\,[0,1[\,,\qquad
n(Z)=\mat{cc}{\sqrt{1-|\cz|^2} & -\bcz \\ \cz & \sqrt{1-|\cz|^2}}\,.
$$
We can parameterize both variables using a ``boost" parameter, thus writing $R=\sinh\eta$ and $\cR=\tanh\eta$.
Finally the derivative term in $Z$ simply reduces to:
$$
\la\Omega|n(Z)^{-1}\pp_tn(Z)|\Omega\ra
\,=\,
i\,\cR^2\,\pp_t\theta
\,=\,
\f12(\bcz\pp_t \cz - \cz\pp_t\bcz)\,.
$$
We can even go further and re-absorb the area factor $A=\la z|z\ra$ into the definition of the complex variable. Indeed since $A$ is real, it factors out of the anti-symmetric combination $\bcz\pp_t \cz - \cz\pp_t\bcz$ which is only sensitive to the argument of $\cz$:
$$
\sqrt{A}\bcz\pp_t(\sqrt{A} \cz) - \sqrt{A}\cz\pp_t(\sqrt{A}\bcz)
\,=\,
A(\bcz\pp_t \cz - \cz\pp_t\bcz)\,.
$$
}
Applying this formula for $z=z^{s,t}_e$ and discarding the total derivatives, we obtain:
\be
S^{(0)}[z^v_e]
\,=\,
\int dt\,
\sum_e A_e\pp_t\xi_e
+A_e\left(
\f{|Z_e^s|^2}{(1+|Z_e^s|^2)}\,\pp_t\theta_e^s
\,+\,
\f{|Z_e^t|^2}{(1+|Z_e^t|^2)}\,\pp_t\theta_e^t
\right)
\,+\,
(\textrm{closure constraints}),
\ee
where $A_e\equiv\,\la z^s_e|z^s_e\ra=\la z^t_e|z^t_e\ra$ denotes the area dual to the edge $e$.
For the details of the Poisson algebra of the twisted geometry variables  $A_e,\xi_e,|Z^{s,t}_e|,\theta^{s,t}_e$, the interested reader can find a thorough analysis in \cite{twisted1,twisted2}. Let us just point out that although the variables $A_e$ are $\SU(2)$-invariant, all the remaining variables $\xi_e,|Z^{s,t}_e|,\theta^{s,t}_e$ have non-trivial Poisson brackets with the closure constraints and are thus not $\SU(2)$-invariant. In order to get $\SU(2)$-invariants from the spinor variables or the twisted geometry variables, you can use cross-ratio observables as introduced in \cite{holotetra}, but we will not go into the details of this construction in the present work.


\begin{thebibliography}{99}
\bibitem{lqg}
Thiemann T., Modern canonical quantum general relativity, Cambridge University Press, Cambridge,
2007;\\
Rovelli C., Quantum gravity, Cambridge University Press, Cambridge, 2004;\\
Ashtekar A.\ and Lewandowski J., Background independent quantum gravity: a status
report, {\it Classical Quantum Gravity} {\bf 21} (2004), R53 [arXiv:gr-qc/0404018]

\bibitem{SFreview}
A. Perez,
{\it The Spin Foam Approach to Quantum Gravity },
arXiv:1205.2019;\\
A. Perez,
{\it The new spin foam models and quantum gravity},
arXiv:1205.0911;\\
E.R. Livine,
{\it A Short and Subjective Introduction to the Spinfoam Framework for Quantum Gravity},
arXiv:1101.5061; \\
C. Rovelli,
{\it Zakopane lectures on loop gravity},
arXiv:1102.3660; \\
A. Perez,
{\it Introduction to Loop Quantum Gravity and Spin Foams},
arXiv:gr-qc/0409061

\bibitem{EPRL}
J. Engle, E.R. Livine, R. Pereira and C. Rovelli,
{\it LQG vertex with finite Immirzi parameter},
Nucl.Phys.B799 (2008) 136-149

\bibitem{FK}
L. Freidel and K. Krasnov,
{\it A New Spin Foam Model for 4d Gravity},
Class.Quant.Grav.25 (2008) 125018

\bibitem{LQCreview}
A. Ashtekar and P. Singh,
{\it Loop Quantum Cosmology: A Status Report},
Class. Quant. Grav. 28 (2011) 213001 [arXiv:1108.0893];\\
M. Bojowald and R. Tavakol,
{\it Loop Quantum Cosmology: Effective theories and oscillating universes},
arXiv:0802.4274;\\
M. Bojowald,
{\it Loop Quantum Cosmology},
Living Rev.Rel.8 (2005) 11 [arXiv:gr-qc/0601085]

\bibitem{LQGcosmo1}
C. Rovelli and F. Vidotto,
{\it Stepping out of Homogeneity in Loop Quantum Cosmology},
Class.Quant.Grav.25 (2008) 225024 [arXiv:0805.4585]

\bibitem{LQGcosmo2}
M.V. Battisti, A. Marciano and C. Rovelli,
{\it Triangulated Loop Quantum Cosmology: Bianchi IX and inhomogenous perturbations},
Phys.Rev.D81 (2010) 064019 [arXiv:0911.2653]

\bibitem{SFcosmo}
E. Bianchi, C. Rovelli and F. Vidotto,
{\it Towards Spinfoam Cosmology},
Phys.Rev.D82 (2010) 084035 [arXiv:1003.3483]

\bibitem{un3}
E.F. Borja, J. Diaz-Polo, I. Garay and E.R. Livine
{\it Dynamics for a 2-vertex Quantum Gravity Model},
Class.Quant.Grav. 27 (2010) 235010 [arXiv:1006.2451]

\bibitem{un5}
E. Borja, L. Freidel, I. Garay and E.R. Livine,
{\it U(N) tools for Loop Quantum Gravity: The Return of the Spinor},
Class.Quant.Grav. 28 (2011) 055005 [arXiv:1010.5451]

\bibitem{SFcosmoLambda}
Eugenio Bianchi, Thomas Krajewski, Carlo Rovelli, Francesca Vidotto,
{\it Cosmological constant in spinfoam cosmology},
Phys.Rev.D83 (2011) 104015 [arXiv:1101.4049]

\bibitem{lqc_def}
M. Bojowald and H.A. Kastrup,
{\it Quantum Symmetry Reduction for Diffeomorphism Invariant Theories of Connections},
 	Class.Quant.Grav. 17 (2000) 3009-3043 [arXiv:hep-th/9907042]

\bibitem{italian}
F. Cianfrani and G. Montani,
{\it A critical analysis of the cosmological implementation of Loop Quantum Gravity},
Mod. Phys. Lett. A27 (2012) 7, 1250032 [arXiv:1201.2329]; \\
F. Cianfrani, A. Marchini and G. Montani,
{\it The picture of the Bianchi I model via gauge fixing in Loop Quantum Gravity},
Europhys. Lett. 99 (2012) 10003 [arXiv:1201.2588]

\bibitem{twisted2}
L. Freidel and S. Speziale,
{\it From twistors to twisted geometries},
Phys.Rev.D82 (2010) 084041 [arXiv:1006.0199]

\bibitem{spinor1}
M. Dupuis and E.R. Livine,
{\it Holomorphic Simplicity Constraints for 4d Spinfoam Models},
arXiv:1104.3683

\bibitem{spinor2}
E.R. Livine and J. Tambornino,
{\it Spinor Representation for Loop Quantum Gravity},
arXiv:1105.3385

\bibitem{twisted1}
L. Freidel and S. Speziale,
{\it Twisted geometries: A geometric parametrisation of SU(2) phase space},
Phys.Rev.D82 (2010) 084040 [arXiv:1001.2748]

\bibitem{un1}
L. Freidel and E.R. Livine,
{\it The Fine Structure of SU(2) Intertwiners from U(N) Representations},
J.Math.Phys. 51 (2010) 082502 [arXiv:0911.3553]

\bibitem{un2}
L. Freidel and E.R. Livine,
{\it U(N) Coherent States for Loop Quantum Gravity},
J.Math.Phys. 52 (2011) 052502 [arXiv:1005.2090]

\bibitem{un4}
M. Dupuis and E.R. Livine,
{\it Revisiting the Simplicity Constraints and Coherent Intertwiners},
Class.Quant.Grav. 28 (2011) 085001 [arXiv:1006.5666]

\bibitem{SFrecursion_simone}
V. Bonzom, E.R. Livine and S. Speziale,
{\it Recurrence relations for spin foam vertices},
Class.Quant.Grav.27 (2010) 125002 [arXiv:0911.2204]

\bibitem{SFrecursion_valentin}
V. Bonzom and L. Freidel,
{\it The Hamiltonian constraint in 3d Riemannian loop quantum gravity},
Class.Quant.Grav.28 (2011) 195006 [arXiv:1101.3524];\\
V. Bonzom,
{\it Spin foam models and the Wheeler-DeWitt equation for the quantum 4-simplex},
Phys.Rev.D84 (2011) 024009 [arXiv:1101.1615]

\bibitem{SFrecursion_final}
V. Bonzom and E.R. Livine,
{\it A new Hamiltonian for the Topological BF phase with spinor networks},
arXiv:1110.3272

\bibitem{un0}
F. Girelli and E.R. Livine,
{\it Reconstructing Quantum Geometry from Quantum Information: Spin Networks as Harmonic
Oscillators},
Class.Quant.Grav. 22 (2005) 3295-3314 [arXiv:gr-qc/0501075]

\bibitem{polyhedron}
E. Bianchi, P. Dona and S. Speziale,
{\it Polyhedra in loop quantum gravity},
Phys.Rev.D83 (2011) 044035 [arXiv:1009.3402]

\bibitem{gluing_bianca}
B. Dittrich and J.P. Ryan,
{\it Phase space descriptions for simplicial 4d geometries},
Class.Quant.Grav.28 (2011) 065006 [arXiv:0807.2806]


\bibitem{AQG}
K. Giesel and T. Thiemann,
{\it Algebraic Quantum Gravity (AQG) I. Conceptual Setup},
Class.Quant.Grav.24 (2007) 2465-2498 [arXiv:gr-qc/0607099]

\bibitem{new_laurent}
L. Freidel, M. Geiller and J. Ziprick,
{\it Continuous formulation of the Loop Quantum Gravity phase space},
arXiv:1110.4833

\bibitem{noncompact}
L. Freidel and E.R. Livine,
{\it Spin Networks for Non-Compact Groups},
J.Math.Phys.44 (2003) 1322-1356 [arXiv:hep-th/0205268]

\bibitem{vmr}
M. V. Battisti, A. Marciano, C. Rovelli
{\it Triangulated Loop Quantum Cosmology: Bianchi IX and inhomogenous perturbations},
Phys.Rev.D81 (2010) 064019 [arXiv:0911.2653]



\bibitem{aps1} A. Ashtekar, T. Paw\-lowski, and P. Singh,
{\it Quantum nature of the big bang},
Phys.Rev.Lett.96 (2006) 141301 [arXiv:gr-qc/0602086];
{\it Quantum Nature of the Big Bang: An Analytical and Numerical Investigation},
Phys. Rev. D{73} (2006) 124038 [arXiv:gr-qc/0604013]

\bibitem{aps3} A. Ashtekar, T. Paw\-lowski, and P. Singh,
{\it Quantum Nature of the Big Bang: Improved dynamics},
Phys.Rev.D74 (2006) 84003 [arXiv:gr-qc/0607039]

\bibitem{tom} E. Bentivegna and T. Paw\-lowski,
{\it Anti-deSitter universe dynamics in LQC},
Phys. Rev. D77, (2008) 124025 [arXiv:0803.4446]

\bibitem{vand} K. Vandersloot,
{\it Loop quantum cosmology and the k = - 1 FRW model},
Phys. Rev. D75 (2007) 23523 [arXiv:gr-qc/0612070]

\bibitem{luc} L.Szulc,
{\it Open FRW model in Loop Quantum Cosmology},
Class.Quant.Grav.24 (2007) 6191 [arXiv:0707.1816]

\bibitem{apsv} A. Ashtekar, T. Paw\-lowski, P. Singh, and K. Vandersloot,
{\it Loop quantum cosmology of k=1 FRW models},
Phys.Rev.D75 (2007) 024035 [arXiv:gr-qc/0612104]

\bibitem{skl} L. Szulc, W. Kami\'nski, and J. Lewandowski,
{\it Closed FRW model in Loop Quantum Cosmology},
Class.Quant.Grav.24 (2007) 2621 [arXiv:gr-qc/0612101]

\bibitem{abl} A. Ashtekar, M. Bojowald, and J. Lewandowski,
{\it Mathematical structure of loop quantum cosmology},
Adv.Theor.Math.Phys.7 (2003) 233 [arXiv:gr-qc/0304074]


\bibitem{bck} M. Bojowald, {\it Loop quantum cosmology and inhomogeneities},
Gen. Rel. Grav. 38 (2006) 1771 [arXiv:gr-qc/0609034]

\bibitem{LS}
E.R. Livine and S. Speziale,
{\it A new spinfoam vertex for quantum gravity},
Phys.Rev.D76 (2007) 084028 [arXiv:0705.0674]

\bibitem{gftreview_razvan}
R. Gurau and J. Ryan,
{\it Colored Tensor Models - a review},
arXiv:1109.4812

\bibitem{gftreview_daniele}
D. Oriti,
{\it The microscopic dynamics of quantum space as a group field theory},
arXiv:1110.5606

\bibitem{EPR}
J. Engle, R. Pereira and C. Rovelli,
{\it The loop-quantum-gravity vertex-amplitude},
Phys.Rev.Lett.99 (2007) 161301 [arXiv:0705.2388]


\bibitem{LS2}
E.R. Livine and S. Speziale,
{\it Consistently Solving the Simplicity Constraints for Spinfoam Quantum Gravity},
Europhys.Lett.81 (2008) 50004



\bibitem{proceeding}
M. Dupuis and E.R. Livine,
{\it Holomorphic Simplicity Constraints for 4d Riemannian Spinfoam Models},
Proceedings of the  Loops'11 conference (May 2011, Madrid, Spain) [arXiv:1111.1125]

\bibitem{EPRL_GFT}
J. Ben Geloun, R. Gurau and V. Rivasseau,
{\it EPRL/FK Group Field Theory},
Europhys.Lett.92 (2010) 60008 [arXiv:1008.0354]

\bibitem{GFTcosmo1}
A. Ashtekar, M. Campiglia and A. Henderson,
{\it Casting Loop Quantum Cosmology in the Spin Foam Paradigm},
Class.Quant.Grav.27 (2010) 135020 [arXiv:1001.5147]

\bibitem{GFTcosmo2}
A. Henderson, C. Rovelli, F. Vidotto and E. Wilson-Ewing,
{\it Local spinfoam expansion in loop quantum cosmology},
Class.Quant.Grav.28 (2011) 025003 [arXiv:1010.0502]

 \bibitem{GFTcosmo3}
G. Calcagni, S. Gielen and D. Oriti,
{\it Two-point functions in (loop) quantum cosmology}, Class.Quant.Grav.28 (2011) 125014
[arXiv:1011.4290]
%
%

\bibitem{frank}
F. Hellmann,
{\it On the Expansions in Spin Foam Cosmology},
Phys. Rev. D 84, 103516 (2011) [arXiv:1105.1334]

\bibitem{holotetra}
L. Freidel, K. Krasnov and E.R Livine,
{\it Holomorphic Factorization for a Quantum Tetrahedron},
Commun.Math.Phys.297 (2010) 45-93 [arXiv:0905.3627]


\end{thebibliography}
\end{document}